\documentclass[twocolumn]{aastex631}

\usepackage{inputenc}
\usepackage{comment}
\usepackage{colortbl}
\usepackage{xspace}
\usepackage{booktabs}
\usepackage{longtable}

\newcommand{\Swift}{\textit{Swift}\xspace}
\newcommand{\Fermi}{\textit{Fermi}\xspace}

\newcommand{\lcolhead}[1]{\makebox[0pt][l]{#1}}

\makeatletter
\renewcommand*\tablecomments[1]{%
  \par\vspace{5pt}\noindent
  #1\par
}
\makeatother

\received{2025 September 30}
\revised{2025 December 5}
\accepted{2025 December 8}

\begin{document}

\title{Classification of a New X-ray Catalog of Likely Counterparts to 4FGL-DR4 Unassociated Gamma-ray Sources Using a Neural Network}

\author[0000-0002-2701-8433]{Kyle D. Neumann}
\affiliation{Department of Astronomy and Astrophysics \\
 Pennsylvania State University,
University Park, PA 16802, USA}

\author[0000-0002-5068-7344]{Abraham D. Falcone}
\affiliation{Department of Astronomy and Astrophysics \\
 Pennsylvania State University,
University Park, PA 16802, USA}

\author[0000-0003-2633-2186]{Stephen DiKerby}
\affiliation{Department of Physics and Astronomy \\
 Michigan State University,
East Lansing, MI 42882, USA}

\author[0009-0004-5158-4322]{Sierra Deppe}
\affiliation{Department of Astronomy and Astrophysics \\
 Pennsylvania State University,
University Park, PA 16802, USA}

\author[0000-0001-7828-7708]{Elizabeth C. Ferrara}
\affiliation{Department of Astronomy \\ University of Maryland, College Park, MD 20742, USA}
\affiliation{Center for Research and Exploration in Space Science and Technology (CRESST) \\
NASA Goddard Space Flight Center, Greenbelt, MD 20771, USA}
\affiliation{Astrophysics Science Division \\
NASA Goddard Space Flight Center, Greenbelt, MD 20771, USA}

\author{Jamie A. Kennea}
\affiliation{Department of Astronomy and Astrophysics \\
 Pennsylvania State University,
University Park, PA 16802, USA}

\author[0000-0003-1673-970X]{Brad Cenko}
\affiliation{Astrophysics Science Division \\ 
NASA Goddard Space Flight Center, Greenbelt, MD 20771, USA}

\author{Eric Grove}
\affiliation{U.S. Naval Research Laboratory \\ 
4555 Overlook Ave., SW Washington, DC 20375, USA}



\begin{abstract}

Our survey of the fourth \Fermi Large Area Telescope catalog (4FGL) unassociated gamma-ray source regions using the X-Ray Telescope (XRT) and Ultraviolet/Optical Telescope (UVOT) aboard the Neil Gehrels \Swift Observatory (\Swift) provides new XRT and UVOT source detections and localizations to help identify potential low-energy counterparts to unassociated \Fermi gamma-ray sources. We present a catalog of 218 singlet and 70 multiplet \Swift X-ray sources detected within the positional uncertainty ellipses of 244 unassociated \Fermi gamma-ray sources from the 4FGL-DR4 catalog, 144 of which are not previously cataloged by \citet{kerby21b}. For each X-ray source, we derive its X-ray flux and photon index, then use simultaneous UVOT observations with optical survey data to estimate its $V$-band magnitude. We use these parameters as inputs for a multi-layer perceptron (MLP) neural network classifier (NNC) trained to classify sources as blazars, pulsars, or ambiguous gamma-ray sources. For the 213 singlet sources with X-ray and optical data, we classify 173 as likely blazars ($P_\mathrm{bzr} > 0.99$) and 6 as likely pulsars ($P_\mathrm{bzr} < 0.01$), with 34 sources yielding ambiguous results. Including 70 multiplet X-ray sources, we increase the number of $P_\mathrm{bzr}>0.99$ to 227 and $P_\mathrm{bzr}<0.01$ to 16. For the subset of these classifications that have been previously studied, a large majority agree with prior classifications, supporting the validity of using this NNC to classify the unknown and newly detected gamma-ray sources.

\end{abstract}


\section{Introduction} \label{sec:intro}

The \Fermi Gamma-ray Space Telescope (\Fermi) has been observing and cataloging the high-energy gamma-ray sky with the Large Area Telescope (LAT) since 2008.
The 4FGL-DR4 catalog is their fourth data release \citep{ballet23} of the fourth catalog \citep{Abdollahi2020}. It includes 7195 gamma-ray sources, of which 4255 are associated with extragalactic blazars or local pulsars, with another 515 being supernova remnants, X-ray binaries, starburst galaxies, and other objects (as of June 2024). This leaves 2425 sources as ``unassociated'' because they lack confident astrophysical classification or lower energy counterparts. Following the pattern of the associated 4FGL sources, these unassociated sources are likely faint blazars and pulsars that could expand their respective catalogs if they can be classified. 

Increasing the samples of known \Fermi blazars and pulsars to fainter levels is crucial to understanding these sources overall. 
While we have significantly expanded all gamma-ray source populations in the last couple of decades, some populations, \Fermi blazars \citep{Ajello20}, have expanded faster than others, \Fermi pulsars \citep{Smith23}.
The unassociated gamma-ray source population contains sources observationally fainter than the majority of associated \Fermi sources for intrinsic (e.g., luminosity) or observational reasons (e.g., redshift, viewing angle). Discovering more \Fermi blazars from this fainter population will help complete the blazar sequence \citep{Ghisellini2008,Ghisellini2017} and test our overall understanding of blazars and jetted AGN as a whole.

Gamma-ray sources are expected to commonly produce lower energy emissions, where even the fainter unassociated gamma-ray sources are seen across the electromagnetic spectrum by numerous surveys and catalogs \citep[e.g.,][]{Kerby2021a,kerby21b,Ulgiati25}. Blazars and pulsars are the most common gamma-ray sources within the 4FGL catalog, and both nearly always produce X-rays if they produce gamma-rays \citep[][respectively]{urry95,romani96}. Blazar jet spectra are dominated by a low-energy peak from radio-to-X-ray produced by synchrotron emission and a high-energy peak from X-ray-to-gamma-ray produced by leptonic processes (e.g., synchrotron self-Compton emission) and/or hadronic processes (e.g., proton synchrotron radiation) \citep{sikora94}. Utilizing pure or hybrid hadronic emission models when modeling blazars allows for high-energy neutrino production, which can help explain the cosmic ray neutrino background \citep{Rodrigues21}. Pulsars primarily emit high-energy radiation from nonthermal processes of their magnetosphere that produce X-ray-to-gamma-ray emission, primarily by synchrotron and curvature radiation or synchro-curvature radiation, which can then be upscattered by self-synchrotron Compton or self-synchro-curvature Compton radiation processes to gamma-ray energies \citep{romani95,Cheng96}. 

Other than X-rays, many gamma-ray sources also produce radio emission. As mentioned, blazars emit radio emission via synchrotron emission in their relativistic jets. While the strength of radio emission varies, the nature of blazar jets requires some amount of radio emission \citep{Boula18}. While radio emission from blazars is almost guaranteed, pulsars do not always emit detectable radio emission. Gamma-ray pulsars overlap with both radio-loud and radio-quiet pulsar samples \citep{Lin16}. The radio emission mechanism for pulsars is quite different from blazars, and it is still under debate, with models ranging from coherent curvature emission to free-electron maser emission \citep{Melrose17}. 

The Neil Gehrels \Swift Observatory (\Swift) \citep{Gehrels2004} systematically surveys all \Fermi unassociated sources not labeled as ``extended'' or ``confused'' with interstellar cloud complexes to detect low-energy counterparts as part of our ongoing observing campaign \citep{Falcone2011}. 
A target exposure time of 4~ks, or more, is reasonable since it produces an X-ray flux limit of approximately $10^{-13}$~erg/s/cm$^2$ \citep{kaur21}, which is an order of magnitude lower than the X-ray fluxes found for 4FGL unassociated sources in \citet{kerby21b}. 
The two primary instruments used to study these gamma-ray targets are the \Swift X-Ray Telescope \citep[XRT, ][]{Burrows2005} and the \Swift UltraViolet Optical Telescope \citep[UVOT, ][]{Roming2005} with observational ranges of 0.3 -- 10.0~keV and 170 -- 600~nm, respectively.

Given the capabilities of \Swift-XRT and the knowledge that nearly all gamma-ray sources emit X-rays as well as gamma-rays, we conclude a likely connection between a 4FGL unassociated source and a single X-ray source within its uncertainty ellipse (hereafter, an X-ray singlet or singlet) in cases where the spurious chance probability of an X-ray detection is low. 
We use the probability of a coincidental 4$\sigma$ X-ray source from \citet{kaur21} to conclude that we expect very few spurious X-ray sources in our sample. In their paper, they used random \Swift-XRT fields across the sky to calculate the spurious source detection density as a function of exposure time. For the typical XRT exposure and field parameters in our survey, $\sim4$~ks and a semi-major axis of $\sim5$\arcmin, they estimate the probability for spurious 4$\sigma$ XRT detections of X-ray sources to be $\lesssim 0.01$ per field \citep{kaur21}. This is a conservative upper limit since this probability only takes into account the chance probability of an X-ray detection in such an error ellipse, regardless of classification. If one considers the likelihood of gamma-ray emission and X-ray emission being produced from two otherwise uncorrelated locations, the probability of chance occurrence would be even smaller. This makes X-ray surveys a great means of locating likely counterparts to unassociated gamma-ray sources, which makes logical sense on the surface since both source types have a low density on the sky and since gamma-ray sources tend to have correlated X-ray emission. 

This chance coincidence probability takes into account the majority of our 4FGL targets, but a small number of our 4FGL target fields deviate from this calculation. We expect this coincidence probability to increase for observations with significantly longer X-ray exposure times and/or larger-than-average 4FGL uncertainty ellipses. 
A deeper exposure and/or a wider 4FGL gamma-ray source uncertainty region can result in multiple detected X-ray sources near a single 4FGL source. While both X-ray sources may be associated with the gamma-ray source, the more likely scenario is that only one of these sources is correlated. Hereafter, we refer to these X-ray sources as multiplets. Due to their increased ambiguity regarding their gamma-ray associations, multiplets are dealt with separately from the X-ray singlets in the work that follows. We also include the multiplets in a separate table to enable further study.

This paper is a continuation of others using \Swift to analyze \Fermi unassociated sources at lower energies \citep[e.g., ][]{Kaur2019,Kerby2021a,kerby21b,Kaur23}. These papers study unassociated gamma-ray sources up to 4FGL-DR2, with ones prior to \citet{kerby21b} ignoring all \Fermi targets with more than one significant X-ray source within their uncertainty ellipses. Similar to \citet{kerby21b}, we analyze sources regardless of the number of nearby X-ray sources. Following this multiwavelength source analysis, we utilize the spectral parameters to classify our sources with a multi-layer perceptron (MLP) neural network classifier (NNC) trained on known gamma-ray blazars and pulsars (see Sec.~5 of \citet{Bishop06} for more details on MLP neural networks). These known sources are separated into the training and validation samples to train the neural network and to verify the accuracy of the NNC, respectively. We use this NNC to classify the test sample, our unassociated sources. While the goal is to discern likely blazars and pulsars from the unassociated source sample, the neural network can also help distinguish non-common gamma-ray sources among our sample.

In Section~\ref{sec:Samples}, we describe our data samples, including gamma-ray, X-ray, and UV/optical observations, and how we manage significant X-ray sources. In Section~\ref{sec:Analysis}, we detail our methods for spectral analysis of X-ray sources with X-ray and UV/optical data. In Section~\ref{sec:NNC}, we explain our methods for training, validating, and utilizing our NNC on the unassociated sources. Finally, in Section~\ref{sec:Results}, we summarize and discuss the results of this catalog and classification analysis.

\section{Observational Program, Targets, and Samples} 
\label{sec:Samples}

\subsection{\textit{Fermi}-LAT Unassociated and \textit{Swift}-XRT Sources}

After 14 years of observations, the \Fermi-LAT 4FGL-DR4 catalog \citep{Abdollahi22,ballet23} now consists of 7194 gamma-ray sources with 2425 of these being unassociated. 
These unassociated sources are the gamma-ray sources that have either failed to be linked with a known low-energy counterpart or lack high-confidence classification (see \citealt{collaboration2019} for the full 4FGL catalog).  
As discussed in \citet{kerby21b}, the 4FGL \Swift-XRT survey is a continuation of the 3FGL \Swift-XRT survey of unassociated sources \citep{Kerby2021a}. This survey produces high-resolution observations at X-ray, UV, and visible energies that can find potential low-energy counterparts for the unassociated sources. 

\Swift has observed 748 of the 2425 4FGL-DR4 unassociated gamma-ray sources with $>4$~ks of exposure time each, as of June 2024. Of the unassociated sources, \Swift has detected significant X-ray sources within 244 of the \Fermi-LAT 95\% uncertainty ellipses. While 218 of these 4FGL targets have only a single X-ray source within their uncertainty ellipse, the remaining 26 have multiple. The methods we use for X-ray data acquisition and source detection are described below.

\Swift-XRT observations are acquired using the HEASARC query interface within 11\arcmin\xspace of all 4FGL unassociated sources. Additionally, we only examine 4FGL unassociated sources with 95\% uncertainty ellipses that have semi-major axes under 10\arcmin. We use these search parameters to prevent the 4FGL uncertainty ellipse from reaching the edges of the 23.6\arcmin\xspace x 23.6\arcmin\xspace field of view for the \Swift-XRT telescope.

The X-ray sources are located by combining \Swift-XRT observations of target fields and analyzing the data using an automated process detailed in \citet{Falcone2011}. An X-ray source is significant if it is within the 95\% uncertainty ellipse of a 4FGL unassociated source, has a signal-to-noise ratio greater than or equal to 4, and is not removed later in the analysis (as detailed below). We use \texttt{XIMAGE} to detect these X-ray sources and the \texttt{XIMAGE} function \texttt{SOSTA} to generate their signal-to-noise ratios. Additionally, we use the script \texttt{xrtcentroid} to calculate the uncertainty radius of each X-ray source. Using this process, we find 288 significant X-ray sources within the uncertainty regions of 244 4FGL-DR4 unassociated sources. For each of these X-ray sources, we give them an internal name of SwXF4-DR4 JHHMMSS.s$\pm$DDMMSS, which acts as a continuation of our naming scheme from \citet{kerby21b}.

During our analysis, we removed 34 X-ray sources due to their angular proximity to Tycho-2 Catalogue stars \citep{hog00}, none of which are included in the 288 significant X-ray sources. For each X-ray source, we calculate a positional uncertainty of $\sim5\arcsec$, so we removed X-ray sources with separations within this error as potential stellar objects. Bright optical sources have the potential to produce a false X-ray source by optical loading. This effect occurs when enough UV/optical photons charge the \Swift-XRT CCD, producing a false X-ray signal. These bright sources are not gamma-ray sources, so we remove them as potential 4FGL counterparts from our analysis and list of sources. If a bright star produces X-rays without a compact stellar companion, it will still be unable to produce detectable gamma-rays. That being said, highly magnetized non-degenerate single stars, such as T Tauri stars, can produce detectable X-rays and gamma-rays \citep{filocomo23}, but they are optically faint and not in the Tycho-2 Catalogue.

\begin{figure}
    \centering
    \includegraphics[width=0.9\linewidth]{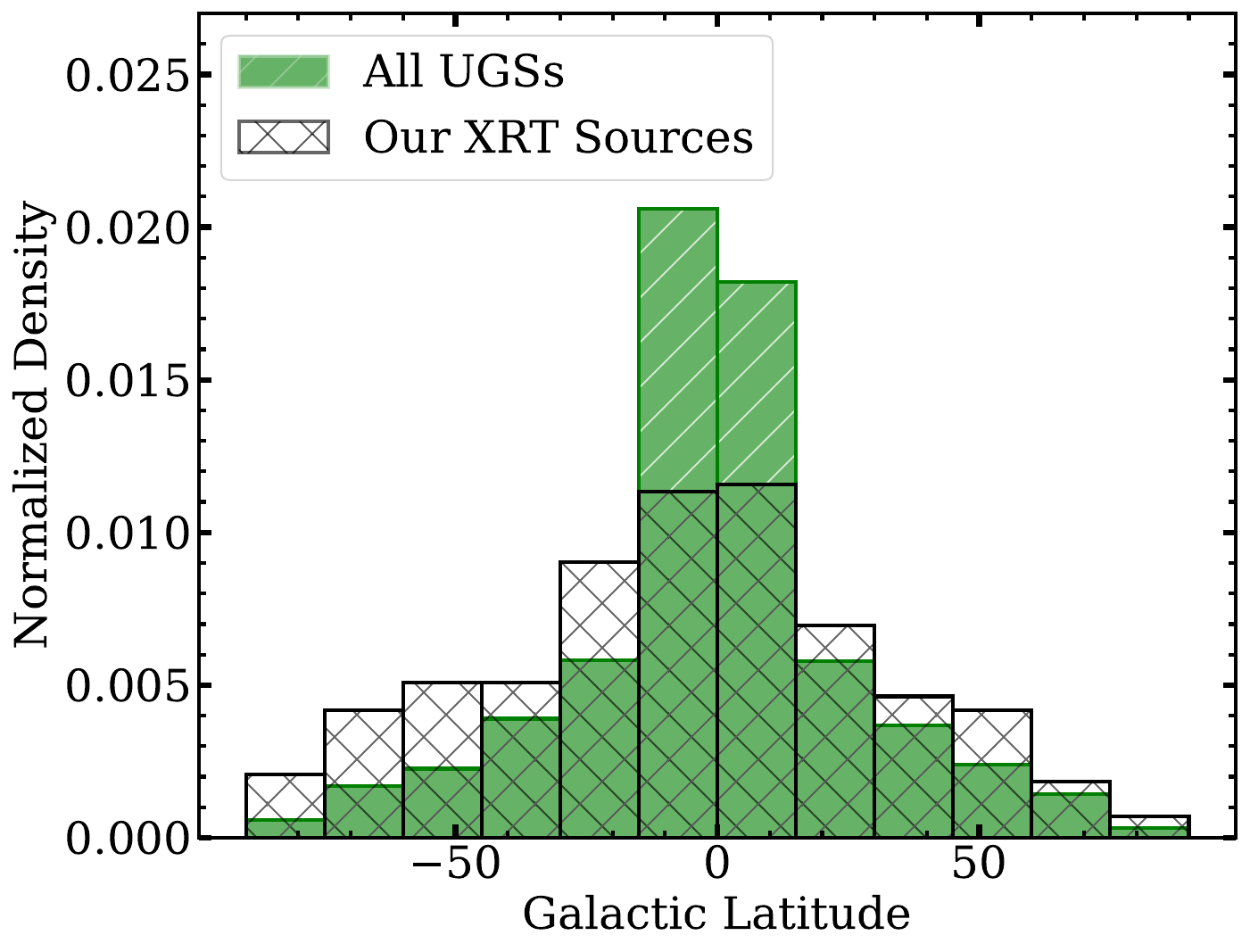}
    \caption{Normalized histogram of the galactic latitudes for unassociated gamma-ray sources. The samples of all unassociated gamma-ray sources (UGSs) are shown in green, and the 4FGL targets with overlapping XRT sources are in black with cross-hatches.}
    \label{fig:glat}
\end{figure}

Fig.~\ref{fig:glat} shows a histogram of galactic latitudes for all unassociated gamma-ray sources and those with our overlapping XRT sources. Comparing these, the 4FGL targets we are specifically studying are more evenly distributed across the sky than the entire sample. This may be due to the galactic plane dimming the detectable X-ray emission from low-latitude sources, which lowers the number of X-ray sources we can detect. Due to the increased number of low-latitude 4FGL targets, this seems to even out the latitude of our targets rather than producing a noticeable dip.


We detect significant X-ray sources within 244 4FGL unassociated sources, and 26 have more than one within their uncertainty ellipse. These 26 4FGL targets have a total of 70 significant X-ray sources with between 2 and 7 in each 95\% uncertainty ellipse. These 4FGL targets and X-ray sources are treated with the same analysis as X-ray singlets. Each significant X-ray source is treated equally without biases towards any. While we analyze all of these X-ray sources equally, only one can be related to the gamma-ray source. We tabulate and discuss these multiplets separately from the singlets due to their more ambiguous nature.

\subsection{\textit{Swift}-UVOT Source Analysis}

For all 4FGL targets with significant X-ray sources in their uncertainty ellipses, we use \Swift-UVOT data to search for a UV/optical counterpart to the X-ray sources. To improve observations of each potential UV/optical counterpart, we first generate UVOT coadded images of each field in each UVOT filter (listed in Table~\ref{tab:UVOTbands}) using \texttt{UVOTIMSUM}. For each X-ray source, we initially place a $\sim5$\arcsec\xspace circular extraction region over their centroid as calculated previously. The radius of this extraction region is set to match the \Swift-UVOT PSF for point sources. Separately, the positional uncertainties of the XRT sources are approximately 5\arcsec, so we search for UV/optical sources in the UVOT image using \texttt{UVOTDETECT} with a 3$\sigma$ detection threshold. If a source is within 5\arcsec\xspace of the X-ray source, we update the centroid of our extraction region to overlap with the UV/optical counterpart better. For these sources, we give them an additional name to signify the UV/optical counterpart following the naming scheme of SwUVF4-DR4 JHHMMSS.s$\pm$DDMMSS based on the UVOT centroid. The background regions are generated by manually choosing a 20\arcsec\xspace region near the extraction region without any overlapping extraneous sources. 


\begin{deluxetable*}{c|cccccc}
\tablecaption{The central wavelengths of the six \textit{Swift}-UVOT bands used to convert \textit{Swift}-UVOT magnitudes to Johnson V magnitudes}
\label{tab:UVOTbands}
\tablewidth{0pt}
\tablehead{UVOT Band & $vv$ & $bb$ & $uu$ & $w1$ & $m2$ & $w2$}
\startdata
Central Wavelength ($\rm{nm}$) & 547 & 439 & 347 & 260 & 225 & 193 \\
\enddata
\end{deluxetable*}

Finally, we use \texttt{UVOTSOURCE} to calculate UVOT AB magnitudes without galactic reddening corrections. Given coadded images, extraction regions, and background regions, this program measures fluxes of both the extraction and background regions to calculate the magnitude of the source using methods described in \citet{breeveld2011updated}. Results of the AB magnitude in each filter for all X-ray sources are detailed in online machine-readable tables. Sources that lack UVOT observations are listed with \texttt{NaN}s. If a source has a UVOT magnitude with an uncertainty of -1.00, this implies the UVOT magnitude is the limiting magnitude from a non-detection rather than the actual source brightness.

\section{Spectral Analysis}
\label{sec:Analysis}

\subsection{X-ray Spectral Analysis}

Using the HEASARC query, we have collected roughly 10,000 \Swift-XRT observations of 4FGL unassociated sources, including those from the observing campaign and serendipitous observations. All observations of targeted fields use the photon counting (PC) mode of the \Swift-XRT, producing two-dimensional X-ray images. These images are summed together for each 4FGL target where we have total X-ray exposure times ranging from 1~ks to 250~ks with a median exposure time of $\sim 6~$ks for those with detected X-ray sources.

We download the level 1 event file for each observation and process it using \texttt{xrtpipeline} v.0.13.7 from the \texttt{HEASoft} software\footnote{\url{https://heasarc.gsfc.nasa.gov/docs/software.html}}. To maintain the quality of our analysis, we only use events graded as good (0 -- 12). We merge the event files and exposure maps of observations over the same 4FGL target using \texttt{Xselect} v.2.5b and \texttt{XIMAGE} v.4.5.1, producing single totaled files of each. For each X-ray source, we also produce a totaled ancillary response file using \texttt{xrtmkarf}. 
The 4FGL catalog contains very few transient gamma-ray sources due to their low statistical significance over long periods of time, so we disregard the time-domain analysis of these observations \citep{Abdollahi2020}. The 4FGL catalog does contain many variable sources, but the faint nature of the unassociated gamma-ray sources makes the study of this difficult.

\begin{figure*}
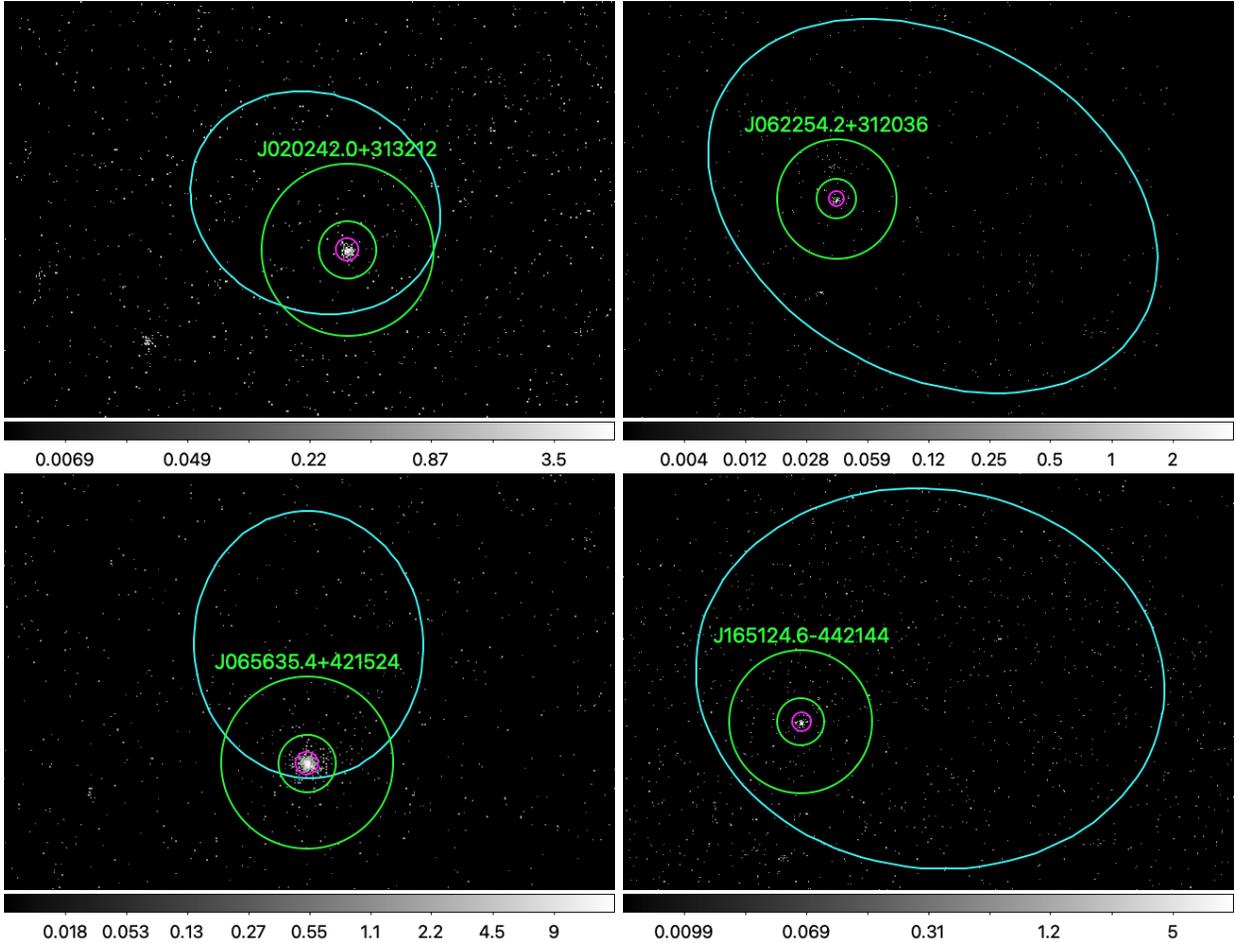

    \centering
    \includegraphics[width=0.45\linewidth]{J0202.7+3133.pdf}
    \includegraphics[width=0.45\linewidth]{J0622.5+3120.pdf}
    \includegraphics[width=0.45\linewidth]{J0656.5+4218.pdf}
    \includegraphics[width=0.45\linewidth]{J1650.9-4420c.pdf}
    \caption{Example of single X-ray sources within 4FGL uncertainty ellipses. The color bar denotes the number of photons per pixel. The cyan ellipses are the 4FGL uncertainty ellipses for 4FGL J0202.7+3133 (\emph{upper left}), 4FGL J0622.5+3120 (\emph{upper right}), 4FGL J0656.5+4218 (\emph{lower left}), and 4FGL J1650.9-4420c (\emph{lower right}). The inner magenta circle is the 20\arcsec\xspace X-ray source extraction region, and the outer green annulus is the 50\arcsec\xspace - 150\arcsec\xspace background extraction region. SwXF4-DR4 names are labeled in green text above the background regions.}
    \label{fig:example}
\end{figure*}

Spectra are produced for source and background regions of each significant X-ray source using \texttt{Xselect}. We use 20\arcsec\xspace circular regions for source regions and annular regions with an inner radius of 50\arcsec\xspace and an outer radius of 150\arcsec\xspace for background regions. These region sizes are chosen to best fit or avoid the PSF of the \Swift-XRT point sources (half-power diameter of 18\arcsec) \citep{capalbi05}. Fig.~\ref{fig:example} shows four examples of these regions overlaying our X-ray observations of the 4FGL targets. Both regions are centered on the X-ray source with exceptions for background regions overlapping with significant or insignificant X-ray sources. These background regions are instead moved to the closest location with no X-ray source within the annulus. 


Two X-ray sources required additional modifications to their background regions. For X-ray source SwXF4-DR4 J073707.2+653454, the point source is located in the extended region produced by NGC 2403, so we generated a background region to include this extended region without the AGN at the core of the galaxy. For X-ray source SwXF4-DR4 J165220.3-452037, it is so bright that the PSF extends into the default background region, so we expand the inner radius beyond the PSF.

To avoid photon pile-up, we check the count rates of the source region. For PC mode, we want to keep the count rate below $\sim0.5$ counts per second as the 2.5~s frames of PC mode would imply a greater than about one photon per frame at greater count rates \citep{capalbi05}. We account for this possibility by generating annular source regions around bright sources instead of simple circular regions. This annular region consists of an inner radius dependent on the count rate and the same 20\arcsec\xspace outer radius from earlier. This process only affected SwXF4-DR4 J165220.3-452037 where we used an inner annular radius of 14\arcsec\xspace to offset the large count rate.

After spectra are generated, we use \texttt{XSPEC} v.12.14.0h \citep{Arnaud1996} to analyze and fit them. We fit the spectra by multiplying the nested models  \texttt{tbabs}, \texttt{cflux}, and \texttt{powerlaw}. \texttt{tbabs} or the Tuebingen-Boulder ISM absorption model calculates the cross section for X-ray absorption given the galactic hydrogen column density, which we look up with the \texttt{FTOOLS} function \texttt{nH} \citep{Wilms2000}. 
\texttt{cflux} is a convolution model that, when paired with \texttt{tbabs}, calculates the unabsorbed flux of an X-ray source between two energies. We use an energy range of 0.3~keV and 10.0~keV to mimic the Swift-XRT observational range. \texttt{powerlaw} is the basic photon power law model of 
\[A(E)\propto E^{-\alpha_X}\]
that calculates the dimensionless X-ray photon index, $\alpha_X$. Because we are using the \texttt{cflux} model, we fix the normalization parameter of \texttt{powerlaw} to a non-zero value. Overall, \texttt{XSPEC} fits the unabsorbed X-ray flux and photon index of an X-ray source. Their uncertainties are jointly measured using the \texttt{steppar} function. The spectral parameters and uncertainties are included in Tables~\ref{tab:xrt_single} and~\ref{tab:xrt_multi}.

We use the Cash Statistic, C-statistic, for fitting instead of the $\chi^2$ statistic due to the dimness of our X-ray sources. C-statistics allow for an accurate fitting when each bin has low counts that more closely match a Poisson distribution rather than the Gaussian distribution of $\chi^2$ statistics \citep{Cash1976}. Following the C-statistics from \citet{Cash1976}, we can approximate the maximum likelihood statistic of the data using Stirling's approximation: \[C=2\sum^N_{i=1}(tm_i)-S_i+S_i(\ln(S_i)-\ln(tm_i)).\] This uses the exposure time, $t$, the predicted count rate, $m_i$, and the observed counts in each bin, $S_i$, to calculate the statistic, C.

\begin{figure*}
    \centering
    \includegraphics[width=0.8\linewidth]{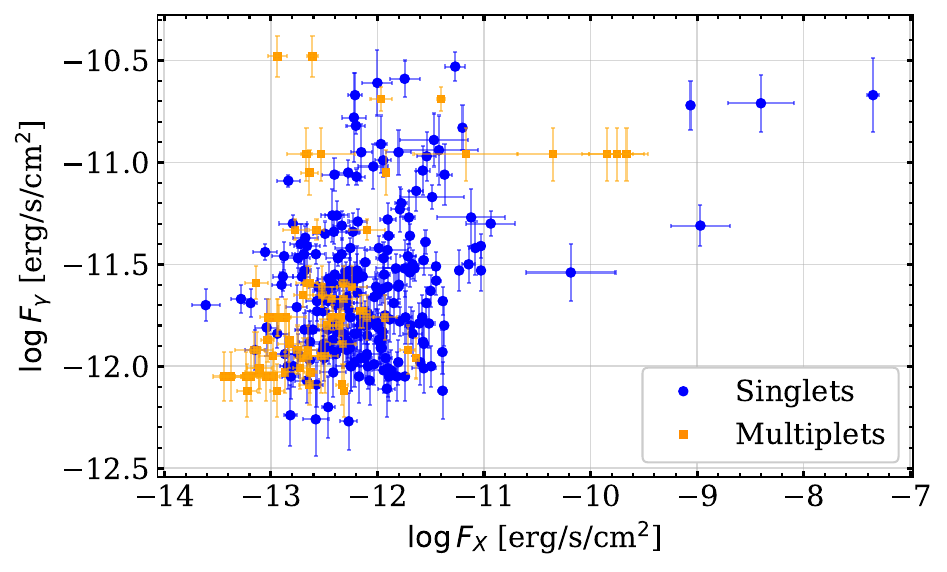}
    \caption{Comparison between the measured gamma-ray flux of the 4FGL target compared to the unabsorbed X-ray flux obtained from spectral fitting with \texttt{XSPEC}. We plot X-ray singlets with blue circles and X-ray multiplets with orange squares. We include error bars from uncertainties given in the 4FGL-DR4 catalog \citep{ballet23} and from the \texttt{XSPEC} fitting for gamma-ray and X-ray fluxes, respectively}
    \label{fig:flux}
\end{figure*}

Fig.~\ref{fig:flux} is a scatter plot comparing the total gamma-ray flux to the unabsorbed X-ray flux of both X-ray singlets and multiplets. 
We use the gamma-ray energy flux from \citet{ballet23}, which spans the observing range of 100~MeV - 100~GeV.


\begin{figure}
    \centering
    \includegraphics[width=0.99\linewidth]{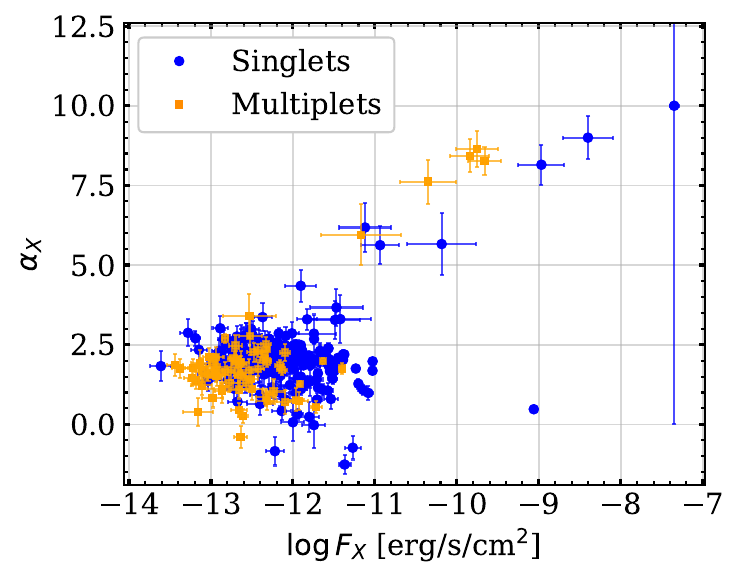}
    \caption{Comparison between the photon index and unabsorbed X-ray flux from spectral fitting with \texttt{XSPEC}. We plot X-ray sources that are the only significant X-ray source in the 4FGL uncertainty ellipse as blue circles. For those with multiple X-ray sources within the 4FGL uncertainty ellipse, we plot those as orange squares. We include error bars from uncertainties given by the \texttt{XSPEC} fitting.}
    \label{fig:pindex}
\end{figure}

Fig.~\ref{fig:pindex} compares the photon index to the unabsorbed X-ray flux obtained from spectral fitting with \texttt{XSPEC}. Fitting with \texttt{powerlaw}, the photon index for 17 X-ray sources is outside of the range of $0 < \alpha_X < 4$ that is expected for blazars and pulsars, the most likely counterparts for the 4FGL unassociated sources. Table~\ref{tab:extremePI} details these 17 sources and their potential counterparts. All of the sources have relatively large hydrogen column densities in relation to the other unassociated sources ($n_H > 3\times10^{21}~\mathrm{cm}^{-2}$), which could affect the fitting software, especially for faint sources at lower energies. 
This makes it difficult to accurately constrain the X-ray flux and photon index without manually forcing our limits in \texttt{XSPEC}. Still, some of these sources may be alternative X-ray sources, leading to unreliable spectral fits with the blazar model.  


\begin{deluxetable*}{lcccl}
\label{tab:extremePI}
\tablewidth{0pt}
\tablecaption{\Swift X-ray sources with extreme photon indices, $\alpha_X < 0$ or $\alpha_X > 4$. Included in main analysis.} 
\tablehead{4FGL$^a$ & SwXF4-DR4 & $n_H^b$ & $\alpha_X^c$ & Possible Counterpart$^d$}
\startdata
J0722.4-2650 & J072219.7-264735 & 0.36 & 4.35 $\pm$ 0.50 & 0.4\arcsec\xspace from Gaia DR3 source, eclipsing binary, $m_G = 12.4$ \\
J0752.0-2931 & J075209.9-293013 & 0.52 & 5.62 $\pm$ 0.59 &  \\
J0902.8-4633 & J090213.2-463039 & 0.68 & 6.18 $\pm$ 0.76 &  \\
J1312.6-6231c* & J131235.7-623625 & 1.19 & -0.40 $\pm$ 0.35 &  \\
J1320.3-6410c & J132016.4-641348 & 1.02 & 8.15 $\pm$ 0.63 &  1.4\arcsec\xspace from Gaia DR3 source, $m_G = 13.9$ \\
J1321.1-6239* & J132036.9-623840 & 1.12 & 8.43 $\pm$ 0.52 &  \\
J1321.1-6239* & J132039.2-623937 & 1.12 & 8.64 $\pm$ 0.57 &  \\
J1321.1-6239* & J132058.8-623446 & 1.12 & 7.62 $\pm$ 0.69 &  \\
J1321.1-6239* & J132138.7-623256 & 1.12 & 8.27 $\pm$ 0.43 &  \\
J1321.1-6239* & J132140.8-623948 & 1.12 & 5.95 $\pm$ 0.96 &  \\
J1638.1-4641c & J163800.8-464012 & 1.61 & -0.84 $\pm$ 0.45 & 1.7\arcsec\xspace from SPICY 39308, YSO candidate \\
J1650.9-4420c & J165124.6-442144 & 1.56 & 10 & 1.7\arcsec\xspace from SPICY 42795, AGB and YSO candidate \\
J1706.5-4023c & J170633.6-402545 & 1.30 & 9.00 $\pm$ 0.67 & 4.7\arcsec\xspace from Gaia DR3 source, YSO candidate, $m_G = 13.2$ \\
J1740.6-2808 & J174042.0-280727 & 0.74 & -1.26 $\pm$ 0.29 & 2.2\arcsec\xspace from CXOU J174042.0-280724, possible LMXB, $m_G = 20.8$ \\
J1855.2+0456 & J185503.0+045947 & 1.15 & -0.73 $\pm$ 0.37 &  \\
J1900.9+0538 & J190054.9+053314 & 1.54 & -0.02 $\pm$ 0.73 &  \\
J2317.6+6036c & J231732.1+604103 & 0.76 & 5.66 $\pm$ 0.96 &  \\
\enddata

\noindent $^a$ 4FGL sources with multiple X-ray sources are marked with an asterisk, *. \\
$^b$ Hydrogen column densities are in units of $10^{22}\mathrm{~cm^{-2}}$ and obtained with the \texttt{nH} function. \\
$^c$ Photon indices of 10 are values given when \texttt{XSPEC} failed to fit the spectral parameters. \\
$^d$ Possible source counterparts include classification and \textit{Gaia} $G$-band magnitude acquired from SIMBAD if available. \\

\end{deluxetable*}

\subsection{V-magnitude Conversions}\label{sec:Vmag_conv}

\Swift-UVOT uses six magnitudes in their observations, yet previous studies use the Johnson $V$ magnitudes as a method to unify optical data in the training and test samples \citep{kerby21b,Kaur23}. Similar to the methods discussed in \citet{kerby21b}, we convert the UVOT magnitudes into $V$-band magnitudes for use in our classification studies using a power-law scaling relation and conversion of their flux densities: 
\[\frac{F_i}{F_V}=\left(\frac{\nu_i}{\nu_V}\right)^{-\alpha_c}=\left(\frac{\lambda_i}{\lambda_V}\right)^{\alpha_c}.\]
We can then insert this equation into a magnitude conversion equation to find
\[m_i-m_V=-2.5\log\frac{F_i}{F_V}=-2.5 \alpha_c \log\frac{\lambda_i}{\lambda_V}.\]
This conversion estimates the $V$-band magnitude given the UVOT magnitude, $m_i$, of the source, the central wavelength of the UVOT filters (given in Table~\ref{tab:UVOTbands}), the $V$-band wavelength (540~nm), and an assumed UV/optical spectral index. Because blazars and other sources have a wide variety of optical spectral indices \citep{zhang23}, we cannot accurately use a single slope for all of our sources. With that in mind, we implement a method to calculate an approximate power-law spectral index utilizing our UVOT data and optical data from numerous catalogs. With the VizieR catalog access tool \citep{Ochsenbein00}, we acquire optical magnitudes from several optical catalogs including the Naval Observatory Merged Astrometric Dataset \citep[NOMAD;][]{Zacharias04}, the Guide Star Catalog \citep{Lasker08}, the Panoramic Survey Telescope and Rapid Response System \citep[Pan-STARRS;][]{Chambers16}, the Galaxy Evolution Explorer \citep[\textit{GALEX};][]{Bianchi17}, the Sloan Digital Sky Survey \citep[SDSS][]{Ahumada20}, the Dark Energy Survey \citep[DES;][]{Abbott21}, and \textit{Gaia} \citep{Gaia21}. 

We first convert the magnitudes into flux density space using the \citet{Bessell1998} conversion
\[\log F_\nu=-\frac{m_\nu+48.6+zp}{2.5}\]
where $zp$ is the zeropoint given by different magnitude systems\footnote{Zeropoints for AB and $V$ magnitudes are $zp=0$.}. We use these flux densities to fit a power-law 
\[\log F_\nu = \alpha_c \log\lambda+b\]
and estimate a more accurate spectral index. This fitting does not include magnitude limits from the UVOT data. For sources lacking enough optical data, we use $\alpha_c=1.5$, which is the average optical spectral slope of blazars measured by \citet{Hu06}. These sources have their UVOT filter marked with an asterisk, *, in Tables~\ref{tab:xrt_single} and~\ref{tab:xrt_multi} (e.g., $vv*$).

To obtain the most accurate estimate of the $V$ magnitude, we use the UVOT detection whose filter has the central wavelength closest to that of the $V$-band. If we have significant detections in all filters, we use the UVOT $vv$ filter due to its nearly identical central wavelength. If we only have magnitude limits for all of our observations, we will follow the same method as above and label these magnitudes as lower limits. Beyond that, all estimations using more distant UVOT filters increase the uncertainty in this estimation as the difference in central wavelengths increases. Tables~\ref{tab:xrt_single} and~\ref{tab:xrt_multi} list this UVOT filter to show the accuracy of our estimation qualitatively.

Three of our X-ray sources have complications with their UV observations. SwXF4-DR4 J073707.2+653454 is within a galaxy with extended UV emission. Rather than identify a single source within this region, we use the magnitude limits of the region. SwXF4-DR4 J131230.2-623431 has several UVOT sources within 5\arcsec\xspace, preventing us from identifying the correct source. Rather than inaccurately choosing the correct source, we regard it as a failed UVOT observation without magnitudes or magnitude limits. SwXF4-DR4 J150425.0+434109 lacks a UV source within 5\arcsec\xspace, but a bright Tycho-2 star is within 10\arcsec\xspace. Unfortunately, the star is bright enough that UV starlight extends into the entire extraction region, so the UV analysis for this source is likely inaccurate. There are an additional five X-ray sources outside of the field of view for all viable UVOT observations resulting in failed UVOT detections, including SwXF4-DR4 J075118.9-002754, J085221.9-251626, J095250.4+135218, J172842.8-513032, and J231732.1+604103.

\begin{figure}
    \centering
    \includegraphics[width=0.99\linewidth]{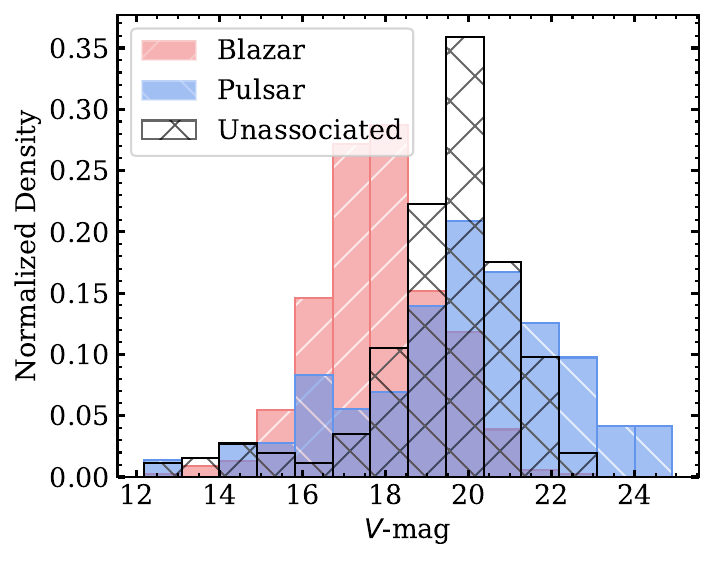}
    \caption{Normalized histogram of the calculated $V$-band magnitudes of our test and training samples. Black cross hatches are our X-ray sources, red hatches are the training gamma-ray blazars, and blue hatches are the gamma-ray pulsars. Magnitude limit values are not included.}
    \label{fig:Vmag}
\end{figure}

Fig.~\ref{fig:Vmag} is a normalized histogram of the predicted $V$-band magnitudes of our unassociated sources and the classified gamma-ray blazars and pulsars from Section~\ref{sec:NNC_samples}. All three samples overlap in magnitude space, but we have the classified blazars skewed towards being optically brighter than the unassociated sources and pulsars. This may imply that 4FGL is biased towards classifying or associating visually brighter blazars as gamma-ray sources over the fainter sources. This would agree with our assumption that many of our unassociated sources are gamma-ray blazars that have failed association. While it may appear that the unassociated sources match pulsars for this parameter, the $V$-band to gamma-ray flux ratios, $\log F_V/F_\gamma$, in Fig.~\ref{fig:params} better show the true distribution of the samples.


\section{Neural Network Classification}
\label{sec:NNC}

\subsection{Training and Test Samples}\label{sec:NNC_samples}

To train our neural network, we need classified gamma-ray blazars and pulsars with gamma-ray, X-ray, and optical data. We use the 4FGL-DR4 catalog to identify potential training sources and acquire gamma-ray data \citep{ballet23}. From the catalog, we use sources classified as BL Lac-type objects (BL Lacs) and flat-spectrum radio quasars (FSRQs) for our blazar sample, and canonical and millisecond pulsars (MSPs) for our pulsar sample. Even though we need as many sources as possible, we exclude blazar candidates of uncertain type (BCUs) and pulsar wind nebulae (PWNe) from our training sample. BCUs have uncertain natures and may not even be blazars as seen with 4FGL J1544.5-1126, a BCU in 4FGL-DR1~\citep{Abdollahi2020}  and a low-mass X-ray binary (LMXB) in DR3~\citep{Abdollahi22}. We exclude PWNe because they have nebular emission mechanisms separate from pulsars that could confuse the NNC \citep{Slane17}. This results in 2310 available gamma-ray blazars and 318 available gamma-ray pulsars. To identify our training sample, we then search for X-ray and optical data on all available sources, which further limits our viable training sources.

With these available gamma-ray blazars and pulsars, we acquire X-ray and optical data from several catalogs and literature searches. For the X-ray data, the first and second Swift X-ray Point Source Catalogs \citep{Evans14,Evans2020} provide X-ray fluxes and photon indices for a significant number of our sources. To further increase our potential training sample, we extend our search into several studies \citep[e.i.][]{Zavlin06,Marelli12,Takeuchi13,SazParkinson2016,Zyuzin2018,Mayer24}. 
Similar to our methods in Section~\ref{sec:Vmag_conv}, we take optical data from the same catalogs and fit a power-law to this data. Rather than use the power-law scaling relation, we simply utilize the fit power-law to estimate the $V$-band flux. We did not perform the same UVOT analysis on these sources as we did for our X-ray sources, so rather than relying on a single data point from other catalogs, we use the best-fit value of all the optical data.
The X-ray and optical data produce the limits on our training sample as we must have data on both to properly train our neural network.

Overall, we identify 993 gamma-ray blazars and 83 gamma-ray pulsars with available gamma-ray, X-ray, and optical data for our training sample. While we are limited to sources that are both X-ray and optically bright, they are mostly representative of the gamma-ray catalog as a whole. Among the blazars, our limited sample consists of nearly identical fraction of subtypes (63\% BL Lac vs 37\% FSRQ) to the total blazars sample from the 4FGL-DR4 catalog (65\% BL Lac vs 35\% FSRQ). Our limited sample of pulsars is slightly less representative with our subtypes (56\% canonical vs 43\% MSP) favoring canonical pulsars over MSPs in relation to the complete gamma-ray pulsar sample (44\% canonical vs 56\% MSP). While our ratio is suboptimal, it still has the necessary sources to represent the gamma-ray pulsars of the 4FGL catalog, with some bias towards canonical pulsars.

We have significantly more blazars than pulsars in our training, so we use the Synthetic Minority Oversampling Technique \citep[SMOTE;][]{Chawla2002} to simulate pulsars and balance the sample. SMOTE utilizes the distribution of parameters for real pulsars to generate ``synthetic'' sources based on the \textit{k} nearest neighbor approach. In this manner, SMOTE introduces random synthetic points connecting real sources in parameter space, effectively creating more sources to train our NNC. Without this equalization of our training sample, the NNC would likely become biased towards our majority class, blazars, as failing to classify the minority class, pulsars, only weakly affects the overall apparent classification accuracy during training \citep[e.g.,][]{Last2017}. We have checked that the distribution of these SMOTE pulsars matches that of our actual pulsars in parameter space, agreeing with previous studies using SMOTE-generated objects \citep{Bethapudi18,Kaur23}.

Similar to the limits put on our training sample, the test sample is limited by X-ray and optical data on our sources, allowing us to test 282 out of 288 of our X-ray sources. Because we identify these sources with X-rays, we have X-ray data on all of them, but 5 of our sources are outside the \Swift-UVOT field of view for all observations, with another being a failed UVOT observation. While we have optical data on 282 sources, many of these sources are non-detections in the UVOT data, producing flux upper limits. We still classify these sources, but we run the classification twice, once at the given magnitude limit and once at a limiting magnitude of $m_V=30$. This produces a range for our classification parameter to avoid assumptions about the optical magnitude.

For our neural network classifier, we utilize the same seven parameters as \citet{kerby21b}. This includes:
\begin{enumerate}
    \item X-ray photon index, $\alpha_X$
    \item gamma-ray photon index, $\Gamma_\gamma$ (\texttt{PL\_Index} in the 4FGL catalog)
    \item the logarithm of gamma-ray flux, $\log F_\gamma$ in $\mathrm{erg~s^{-1}~cm^{-2}}$ (\texttt{Energy\_Flux} in the 4FGL catalog)
    \item the logarithm of the X-ray to gamma-ray flux ratio, $\log F_X/F_\gamma$
    \item the logarithm of the $V$-band to gamma-ray flux ratio, $\log F_V/F_\gamma$
    \item the logarithm of the year-to-year gamma-ray variability, $\log~$Var (\texttt{Variability\_Index} in the 4FGL catalog)
    \item the logarithm of the gamma-ray curvature significance, $\log\mathrm{Sig_{curv}}$ (\texttt{PLEC\_SigCurv} in the 4FGL catalog). 
\end{enumerate}
All sources within our training and test samples have these seven parameters or estimates on them. For consistency when applying the NNC to the data, we use the gamma-ray photon index provided by a power-law for our NNC, yet a log-parabola provides an improved overall fit to the gamma-ray data for some gamma-ray blazars. The differences between the power-law and log-parabola fits are minimal and lie within error bars for the input parameters to the neural network, allowing us to use one consistent set of fit parameters, the power-law fit. Because the distance of our sources may vary wildly, we use flux ratios for $F_X$ and $F_V$ rather than their given values to limit the distance dependence to $F_\gamma$, alone. Regarding X-ray flux, we fit our X-ray spectra to an unabsorbed X-ray flux within 0.3~keV and 10~keV, which extended to our training sample. For sources with only observed X-ray flux or an X-ray flux outside this energy band, we recalculate the unabsorbed flux using the Portable, Interactive Multi-Mission Simulator (\texttt{PIMMS}) and the fit X-ray photon index. To calculate $F_V$, we convert our $V$-band magnitude to an approximate flux with the conversion from \citet{Bessell1998}
\[\log F_V = \log \nu F_\nu = -\frac{m_V-48.6}{2.5}+\log\nu,\]
where $\nu=\frac{c}{540~\mathrm{nm}} \approx 5.56\times10^{14}$~Hz.
This gets us our approximation for $\log F_V$ for our parameters. Finally, when converting the $\mathrm{Sig_{curv}}$ to log-scale, many gamma-ray sources in our samples have values of zero. Because the log of zero is negative infinity, we equate this result to -4 in our analysis. Unlike previous 4FGL data releases, 4FGL-DR4 implemented Gaussian priors on the curvature parameters, which resulted in poor spectral fits becoming more apparent. Rather than use the $\mathrm{Sig_{curv}}$ from these poor fits, the 4FGL-DR4 catalog sets $\mathrm{Sig_{curv}}=0$. This is especially the case for blazars because they are more capable of having gamma-ray spectra without a curvature or an exponential cutoff \citep{Ackermann2015}. 

\begin{figure*}
    \centering
    \includegraphics[width=0.9\linewidth]{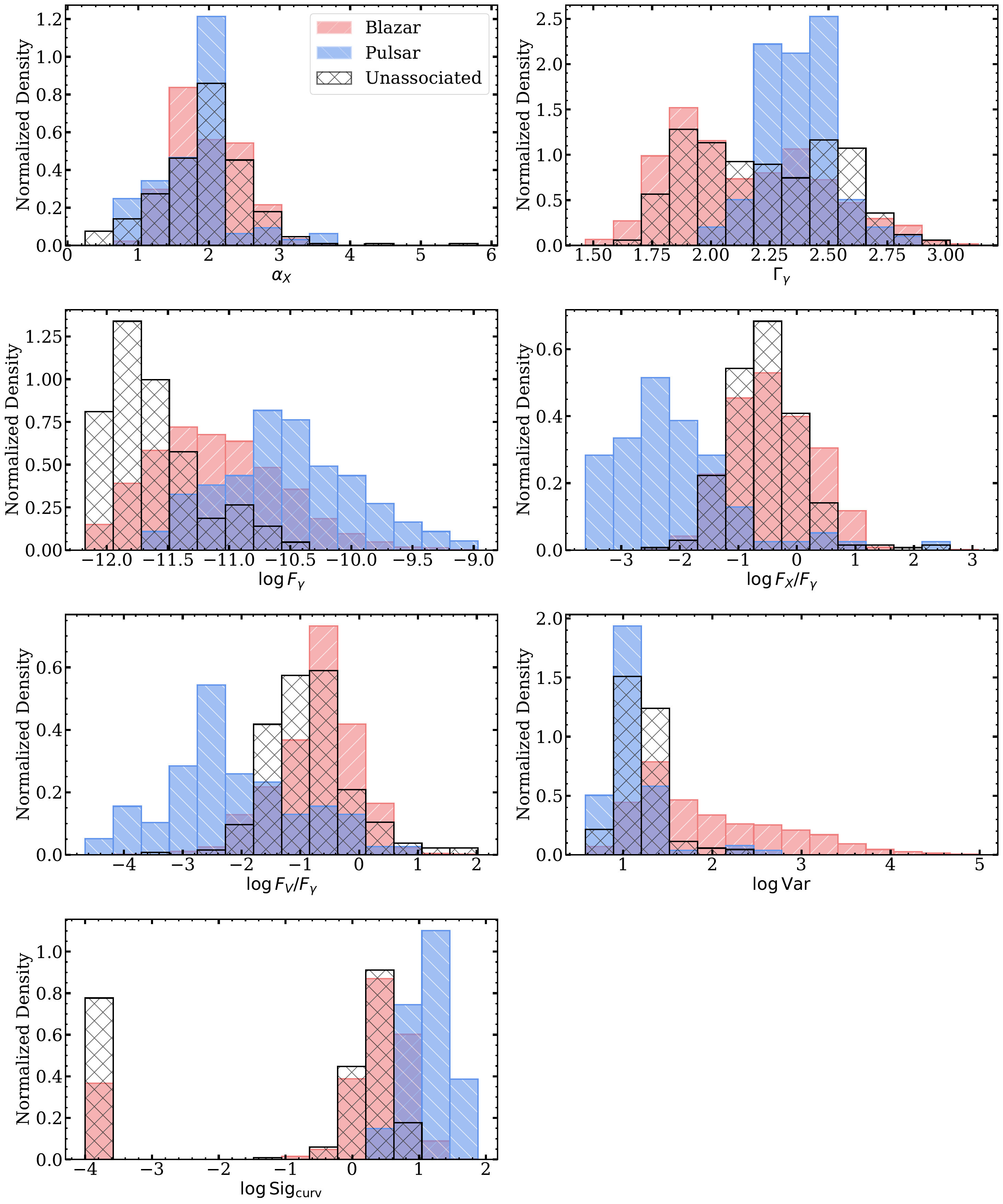}
    \caption{Normalized histograms of classification parameters for our training and test samples. Black cross-hatches are our X-ray sources, red hatches are the training gamma-ray blazars, and blue hatches are the gamma-ray pulsars.}
    \label{fig:params}
\end{figure*}

Fig.~\ref{fig:params} shows the normalized histograms of our seven parameters for our samples. Comparing the parameters of different samples, we see both overlapping and distinct characteristics from our sources. Among all but one training parameter, there are clear distributions to help distinguish blazars from pulsars. As expected, we see our unassociated sources show less gamma-ray flux, resulting in weaker fits for the variability index and curvature. While both are dependent on the intrinsic properties of the source, low-number statistics caused by a fainter (intrinsic or distance-dependent) source can inhibit our ability to detect either parameter.


\subsection{Network Design}

Our NNC is designed to calculate the blazar probability, $P_\mathrm{bzr}$, of unassociated sources based on how they compare to known blazars and pulsars. In this scheme, we train the NNC to output $P_\mathrm{bzr}=1$ for sources that it sees as 100\% blazars and $P_\mathrm{bzr}=0$ for sources it sees as 0\% blazars. This $P_\mathrm{bzr}=0$ is trained on the known pulsars rather than all non-blazar gamma-ray sources to better predict between the two. The NNC has seven input nodes (one for each rescaled training parameter), one hidden layer with four neurons, and one output node that returns the blazar probability. Varying the number of hidden layers and neurons did not result in significant improvement, so we kept the layout from our previous NNC \citep{kerby21b}. We use the \texttt{MLPClassifier} from the \texttt{scikit-learn} Python package with the ``logistic'' activation function and ``adam'' optimizer \citep{Kingma2014} to train the NNC. We scale our input parameters using \texttt{StandardScaler} to standardize the mean of each parameter to zero with a variance of one. To verify the accuracy of our NNC, we took a validation subsample from our training sample with the random selection method \texttt{StratifiedShuffleSplit} to evenly transfer 20\% ($\sim398$ sources) of training blazars and pulsars from the training to validation samples.

We use the Log-Loss error as our cost function to determine when to stop training the NNC. This parameter measures the error in binary classification methods. For a binary classification scheme with true classifications of $y_i=(0,1)$ and predicted classification probabilities of $p_i\in[0,1]$, the Log-Loss parameter is calculated with 
\[L_{\log}=-\sum_i(y_i\log(p_i)+(1-y_i)\log(1-p_i)). \]

\begin{figure}
    \centering
    \includegraphics[width=0.99\linewidth]{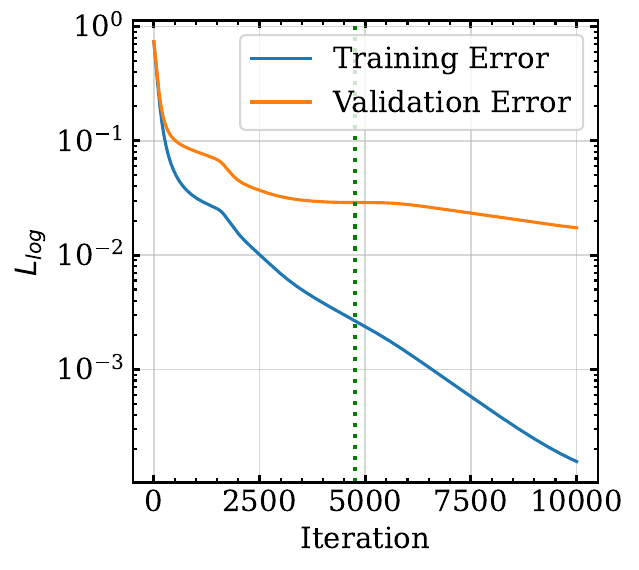}
    \caption{Evolution of $L_{\log}$ parameter over iterative steps of the NNC of the training sample (blue) and the validation sample (orange). The dotted green line represents the iteration of the local minimum, which is when we stop training the neural network.}
    \label{fig:logloss}
\end{figure}

Fig.~\ref{fig:logloss} shows the $L_{\log}$ calculated for the training and validation samples for each iterative step of our neural network training. As the NNC is training, the $L_{\log}$ will ideally decrease for both the training and validation samples. At a certain point, the validation error will reach a local minimum before increasing, while the training error will continue to decrease. This implies that the NNC has begun to overfit the training sample and will lose accuracy, so we stop training at this point. This can be seen in Fig.~\ref{fig:logloss} as we find a local minimum at iteration 4755 of training.


\begin{figure}
    \centering
    \includegraphics[width=0.99\linewidth]{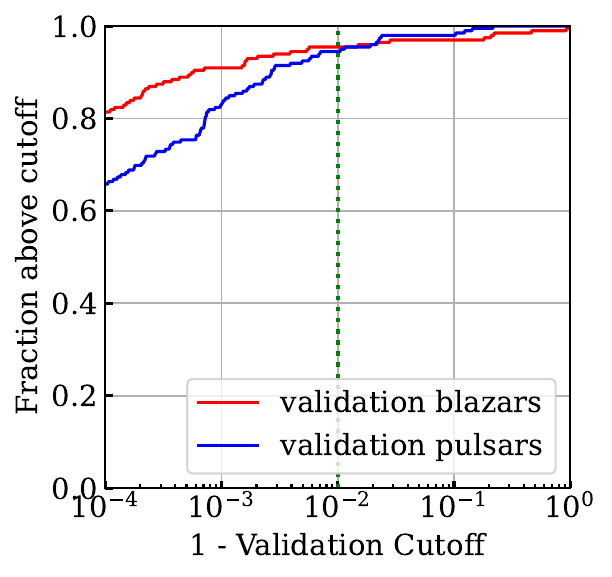}
    \caption{Fraction of validation blazars (red) and pulsars (blue) successfully classified given different cutoff values for $P_\mathrm{bzr}$. The dotted green line is the cutoff value we use for our analysis, where classifications of $P_\mathrm{bzr}>0.99$ are considered blazars and $P_\mathrm{bzr} < 0.01$ are considered pulsars.}
    \label{fig:validation}
\end{figure}

After we find the ideal iteration of our NNC to use, we input the validation sample and record the output blazar probabilities as seen in Fig.~\ref{fig:validation}. With a default 50\% cutoff where probabilities of $P_\mathrm{bzr} > 0.5$ are considered blazars, our NNC has an overall accuracy of 99\% for validation blazars and 100\% for validation pulsars. Our NNC has improved with the updated data releases from the previous catalogs \citep{Kaur2019,Kerby2021a,kerby21b}, and even requiring a validation score above 99\%, we still accurately classify 95\% of validation blazars and pulsars, an increase of 5\% from \citet{kerby21b}. The remaining 5\% of validation sources received ambiguous results ($0.01 < P_\mathrm{bzr} < 0.99$), rather than being classified as the opposing class. Going further, we can also now classify pulsars beyond 99.9\% confidence, where 82.9\% of validation pulsars can be accurately classified at this confidence level rather than the 0\% seen by \citet{kerby21b}. While this shows an improvement since our previous NNCs, we still utilize the 99\% validation cutoff for our purposes. 


\subsection{Results}

With the NNC trained and validated, we input our test sample of 282 X-ray sources for both the primary, $P_\mathrm{bzr}$, and limiting magnitude, $P_\mathrm{lim}$, iterations, which we show in Tables~\ref{tab:class_single} and~\ref{tab:class_multi} for singlets and multiplets, respectively. We find that 227 sources have $P_\mathrm{bzr}>0.99$ and 16 others have $P_\mathrm{bzr}<0.01$, where the remaining 39 sources received $0.01 \leq P_\mathrm{bzr} \leq 0.99$. When we utilize a limiting magnitude of $m_V=30$ for the 65 sources lacking $3\sigma$ detection, 13 more sources increase their $P_\mathrm{lim}>0.99$, and seven sources initially below 0.01 increase their values to $P_\mathrm{lim} > 0.01$. Of the 213 singlet sources with UVOT data, 173 are likely blazars ($P_\mathrm{bzr}>0.99$), and 6 are likely pulsars ($P_\mathrm{bzr}<0.01$), regardless of limiting magnitude. Three more sources started with $P_\mathrm{bzr}<0.01$ but increased their blazar probability when retested with the limiting magnitude. We find ambiguous results ($0.01 \leq P_\mathrm{bzr} \leq 0.99$) for 31 singlets, which decreases to 23 when the limiting magnitude is taken. While sources go from being ambiguous with their given magnitude limit to ``classified'' with our limiting magnitude, we still consider their results as ambiguous.

In cases where multiplets are present, the situation becomes more ambiguous as to whether or not one of the X-ray sources or multiple X-ray sources are related to the gamma-ray source.
At best, the NNC can only produce accurate and viable blazar probabilities for a single source within the field, and all other sources are coincident with the gamma-ray emission. While the NNC results may be accurate for the physical source, we cannot assume the same level of confidence for the multiplets as we do for the singlets.

\section{Discussion}
\label{sec:Results}

\subsection{Summary}

Using \Swift, we detect significant X-ray sources within the uncertainty ellipse for 244 4FGL-DR4 unassociated gamma-ray sources. With publicly available data from \Fermi-LAT, \Swift-XRT, and \Swift-UVOT, we compile the results from X-ray and UV spectral analysis for all significant X-ray sources. We present these data in Tables~\ref{tab:xrt_single} and~\ref{tab:xrt_multi} for \Fermi-LAT sources with single and multiple X-ray sources in their uncertainty ellipses, respectively. Extended versions of these tables and individual UVOT magnitudes of X-ray sources can be found in the attached machine-readable tables. The 218 X-ray sources described in Table~\ref{tab:xrt_single} are the likely lower-energy counterparts of the 4FGL gamma-ray sources as they are the only X-ray source within their uncertainty ellipses. The spectra of these X-ray sources is thus related to the lower-energy emission processes of these sources that also produce high-energy gamma-rays seen by \Fermi. The case is not as clear for the 26 4FGL targets with multiple X-ray sources, but we analyze them using the same methods.

When we input the viable X-ray sources into our NNC, we identify 173 likely blazars $(P_\mathrm{bzr}>0.99)$ and 6 likely pulsars $(P_\mathrm{bzr}>0.99)$ among the singlets. Eleven more sources increase their blazar probability to above the 99\% cutoff when a limiting magnitude of $m_V=30$ is used in place of a lower magnitude limit. The NNC results are fully shown in Tables~\ref{tab:class_single} and~\ref{tab:class_multi} for singlets and multiplets, respectively. 

Further analysis of our 173 likely blazars may yield additional \Fermi blazars to the blazar sequence. Our sources have characteristically lower gamma-ray fluxes than the known samples, with potential implications for fainter luminosities that may further deepen the flux limit of the blazar sequence.

Studying the soft Galactic unassociated (SGU) source Flag (14) in the 4FGL-DR4 data table, we see that of our 218 singlets, 19 are likely associated with SGUs. When we compare these sources to their NNC classifications, 16 are identified as likely pulsars or have received ambiguous results. Expanding this observation to multiplets, four gamma-ray sources are flagged as SGUs, with three of their X-ray multiplets identified as potential pulsars in our NNC.

\subsection{Classification Comparison}\label{sec:simbad}

After calculating blazar probabilities for our entire sample of potential unassociated sources, we utilize the SIMBAD Astronomical Database \citep{Wenger00} to search for potential classified counterparts to our X-ray/UV/optical sources. If we identify a UVOT counterpart to our X-ray source, we utilize the more accurate UVOT coordinates; otherwise, we use the XRT coordinates. With a maximum offset of 5\arcsec\xspace, we locate potential classified counterparts to 109 of our 288 X-ray sources, which we list alongside our NNC results in Tables~\ref{tab:class_single} and~\ref{tab:class_multi}. Of these 109 sources, only SwXF4-DR4 J085905.3-473039 and SwXF4-DR4 J085926.4-434525 are coincident with multiple potential counterparts, and both X-ray sources may be associated with star-forming regions or the young stellar objects within. 

When we compare these potential counterparts to the NNC blazar probabilities, we see that many match our expectations. 
Among the singlets likely associated with blazars or blazar candidates, 36 out of 37 result in $P_\mathrm{bzr}>0.99$, with one blazar candidate having a $P_\mathrm{bzr}$ = 0.989. 
When we expand this to include any galaxy-related sources (e.g., AGN candidates, QSOs, unclassified galaxies), now three have ambiguous results, and a fourth is classified as a likely pulsar out of 59 total singlets. While none of the potential classified counterparts are canonical pulsars or MSPs, we have many stellar classifications of our potential counterparts. Of the 24 classified stellar singlets, 13 received ambiguous classification results from the NNC. The NNC was successfully able to identify some of them as not belonging to any source from the training sample, rather than producing false classifications.

Expanding this comparison to classified X-ray multiplets, we see that they also have classifications that match our results, despite their ambiguity. Among the multiplets likely associated with blazars or blazar candidates, both have $P_\mathrm{bzr}>0.99$. Including all galaxy-related sources, all 16 multiplets received $P_\mathrm{bzr}>0.99$. 

\subsubsection{X-ray Sources of Note}

We catalog 288 X-ray sources within the uncertainty ellipses of 4FGL unassociated gamma-ray sources, and some of them have uncommon classifications. 
While we anticipate that many of our sources will be blazars and pulsars, some X-ray sources may be linked to classified objects that are not typically detected with gamma-rays, such as X-ray binaries and Wolf-Rayet stars.

SwXF4-DR4 J165220.3-452037 is the only likely \Swift X-ray counterpart to 4FGL J1652.2-4516 and is associated with the X-ray transient XTE J1652-453 with a separation of 2.86\arcsec\xspace \citep{markwardt09a}. This X-ray transient was discovered in 2009 during its only observed outburst and analysis supports an X-ray binary system containing a black hole \citep{markwardt09b,hiemstra11}, but it has since entered a period of X-ray quiescence. XTE J1652-453 is the only X-ray source within the uncertainty ellipse of 4FGL J1652.2-4516, yet its relationship is more ambiguous than other single X-ray source targets. XTE J1652-453 is a transient that went into X-ray outburst for $\sim145~$days, but the 4FGL catalog does not usually contain short duration, non-recurrent transients due to its long duration \citep{Abdollahi2020}. No increase in gamma-ray flux was detected during this outburst, but the large uncertainty on the gamma-ray flux may hide this effect. If these sources are associated, they could add to the limited sample of gamma-ray binaries.

SwXF4-DR4 J073733.7+653306 and SwXF4-DR4 J073707.2+653454 are the only two X-ray sources in 4FGL J0737.4+6535, and both of them are associated with X-ray binaries. These two sources are associated with the high-mass X-ray binary (HMXB) 2XMM J073733.4+653308 (1.86\arcsec\xspace offset) and the X-ray binary 2XMM J073707.3+653456 (2.20\arcsec\xspace offset), respectively, in NGC 2403 \citep{mineo12,binder15}. The classification of 2XMM J073707.3+653456 is unclear as it is classified as a HMXB in \citet{mineo12} but a LMXB in \citet{binder15}. While it is unclear which source may have an association with the gamma-ray source 4FGL J0737.4+6535, either association increases the sample of gamma-ray binaries. 

SwXF4-DR4 J174042.0-280727 is likely associated with the X-ray source CXOU J174042.0-280724 with a separation of 2.20\arcsec\xspace. While this connection may seem unremarkable on its own, the detection of a hard, absorbed X-ray spectrum and a hard X-ray outburst strengthens its identification as a low-energy counterpart to the fast, hard X-ray transient IGR J17407-2808 \citep{heinke09,romano11}. Current evidence suggests it may be an unusual LMXB \citep{romano16,ducci23}. Notably, SwXF4-DR4 J174042.0-280727 is the only X-ray source within the uncertainty ellipse of 4FGL J1740.6-2808, supporting a potential association. If CXOU J174042.0-280724/IGR J17407-2808 is ultimately classified as a LMXB, its connection with 4FGL J1740.6-2808 may classify it as a gamma-ray binary.

SwXF4-DR4 J101626.1-572804 is likely associated with the evolved, massive star, WR 17-1 with a separation of 1.99\arcsec\xspace \citep{shara09}, and it is the only X-ray source within 4FGL J1016.2-5729c. While a lone Wolf-Rayet star likely cannot produce gamma-rays, a Wolf-Rayet star in a binary system can produce gamma-rays as seen with Cygnus X-3 \citep{dudus13}. 

SwXF4-DR4 J173411.3-293115 of 4FGL J1734.0-2933 may be associated with the cataclysmic binary CXOGBS J173411.3-293117 with a separation of 3.78\arcsec\xspace \citep{torres14}. This source is the only significant X-ray source within 4FGL J1734.0-2933. \citet{torres14} analyzed the optical spectrum of the binary and found it to be accretion-dominated.

SwXF4-DR4 J132049.9-623749 is one of seven X-ray sources within 4FGL J1321.1-6239, and it is likely associated with the nova V1047 Cen with a separation of 0.31\arcsec\xspace \citep{liller05}. While this nova erupted in 2005, it appears to have rebrightened in 2019 by 2.5 magnitude in optical bands \citep{mroz19,aydi22}, but this did not translate into an X-ray rebrightening. \Swift observed both this initial eruption in 2005 and rebrightening in 2019, yet we detect no significant X-ray source in the summed field after 2006. 

\subsection{Potential Radio Counterparts}

Other than potential classified counterparts, we also attempt to identify potential radio counterparts for our unassociated sources, the results of which we present in Tables~\ref{tab:radio_single} and~\ref{tab:radio_multi}. Given the relativistic jets associated with high-energy sources, we expect that many of our gamma-ray sources should also have radio emission at some level \citep{Ackermann11}. We search the Very Large Array Sky Survey (2 - 4~GHz)\citep[VLASS;][]{Gordon21} and the Rapid ASKAP Continuum Survey (0.75 - 1.5~GHz)\citep[RACS;][]{Hale21} radio surveys for radio sources within 5\arcsec\xspace of our X-ray (or UV/optical) source.  If we locate sources from both surveys, we have a preference for the VLASS data due to its higher angular resolution in comparison to RACS. 

For our 288 X-ray sources, we identify 144 potential radio counterparts, of which 130 are X-ray singlets. In addition to the radio flux density, Tables~\ref{tab:radio_single} and~\ref{tab:radio_multi} also include estimates on the radio loudness parameter, $R$, of each source by calculating the ratio of the radio to $V$-band flux densities, $F_\mathrm{rad}/F_\nu$. For sources with a radio counterpart and measured or lower limits for the UVOT magnitude, we can produce estimates on $R$. Following the definition of radio-loud populations from \citet{Kellerman89}, we find that among sources with potential radio counterparts, only SwXF4-DR4 J141546.0-150228 would be considered radio quiet, but this may be an error as the listed integrated flux density in the VLASS Quick Look catalog is 0.0~mJy.

\subsection{Comparison to \citet{kerby21b} and Others}

This paper seeks to be a 4FGL-DR4 continuation of the 4FGL-DR2 data in \citet{kerby21b}, but there are some differences. Between the two catalogs, we expect to share many X-ray sources. Of their 205 X-ray/UV sources within 4FGL unassociated targets, we share an overall 144 (50\%) that are labeled in Tables~\ref{tab:xrt_single} and~\ref{tab:xrt_multi}. All of these sources, except for one, match their SwXF4 counterparts within their expected positional uncertainty. SwXF4-DR4 J012621.9-674624 is the one exception with an offset from its \citet{kerby21b} counterpart, SwXF4 J012621.6-674638, of 14\arcsec. It is unclear how this offset originated, but it can also be seen in \citet{stroh13} where source 3 of 4FGL J0126.3-6746 is visually offset from an apparent center.

In addition to the X-ray sources seen in both samples, there are 61 sources in \citet{kerby21b} that are no longer part of our unassociated sample. 42 of these sources are no longer included because their 4FGL target became associated with a low-energy counterpart between 4FGL-DR2 and 4FGL-DR4. Twelve sources have a lower signal-to-noise ratio than our cutoff of 4 due to new observations or shifts in processing. Two sources are now within the positional uncertainty of Tycho-2 sources. SwXF4 J112624.9-500808 is likely associated with GRB 140719A, so it cannot be the low-energy counterpart for 4FGL J1126.0-5007. Finally, five sources are no longer within the uncertainty ellipse of the 4FGL targets. Looking into \citet{kerby21b}, their uncertainty ellipses are about 90\degr\xspace rotationally offset from the true orientation of the ellipses. When we corrected this offset, this caused five sources from \citet{kerby21b} to move outside the uncertainty ellipses.

Other than sources common to or removed since \citet{kerby21b}, we identify 144 new X-ray sources within the uncertainty regions of 4FGL gamma-ray sources, with an overall 105 4FGL targets new to our analysis. This includes 94 new X-ray singlets likely associated with their 4FGL targets.

Additionally, we analyze and process optical data differently. For our X-ray sources, we estimate the $V$-band magnitude differently, where \citet{kerby21b} assumed a fixed optical slope of $\alpha_c=0.5$, and we measure an approximate optical slope using UVOT and optical catalogs from VizieR for a more accurate $V$-magnitude estimate. Additionally, rather than analyzing just the measured UVOT magnitude limit for non-detected sources, we now analyze the NNC results for both this magnitude limit and a set limit of $m_V=30$ to study the range of blazar probabilities. 

When we compare the classifications of sources found in the analyses of both our and \citet{kerby21b}, a significant majority match. Of the 143 X-ray sources classified in both analyses, 106 X-ray singlets and 18 multiplets agree in classification (including ambiguous results). There are an additional nine singlets and one multiplet that received ambiguous results previously and are now likely blazars, along with one new likely pulsar following this route. The remaining sources received ambiguous results here, where only two sources go from being classified as likely blazars in \citep{kerby21b} to likely pulsars here.

Other than \citet{kerby21b}, we also compare our classification results with \citet{Zhu24},\citet{Yang24}, and \citet{Pathania26}, all of whom use machine-learning (ML) algorithms to classify 4FGL unassociated gamma-ray sources. 

\citet{Zhu24} utilizes a combination of all source gamma-ray parameters and four ML algorithms, with a voting ensemble classifier, to classify unassociated gamma-ray sources at high and low galactic latitudes. They classified all high latitude sources as AGN-like or PSR-like with no ambiguous results, and all low latitude sources as AGN-like, PSR-like, or ``other''. These ``other'' sources are trained on actual non-blazar/pulsar sources rather than classification uncertainty, so they cannot be accurately compared to our ambiguous classification results. Unlike our catalog, they focus solely on the gamma-ray source, so we will compare all singlets and multiplets to their respective 4FGL counterpart. Comparing only their AGN-like and PSR-like sources, 162 out of 183 X-ray singlets and 54 out of 57 multiplets, related to 22 4FGL sources, have matching classifications. Only one of their AGN-like sources was classified as a likely pulsar, here. Among the shared sources with ``other'' classifications, 14 out of 23 singlets and 9 out of 13 multiplets, related to four 4FGL sources, receive ambiguous results in our analysis.

\citet{Yang24} use multiwavelength observations of potential X-ray counterparts of 4FGL unassociated gamma-ray sources to classify sources with a random forest (RF) ML algorithm. Rather than classify them as likely blazars or pulsars, \citet{Yang24} classifies sources as AGN, low-mass stars, high-mass stars, YSOs, cataclysmic variables, HMXBs, LMXBs, or isolated neutron stars. Overall, we share 12 classified X-ray sources with seven having non-ambiguous classifications in our work. Four of these seven sources have matching classifications and another four receiving ambiguous results in our work are classified as non-AGN/neutron star-like in \citet{Yang24}. One of our likely blazars was classified as a neutron star, one of their neutron stars received ambiguous results from our analysis, and the remaining two sources classified in our work receive non-AGN/neutron star-like classifications in \citet{Yang24}.

\citet{Pathania26} use eight gamma-ray features as input parameters into two ML algorithms, RF and Extreme Gradient Boosting (XGB), to classify unassociated sources as pulsar or blazars. We share 219 4FGL targets between 197 singlets and 63 multiplets. 148 singlets and 51 multiplets have matching classifications, and the remaining sources received ambiguous results in one or both classifications.

The vast majority of our X-ray sources have matching results to these other ML classifiers, regardless of algorithm or input parameters. These results help validate the use of this NNC to classify unassociated gamma-ray sources.


\subsection{Next Steps}
As these 4FGL sources and others receive more \Swift observations, we will continue to improve the analysis of our current sources while also discovering previously missed potential low-energy counterparts. \Swift plans on continuing follow-up of the 2425 4FGL-DR4 unassociated sources into the future. More exposure time will ideally reveal dimmer X-ray sources. Additionally, many of these sources currently lack \Swift observations. Continuing forward, we expect more of these unobserved 4FGL targets to host detectable X-ray sources for future analysis.

To classify our sample of new likely blazars and pulsars, we apply a multilayer perceptron NNC approach. While a direct comparison using other ML algorithms with our input parameters is beyond the scope of the current work, we found in prior studies that this NNC approach was better at discriminating between source types when compared to decision tree and RF techniques \citep{Kerby2021a,kerby21b}. Future work could include further comparison to other classification methods.

Utilizing the classification results of our NNC, we plan to study the sample of likely blazars further by creating a new NNC to specifically classify the blazars as FSRQ or BL Lac, like previously done by \citet{Kaur23} for an earlier catalog. Further still, proving the validity of our association and classification of 4FGL-DR4 unassociated gamma-ray sources increases the sample of gamma-ray blazars and pulsars.
Additional discoveries can help fill in the completeness of the blazer sequence \citep{Ghisellini2008,Ghisellini2017} by including possibly lower luminosity blazars to the associated sources in the \Fermi blazar catalog \citep{Ajello20}. Additional discoveries of pulsars benefit the smaller population of \Fermi pulsars \citep{Smith23}. With these sources classified, we can then identify other, more uncommon gamma-ray sources that do not match blazars or pulsars.


\begin{acknowledgments}
\section*{Acknowledgements}
The authors would like to express their gratitude to the \Swift Science Operations Team for their dedicated efforts in scheduling the \Swift observations of the 4FGL unassociated gamma-ray sources.

This research has made use of data and/or software provided by the High Energy Astrophysics Science Archive Research Center (HEASARC), which is a service of the Astrophysics Science Division at NASA/GSFC. We gratefully acknowledge the support of NASA grants 80NSSC20K1526 and 80NSSC22K1588, and J. Eric Grove is supported for Fermi-LAT work under NASA contract S-15633Y.


\end{acknowledgments}

%

\vspace{5mm}
\facilities{Swift(XRT and UVOT)}


\software{\texttt{Astropy} \citep{astropy22}, \texttt{FTOOLS} \citep{FTOOLS}, \texttt{Matplotlib} \citep{Matplotlib}, \texttt{NumPy} \citep{NumPy}, \texttt{scikit-learn} \citep{Pedregosa11}}





\section*{Data Availability}
All data discussed in this catalog is available as CDS Machine-Readable Tables (MRT) in the online journal as supplementary material. A description of the columns is included as Table~\ref{tab:mrt}. 


\bibliography{4FGL}{}

@book{Bishop06,
  author = {Bishop, Christopher M.},
  title = {Pattern Recognition and Machine Learning},
  year = {2006},
  publisher = {Springer},
  address = {New York},
  isbn = {0387310738},
  series = {Information Science and Statistics}
}

@ARTICLE{Pathania26,
       author = {{Pathania}, A. and {Singh}, K.~K. and {Singh}, S.~K. and {Tolamatti}, A. and {Singh}, B.~B. and {Yadav}, K.~K.},
        title = "{Identification of gamma ray pulsar candidates in the Fermi-LAT 4FGL-DR4 unassociated sources using supervised machine learning}",
      journal = {Astroparticle Physics},
         year = 2026,
        month = feb,
       volume = {175},
          eid = {103185},
        pages = {103185},
          doi = {10.1016/j.astropartphys.2025.103185},
archivePrefix = {arXiv},
       eprint = {2510.08654},
 primaryClass = {astro-ph.HE},
       adsurl = {https://ui.adsabs.harvard.edu/abs/2026APh...17503185P}
}

@ARTICLE{Zhu24,
       author = {{Zhu}, K.~R. and {Chen}, J.~M. and {Zheng}, Y.~G. and {Zhang}, L.},
        title = "{Classifications of Fermi-LAT unassociated sources in multiple machine learning methods}",
      journal = {\mnras},
         year = 2024,
        month = jan,
       volume = {527},
       number = {2},
        pages = {1794-1812},
          doi = {10.1093/mnras/stad2813},
archivePrefix = {arXiv},
       eprint = {2311.03678},
 primaryClass = {astro-ph.HE},
       adsurl = {https://ui.adsabs.harvard.edu/abs/2024MNRAS.527.1794Z}
}

@ARTICLE{Yang24,
       author = {{Yang}, Hui and {Hare}, Jeremy and {Kargaltsev}, Oleg},
        title = "{A Multiwavelength Machine-learning Approach to Classifying X-Ray Sources in the Fields of Unidentified 4FGL-DR4 Sources}",
      journal = {\apj},
         year = 2024,
        month = aug,
       volume = {971},
       number = {2},
          eid = {180},
        pages = {180},
          doi = {10.3847/1538-4357/ad543e},
archivePrefix = {arXiv},
       eprint = {2403.05068},
 primaryClass = {astro-ph.HE},
       adsurl = {https://ui.adsabs.harvard.edu/abs/2024ApJ...971..180Y}
}

@article{Bethapudi18,
title = {Separation of pulsar signals from noise using supervised machine learning algorithms},
journal = {Astronomy and Computing},
volume = {23},
pages = {15-26},
year = {2018},
issn = {2213-1337},
doi = {https://doi.org/10.1016/j.ascom.2018.02.002},
url = {https://www.sciencedirect.com/science/article/pii/S2213133717301397},
author = {S. Bethapudi and S. Desai},
}

@INCOLLECTION{Slane17,
       author = {{Slane}, Patrick},
        title = "{Pulsar Wind Nebulae}",
    booktitle = {Handbook of Supernovae},
         year = 2017,
       editor = {{Alsabti}, Athem W. and {Murdin}, Paul},
        pages = {2159},
          doi = {10.1007/978-3-319-21846-5_95},
       adsurl = {https://ui.adsabs.harvard.edu/abs/2017hsn..book.2159S}
}

@article{Abdollahi22,
doi = {10.3847/1538-4365/ac6751},
url = {https://doi.org/10.3847/1538-4365/ac6751},
year = {2022},
month = {jun},
publisher = {The American Astronomical Society},
volume = {260},
number = {2},
pages = {53},
author = {Abdollahi, S. and Acero, F. and Baldini, L. and Ballet, J. and Bastieri, D. and Bellazzini, R. and Berenji, B. and Berretta, A. and Bissaldi, E. and Blandford, R. D. and Bloom, E. and Bonino, R. and Brill, A. and Britto, R. J. and Bruel, P. and Burnett, T. H. and Buson, S. and Cameron, R. A. and Caputo, R. and Caraveo, P. A. and Castro, D. and Chaty, S. and Cheung, C. C. and Chiaro, G. and Cibrario, N. and Ciprini, S. and Coronado-Blázquez, J. and Crnogorcevic, M. and Cutini, S. and D’Ammando, F. and De Gaetano, S. and Digel, S. W. and Di Lalla, N. and Dirirsa, F. and Di Venere, L. and Domínguez, A. and Fallah Ramazani, V. and Fegan, S. J. and Ferrara, E. C. and Fiori, A. and Fleischhack, H. and Franckowiak, A. and Fukazawa, Y. and Funk, S. and Fusco, P. and Galanti, G. and Gammaldi, V. and Gargano, F. and Garrappa, S. and Gasparrini, D. and Giacchino, F. and Giglietto, N. and Giordano, F. and Giroletti, M. and Glanzman, T. and Green, D. and Grenier, I. A. and Grondin, M.-H. and Guillemot, L. and Guiriec, S. and Gustafsson, M. and Harding, A. K. and Hays, E. and Hewitt, J. W. and Horan, D. and Hou, X. and Jóhannesson, G. and Karwin, C. and Kayanoki, T. and Kerr, M. and Kuss, M. and Landriu, D. and Larsson, S. and Latronico, L. and Lemoine-Goumard, M. and Li, J. and Liodakis, I. and Longo, F. and Loparco, F. and Lott, B. and Lubrano, P. and Maldera, S. and Malyshev, D. and Manfreda, A. and Martí-Devesa, G. and Mazziotta, M. N. and Mereu, I. and Meyer, M. and Michelson, P. F. and Mirabal, N. and Mitthumsiri, W. and Mizuno, T. and Moiseev, A. A. and Monzani, M. E. and Morselli, A. and Moskalenko, I. V. and Negro, M. and Nuss, E. and Omodei, N. and Orienti, M. and Orlando, E. and Paneque, D. and Pei, Z. and Perkins, J. S. and Persic, M. and Pesce-Rollins, M. and Petrosian, V. and Pillera, R. and Poon, H. and Porter, T. A. and Principe, G. and Rainò, S. and Rando, R. and Rani, B. and Razzano, M. and Razzaque, S. and Reimer, A. and Reimer, O. and Reposeur, T. and Sánchez-Conde, M. and Saz Parkinson, P. M. and Scotton, L. and Serini, D. and Sgrò, C. and Siskind, E. J. and Smith, D. A. and Spandre, G. and Spinelli, P. and Sueoka, K. and Suson, D. J. and Tajima, H. and Tak, D. and Thayer, J. B. and Thompson, D. J. and Torres, D. F. and Troja, E. and Valverde, J. and Wood, K. and Zaharijas, G.},
title = {Incremental Fermi Large Area Telescope Fourth Source Catalog},
journal = {The Astrophysical Journal Supplement Series}
}

@ARTICLE{Ajello20,
       author = {{Ajello}, M. and {Angioni}, R. and {Axelsson}, M. and {Ballet}, J. and {Barbiellini}, G. and {Bastieri}, D. and {Becerra Gonzalez}, J. and {Bellazzini}, R. and {Bissaldi}, E. and {Bloom}, E.~D. and {Bonino}, R. and {Bottacini}, E. and {Bruel}, P. and {Buson}, S. and {Cafardo}, F. and {Cameron}, R.~A. and {Cavazzuti}, E. and {Chen}, S. and {Cheung}, C.~C. and {Ciprini}, S. and {Costantin}, D. and {Cutini}, S. and {D'Ammando}, F. and {de la Torre Luque}, P. and {de Menezes}, R. and {de Palma}, F. and {Desai}, A. and {Di Lalla}, N. and {Di Venere}, L. and {Dom{\'\i}nguez}, A. and {Dirirsa}, F. Fana and {Ferrara}, E.~C. and {Finke}, J. and {Franckowiak}, A. and {Fukazawa}, Y. and {Funk}, S. and {Fusco}, P. and {Gargano}, F. and {Garrappa}, S. and {Gasparrini}, D. and {Giglietto}, N. and {Giordano}, F. and {Giroletti}, M. and {Green}, D. and {Grenier}, I.~A. and {Guiriec}, S. and {Harita}, S. and {Hays}, E. and {Horan}, D. and {Itoh}, R. and {J{\'o}hannesson}, G. and {Kovac'evic'}, M. and {Krauss}, F. and {Kreter}, M. and {Kuss}, M. and {Larsson}, S. and {Leto}, C. and {Li}, J. and {Liodakis}, I. and {Longo}, F. and {Loparco}, F. and {Lott}, B. and {Lovellette}, M.~N. and {Lubrano}, P. and {Madejski}, G.~M. and {Maldera}, S. and {Manfreda}, A. and {Mart{\'\i}-Devesa}, G. and {Massaro}, F. and {Mazziotta}, M.~N. and {Mereu}, I. and {Meyer}, M. and {Migliori}, G. and {Mirabal}, N. and {Mizuno}, T. and {Monzani}, M.~E. and {Morselli}, A. and {Moskalenko}, I.~V. and {Negro}, M. and {Nemmen}, R. and {Nuss}, E. and {Ojha}, L.~S. and {Ojha}, R. and {Omodei}, N. and {Orienti}, M. and {Orlando}, E. and {Ormes}, J.~F. and {Paliya}, V.~S. and {Pei}, Z. and {Pe{\~n}a-Herazo}, H. and {Persic}, M. and {Pesce-Rollins}, M. and {Petrov}, L. and {Piron}, F. and {Poon}, H. and {Principe}, G. and {Rain{\`o}}, S. and {Rando}, R. and {Rani}, B. and {Razzano}, M. and {Razzaque}, S. and {Reimer}, A. and {Reimer}, O. and {Schinzel}, F.~K. and {Serini}, D. and {Sgr{\`o}}, C. and {Siskind}, E.~J. and {Spandre}, G. and {Spinelli}, P. and {Suson}, D.~J. and {Tachibana}, Y. and {Thompson}, D.~J. and {Torres}, D.~F. and {Torresi}, E. and {Troja}, E. and {Valverde}, J. and {van Zyl}, P. and {Yassine}, M.},
        title = "{The Fourth Catalog of Active Galactic Nuclei Detected by the Fermi Large Area Telescope}",
      journal = {\apj},
         year = 2020,
        month = apr,
       volume = {892},
       number = {2},
          eid = {105},
        pages = {105},
          doi = {10.3847/1538-4357/ab791e},
archivePrefix = {arXiv},
       eprint = {1905.10771},
 primaryClass = {astro-ph.HE},
       adsurl = {https://ui.adsabs.harvard.edu/abs/2020ApJ...892..105A}
}

@ARTICLE{Smith23,
       author = {{Smith}, D.~A. and {Abdollahi}, S. and {Ajello}, M. and {Bailes}, M. and {Baldini}, L. and {Ballet}, J. and {Baring}, M.~G. and {Bassa}, C. and {Gonzalez}, J. Becerra and {Bellazzini}, R. and {Berretta}, A. and {Bhattacharyya}, B. and {Bissaldi}, E. and {Bonino}, R. and {Bottacini}, E. and {Bregeon}, J. and {Bruel}, P. and {Burgay}, M. and {Burnett}, T.~H. and {Cameron}, R.~A. and {Camilo}, F. and {Caputo}, R. and {Caraveo}, P.~A. and {Cavazzuti}, E. and {Chiaro}, G. and {Ciprini}, S. and {Clark}, C.~J. and {Cognard}, I. and {Corongiu}, A. and {Orestano}, P. Cristarella and {Crnogorcevic}, M. and {Cuoco}, A. and {Cutini}, S. and {D'Ammando}, F. and {de Angelis}, A. and {DeCesar}, M.~E. and {De Gaetano}, S. and {de Menezes}, R. and {Deneva}, J. and {de Palma}, F. and {Di Lalla}, N. and {Dirirsa}, F. and {Di Venere}, L. and {Dom{\'\i}nguez}, A. and {Dumora}, D. and {Fegan}, S.~J. and {Ferrara}, E.~C. and {Fiori}, A. and {Fleischhack}, H. and {Flynn}, C. and {Franckowiak}, A. and {Freire}, P.~C.~C. and {Fukazawa}, Y. and {Fusco}, P. and {Galanti}, G. and {Gammaldi}, V. and {Gargano}, F. and {Gasparrini}, D. and {Giacchino}, F. and {Giglietto}, N. and {Giordano}, F. and {Giroletti}, M. and {Green}, D. and {Grenier}, I.~A. and {Guillemot}, L. and {Guiriec}, S. and {Gustafsson}, M. and {Harding}, A.~K. and {Hays}, E. and {Hewitt}, J.~W. and {Horan}, D. and {Hou}, X. and {Jankowski}, F. and {Johnson}, R.~P. and {Johnson}, T.~J. and {Johnston}, S. and {Kataoka}, J. and {Keith}, M.~J. and {Kerr}, M. and {Kramer}, M. and {Kuss}, M. and {Latronico}, L. and {Lee}, S. -H. and {Li}, D. and {Li}, J. and {Limyansky}, B. and {Longo}, F. and {Loparco}, F. and {Lorusso}, L. and {Lovellette}, M.~N. and {Lower}, M. and {Lubrano}, P. and {Lyne}, A.~G. and {Maan}, Y. and {Maldera}, S. and {Manchester}, R.~N. and {Manfreda}, A. and {Marelli}, M. and {Mart{\'\i}-Devesa}, G. and {Mazziotta}, M.~N. and {McEnery}, J.~E. and {Mereu}, I. and {Michelson}, P.~F. and {Mickaliger}, M. and {Mitthumsiri}, W. and {Mizuno}, T. and {Moiseev}, A.~A. and {Monzani}, M.~E. and {Morselli}, A. and {Negro}, M. and {Nemmen}, R. and {Nieder}, L. and {Nuss}, E. and {Omodei}, N. and {Orienti}, M. and {Orlando}, E. and {Ormes}, J.~F. and {Palatiello}, M. and {Paneque}, D. and {Panzarini}, G. and {Parthasarathy}, A. and {Persic}, M. and {Pesce-Rollins}, M. and {Pillera}, R. and {Poon}, H. and {Porter}, T.~A. and {Possenti}, A. and {Principe}, G. and {Rain{\`o}}, S. and {Rando}, R. and {Ransom}, S.~M. and {Ray}, P.~S. and {Razzano}, M. and {Razzaque}, S. and {Reimer}, A. and {Reimer}, O. and {Renault-Tinacci}, N. and {Romani}, R.~W. and {S{\'a}nchez-Conde}, M. and {Parkinson}, P.~M. Saz and {Scotton}, L. and {Serini}, D. and {Sgr{\`o}}, C. and {Shannon}, R. and {Sharma}, V. and {Shen}, Z. and {Siskind}, E.~J. and {Spandre}, G. and {Spinelli}, P. and {Stappers}, B.~W. and {Stephens}, T.~E. and {Suson}, D.~J. and {Tabassum}, S. and {Tajima}, H. and {Tak}, D. and {Theureau}, G. and {Thompson}, D.~J. and {Tibolla}, O. and {Torres}, D.~F. and {Valverde}, J. and {Venter}, C. and {Wadiasingh}, Z. and {Wang}, N. and {Wang}, N. and {Wang}, P. and {Weltevrede}, P. and {Wood}, K. and {Yan}, J. and {Zaharijas}, G. and {Zhang}, C. and {Zhu}, W.},
        title = "{The Third Fermi Large Area Telescope Catalog of Gamma-Ray Pulsars}",
      journal = {\apj},
         year = 2023,
        month = dec,
       volume = {958},
       number = {2},
          eid = {191},
        pages = {191},
          doi = {10.3847/1538-4357/acee67},
archivePrefix = {arXiv},
       eprint = {2307.11132},
 primaryClass = {astro-ph.HE},
       adsurl = {https://ui.adsabs.harvard.edu/abs/2023ApJ...958..191S}
}

@ARTICLE{Ulgiati25,
       author = {{Ulgiati}, A. and {Paiano}, S. and {Pintore}, F. and {Russell}, T.~D. and {Sbarufatti}, B. and {Pinto}, C. and {Ambrosi}, E. and {D'A{\`\i}}, A. and {Cusumano}, G. and {Del Santo}, M.},
        title = "{Search for the multi-wavelength counterparts to extragalactic unassociated Fermi {\ensuremath{\gamma}}-ray sources}",
      journal = {\aap},
         year = 2025,
        month = feb,
       volume = {694},
          eid = {A176},
        pages = {A176},
          doi = {10.1051/0004-6361/202452366},
archivePrefix = {arXiv},
       eprint = {2412.19314},
 primaryClass = {astro-ph.HE},
       adsurl = {https://ui.adsabs.harvard.edu/abs/2025A&A...694A.176U}
}

@ARTICLE{Ackermann11,
       author = {{Ackermann}, M. and {Ajello}, M. and {Allafort}, A. and {Angelakis}, E. and {Axelsson}, M. and {Baldini}, L. and {Ballet}, J. and {Barbiellini}, G. and {Bastieri}, D. and {Bellazzini}, R. and {Berenji}, B. and {Blandford}, R.~D. and {Bloom}, E.~D. and {Bonamente}, E. and {Borgland}, A.~W. and {Bouvier}, A. and {Bregeon}, J. and {Brez}, A. and {Brigida}, M. and {Bruel}, P. and {Buehler}, R. and {Buson}, S. and {Caliandro}, G.~A. and {Cameron}, R.~A. and {Cannon}, A. and {Caraveo}, P.~A. and {Casandjian}, J.~M. and {Cavazzuti}, E. and {Cecchi}, C. and {Charles}, E. and {Chekhtman}, A. and {Cheung}, C.~C. and {Ciprini}, S. and {Claus}, R. and {Cohen-Tanugi}, J. and {Cutini}, S. and {de Palma}, F. and {Dermer}, C.~D. and {Silva}, E. do Couto e. and {Drell}, P.~S. and {Dubois}, R. and {Dumora}, D. and {Escande}, L. and {Favuzzi}, C. and {Fegan}, S.~J. and {Focke}, W.~B. and {Fortin}, P. and {Frailis}, M. and {Fuhrmann}, L. and {Fukazawa}, Y. and {Fusco}, P. and {Gargano}, F. and {Gasparrini}, D. and {Gehrels}, N. and {Giglietto}, N. and {Giommi}, P. and {Giordano}, F. and {Giroletti}, M. and {Glanzman}, T. and {Godfrey}, G. and {Grandi}, P. and {Grenier}, I.~A. and {Guiriec}, S. and {Hadasch}, D. and {Hayashida}, M. and {Hays}, E. and {Healey}, S.~E. and {J{\'o}hannesson}, G. and {Johnson}, A.~S. and {Kamae}, T. and {Katagiri}, H. and {Kataoka}, J. and {Kn{\"o}dlseder}, J. and {Kuss}, M. and {Lande}, J. and {Lee}, S. -H. and {Longo}, F. and {Loparco}, F. and {Lott}, B. and {Lovellette}, M.~N. and {Lubrano}, P. and {Makeev}, A. and {Max-Moerbeck}, W. and {Mazziotta}, M.~N. and {McEnery}, J.~E. and {Mehault}, J. and {Michelson}, P.~F. and {Mizuno}, T. and {Monte}, C. and {Monzani}, M.~E. and {Morselli}, A. and {Moskalenko}, I.~V. and {Murgia}, S. and {Naumann-Godo}, M. and {Nishino}, S. and {Nolan}, P.~L. and {Norris}, J.~P. and {Nuss}, E. and {Ohsugi}, T. and {Okumura}, A. and {Omodei}, N. and {Orlando}, E. and {Ormes}, J.~F. and {Ozaki}, M. and {Paneque}, D. and {Pavlidou}, V. and {Pelassa}, V. and {Pepe}, M. and {Pesce-Rollins}, M. and {Pierbattista}, M. and {Piron}, F. and {Porter}, T.~A. and {Rain{\`o}}, S. and {Razzano}, M. and {Readhead}, A. and {Reimer}, A. and {Reimer}, O. and {Richards}, J.~L. and {Romani}, R.~W. and {Sadrozinski}, H.~F. -W. and {Scargle}, J.~D. and {Sgr{\`o}}, C. and {Siskind}, E.~J. and {Smith}, P.~D. and {Spandre}, G. and {Spinelli}, P. and {Strickman}, M.~S. and {Suson}, D.~J. and {Takahashi}, H. and {Tanaka}, T. and {Taylor}, G.~B. and {Thayer}, J.~G. and {Thayer}, J.~B. and {Thompson}, D.~J. and {Torres}, D.~F. and {Tosti}, G. and {Tramacere}, A. and {Troja}, E. and {Vandenbroucke}, J. and {Vianello}, G. and {Vitale}, V. and {Waite}, A.~P. and {Wang}, P. and {Winer}, B.~L. and {Wood}, K.~S. and {Yang}, Z. and {Ziegler}, M.},
        title = "{The Radio/Gamma-Ray Connection in Active Galactic Nuclei in the Era of the Fermi Large Area Telescope}",
      journal = {\apj},
         year = 2011,
        month = nov,
       volume = {741},
       number = {1},
          eid = {30},
        pages = {30},
          doi = {10.1088/0004-637X/741/1/30},
archivePrefix = {arXiv},
       eprint = {1108.0501},
 primaryClass = {astro-ph.CO},
       adsurl = {https://ui.adsabs.harvard.edu/abs/2011ApJ...741...30A}
}

@article{Pedregosa11,
author = {Pedregosa, Fabian and Varoquaux, Gaël and Gramfort, Alexandre and Michel, Vincent and Thirion, Bertrand and Grisel, Olivier and Blondel, Mathieu and Prettenhofer, Peter and Weiss, Ron and Dubourg, Vincent and Vanderplas, Jake and Passos, Alexandre and Cournapeau, David and Brucher, Matthieu and Perrot, Matthieu and Duchesnay, Édouard},
journal = {Journal of Machine Learning Research},
pages = {2825--2830},
title = {{Scikit-learn: Machine Learning in Python}},
url = {https://jmlr.csail.mit.edu/papers/v12/pedregosa11a.html},
volume = {12},
year = {2011}
}

@INPROCEEDINGS{Melrose17,
       author = {{Melrose}, D.~B. and {Rafat}, M.~Z.},
        title = "{Pulsar radio emission mechanism: Why no consensus?}",
    booktitle = {Journal of Physics Conference Series},
         year = 2017,
       series = {Journal of Physics Conference Series},
       volume = {932},
        month = dec,
    publisher = {IOP},
          eid = {012011},
        pages = {012011},
          doi = {10.1088/1742-6596/932/1/012011},
       adsurl = {https://ui.adsabs.harvard.edu/abs/2017JPhCS.932a2011M}
}

@ARTICLE{Lin16,
       author = {{Lin}, Lupin Chun-Che},
        title = "{Radio-quiet Gamma-ray Pulsars}",
      journal = {Journal of Astronomy and Space Sciences},
         year = 2016,
        month = sep,
       volume = {33},
        pages = {147-166},
          doi = {10.5140/JASS.2016.33.3.147},
       adsurl = {https://ui.adsabs.harvard.edu/abs/2016JASS...33..147L}
}

@ARTICLE{Boula18,
       author = {{Boula}, Stella and {Petropoulou}, Maria and {Mastichiadis}, Apostolos},
        title = "{On the Connection of Radio and {\ensuremath{\gamma}}-Ray Emission in Blazars}",
      journal = {Galaxies},
         year = 2018,
        month = dec,
       volume = {7},
       number = {1},
          eid = {3},
        pages = {3},
          doi = {10.3390/galaxies7010003},
archivePrefix = {arXiv},
       eprint = {1901.08793},
 primaryClass = {astro-ph.HE},
       adsurl = {https://ui.adsabs.harvard.edu/abs/2018Galax...7....3B}
}

@ARTICLE{Cheng96,
       author = {{Cheng}, K.~S. and {Zhang}, J.~L.},
        title = "{General Radiation Formulae for a Relativistic Charged Particle Moving in Curved Magnetic Field Lines: The Synchrocurvature Radiation Mechanism}",
      journal = {\apj},
         year = 1996,
        month = may,
       volume = {463},
        pages = {271},
          doi = {10.1086/177239},
       adsurl = {https://ui.adsabs.harvard.edu/abs/1996ApJ...463..271C}
}

@ARTICLE{Rodrigues21,
       author = {{Rodrigues}, Xavier and {Heinze}, Jonas and {Palladino}, Andrea and {van Vliet}, Arjen and {Winter}, Walter},
        title = "{Active Galactic Nuclei Jets as the Origin of Ultrahigh-Energy Cosmic Rays and Perspectives for the Detection of Astrophysical Source Neutrinos at EeV Energies}",
      journal = {\prl},
         year = 2021,
        month = may,
       volume = {126},
       number = {19},
          eid = {191101},
        pages = {191101},
          doi = {10.1103/PhysRevLett.126.191101},
archivePrefix = {arXiv},
       eprint = {2003.08392},
 primaryClass = {astro-ph.HE},
       adsurl = {https://ui.adsabs.harvard.edu/abs/2021PhRvL.126s1101R}
}

@ARTICLE{Kellerman89,
       author = {{Kellermann}, K.~I. and {Sramek}, R. and {Schmidt}, M. and {Shaffer}, D.~B. and {Green}, R.},
        title = "{VLA Observations of Objects in the Palomar Bright Quasar Survey}",
      journal = {\aj},
         year = 1989,
        month = oct,
       volume = {98},
        pages = {1195},
          doi = {10.1086/115207},
       adsurl = {https://ui.adsabs.harvard.edu/abs/1989AJ.....98.1195K}
}

@ARTICLE{Hale21,
       author = {{Hale}, Catherine L. and {McConnell}, D. and {Thomson}, A.~J.~M. and {Lenc}, E. and {Heald}, G.~H. and {Hotan}, A.~W. and {Leung}, J.~K. and {Moss}, V.~A. and {Murphy}, T. and {Pritchard}, J. and {Sadler}, E.~M. and {Stewart}, A.~J. and {Whiting}, M.~T.},
        title = "{The Rapid ASKAP Continuum Survey Paper II: First Stokes I Source Catalogue Data Release}",
      journal = {\pasa},
         year = 2021,
        month = nov,
       volume = {38},
          eid = {e058},
        pages = {e058},
          doi = {10.1017/pasa.2021.47},
archivePrefix = {arXiv},
       eprint = {2109.00956},
 primaryClass = {astro-ph.GA},
       adsurl = {https://ui.adsabs.harvard.edu/abs/2021PASA...38...58H}
}

@ARTICLE{Gordon21,
       author = {{Gordon}, Yjan A. and {Boyce}, Michelle M. and {O'Dea}, Christopher P. and {Rudnick}, Lawrence and {Andernach}, Heinz and {Vantyghem}, Adrian N. and {Baum}, Stefi A. and {Bui}, Jean-Paul and {Dionyssiou}, Mathew and {Safi-Harb}, Samar and {Sander}, Isabel},
        title = "{A Quick Look at the 3 GHz Radio Sky. I. Source Statistics from the Very Large Array Sky Survey}",
      journal = {\apjs},
         year = 2021,
        month = aug,
       volume = {255},
       number = {2},
          eid = {30},
        pages = {30},
          doi = {10.3847/1538-4365/ac05c0},
archivePrefix = {arXiv},
       eprint = {2102.11753},
 primaryClass = {astro-ph.GA},
       adsurl = {https://ui.adsabs.harvard.edu/abs/2021ApJS..255...30G}
}

@ARTICLE{Wenger00,
       author = {{Wenger}, M. and {Ochsenbein}, F. and {Egret}, D. and {Dubois}, P. and {Bonnarel}, F. and {Borde}, S. and {Genova}, F. and {Jasniewicz}, G. and {Lalo{\"e}}, S. and {Lesteven}, S. and {Monier}, R.},
        title = "{The SIMBAD astronomical database. The CDS reference database for astronomical objects}",
      journal = {\aaps},
         year = 2000,
        month = apr,
       volume = {143},
        pages = {9-22},
          doi = {10.1051/aas:2000332},
archivePrefix = {arXiv},
       eprint = {astro-ph/0002110},
 primaryClass = {astro-ph},
       adsurl = {https://ui.adsabs.harvard.edu/abs/2000A&AS..143....9W}
}

@ARTICLE{Hu06,
       author = {{Hu}, Shao Ming and {Zhao}, G. and {Guo}, H.~Y. and {Zhang}, X. and {Zheng}, Y.~G.},
        title = "{The optical spectral slope variability of 17 blazars}",
      journal = {\mnras},
         year = 2006,
        month = sep,
       volume = {371},
       number = {3},
        pages = {1243-1250},
          doi = {10.1111/j.1365-2966.2006.10721.x},
archivePrefix = {arXiv},
       eprint = {astro-ph/0609009},
 primaryClass = {astro-ph},
       adsurl = {https://ui.adsabs.harvard.edu/abs/2006MNRAS.371.1243H}
}

@ARTICLE{Ahumada20,
       author = {{Ahumada}, Romina and {Allende Prieto}, Carlos and {Almeida}, Andr{\'e}s and {Anders}, Friedrich and {Anderson}, Scott F. and {Andrews}, Brett H. and {Anguiano}, Borja and {Arcodia}, Riccardo and {Armengaud}, Eric and {Aubert}, Marie and {Avila}, Santiago and {Avila-Reese}, Vladimir and {Badenes}, Carles and {Balland}, Christophe and {Barger}, Kat and {Barrera-Ballesteros}, Jorge K. and {Basu}, Sarbani and {Bautista}, Julian and {Beaton}, Rachael L. and {Beers}, Timothy C. and {Benavides}, B. Izamar T. and {Bender}, Chad F. and {Bernardi}, Mariangela and {Bershady}, Matthew and {Beutler}, Florian and {Bidin}, Christian Moni and {Bird}, Jonathan and {Bizyaev}, Dmitry and {Blanc}, Guillermo A. and {Blanton}, Michael R. and {Boquien}, M{\'e}d{\'e}ric and {Borissova}, Jura and {Bovy}, Jo and {Brandt}, W.~N. and {Brinkmann}, Jonathan and {Brownstein}, Joel R. and {Bundy}, Kevin and {Bureau}, Martin and {Burgasser}, Adam and {Burtin}, Etienne and {Cano-D{\'\i}az}, Mariana and {Capasso}, Raffaella and {Cappellari}, Michele and {Carrera}, Ricardo and {Chabanier}, Sol{\`e}ne and {Chaplin}, William and {Chapman}, Michael and {Cherinka}, Brian and {Chiappini}, Cristina and {Doohyun Choi}, Peter and {Chojnowski}, S. Drew and {Chung}, Haeun and {Clerc}, Nicolas and {Coffey}, Damien and {Comerford}, Julia M. and {Comparat}, Johan and {da Costa}, Luiz and {Cousinou}, Marie-Claude and {Covey}, Kevin and {Crane}, Jeffrey D. and {Cunha}, Katia and {Ilha}, Gabriele da Silva and {Dai}, Yu Sophia and {Damsted}, Sanna B. and {Darling}, Jeremy and {Davidson}, Jr., James W. and {Davies}, Roger and {Dawson}, Kyle and {De}, Nikhil and {de la Macorra}, Axel and {De Lee}, Nathan and {Queiroz}, Anna B{\'a}rbara de Andrade and {Deconto Machado}, Alice and {de la Torre}, Sylvain and {Dell'Agli}, Flavia and {du Mas des Bourboux}, H{\'e}lion and {Diamond-Stanic}, Aleksandar M. and {Dillon}, Sean and {Donor}, John and {Drory}, Niv and {Duckworth}, Chris and {Dwelly}, Tom and {Ebelke}, Garrett and {Eftekharzadeh}, Sarah and {Davis Eigenbrot}, Arthur and {Elsworth}, Yvonne P. and {Eracleous}, Mike and {Erfanianfar}, Ghazaleh and {Escoffier}, Stephanie and {Fan}, Xiaohui and {Farr}, Emily and {Fern{\'a}ndez-Trincado}, Jos{\'e} G. and {Feuillet}, Diane and {Finoguenov}, Alexis and {Fofie}, Patricia and {Fraser-McKelvie}, Amelia and {Frinchaboy}, Peter M. and {Fromenteau}, Sebastien and {Fu}, Hai and {Galbany}, Llu{\'\i}s and {Garcia}, Rafael A. and {Garc{\'\i}a-Hern{\'a}ndez}, D.~A. and {Garma Oehmichen}, Luis Alberto and {Ge}, Junqiang and {Geimba Maia}, Marcio Antonio and {Geisler}, Doug and {Gelfand}, Joseph and {Goddy}, Julian and {Gonzalez-Perez}, Violeta and {Grabowski}, Kathleen and {Green}, Paul and {Grier}, Catherine J. and {Guo}, Hong and {Guy}, Julien and {Harding}, Paul and {Hasselquist}, Sten and {Hawken}, Adam James and {Hayes}, Christian R. and {Hearty}, Fred and {Hekker}, S. and {Hogg}, David W. and {Holtzman}, Jon A. and {Horta}, Danny and {Hou}, Jiamin and {Hsieh}, Bau-Ching and {Huber}, Daniel and {Hunt}, Jason A.~S. and {Ider Chitham}, J. and {Imig}, Julie and {Jaber}, Mariana and {Jimenez Angel}, Camilo Eduardo and {Johnson}, Jennifer A. and {Jones}, Amy M. and {J{\"o}nsson}, Henrik and {Jullo}, Eric and {Kim}, Yerim and {Kinemuchi}, Karen and {Kirkpatrick}, IV, Charles C. and {Kite}, George W. and {Klaene}, Mark and {Kneib}, Jean-Paul and {Kollmeier}, Juna A. and {Kong}, Hui and {Kounkel}, Marina and {Krishnarao}, Dhanesh and {Lacerna}, Ivan and {Lan}, Ting-Wen and {Lane}, Richard R. and {Law}, David R. and {Le Goff}, Jean-Marc and {Leung}, Henry W. and {Lewis}, Hannah and {Li}, Cheng and {Lian}, Jianhui and {Lin}, Lihwai and {Long}, Dan and {Longa-Pe{\~n}a}, Pen{\'e}lope and {Lundgren}, Britt and {Lyke}, Brad W. and {Mackereth}, J. Ted and {MacLeod}, Chelsea L. and {Majewski}, Steven R. and {Manchado}, Arturo and {Maraston}, Claudia and {Martini}, Paul and {Masseron}, Thomas and {Masters}, Karen L. and {Mathur}, Savita and {McDermid}, Richard M. and {Merloni}, Andrea and {Merrifield}, Michael and {M{\'e}sz{\'a}ros}, Szabolcs and {Miglio}, Andrea and {Minniti}, Dante and {Minsley}, Rebecca and {Miyaji}, Takamitsu and {Mohammad}, Faizan Gohar and {Mosser}, Benoit and {Mueller}, Eva-Maria and {Muna}, Demitri and {Mu{\~n}oz-Guti{\'e}rrez}, Andrea and {Myers}, Adam D. and {Nadathur}, Seshadri and {Nair}, Preethi and {Nandra}, Kirpal and {Correa do Nascimento}, Janaina and {Nevin}, Rebecca Jean and {Newman}, Jeffrey A. and {Nidever}, David L. and {Nitschelm}, Christian and {Noterdaeme}, Pasquier and {O'Connell}, Julia E. and {Olmstead}, Matthew D. and {Oravetz}, Daniel and {Oravetz}, Audrey and {Osorio}, Yeisson and {Pace}, Zachary J. and {Padilla}, Nelson and {Palanque-Delabrouille}, Nathalie and {Palicio}, Pedro A.},
        title = "{The 16th Data Release of the Sloan Digital Sky Surveys: First Release from the APOGEE-2 Southern Survey and Full Release of eBOSS Spectra}",
      journal = {\apjs},
         year = 2020,
        month = jul,
       volume = {249},
       number = {1},
          eid = {3},
        pages = {3},
          doi = {10.3847/1538-4365/ab929e},
archivePrefix = {arXiv},
       eprint = {1912.02905},
 primaryClass = {astro-ph.GA},
       adsurl = {https://ui.adsabs.harvard.edu/abs/2020ApJS..249....3A}
}

@ARTICLE{Chambers16,
       author = {{Chambers}, K.~C. and {Magnier}, E.~A. and {Metcalfe}, N. and {Flewelling}, H.~A. and {Huber}, M.~E. and {Waters}, C.~Z. and {Denneau}, L. and {Draper}, P.~W. and {Farrow}, D. and {Finkbeiner}, D.~P. and {Holmberg}, C. and {Koppenhoefer}, J. and {Price}, P.~A. and {Rest}, A. and {Saglia}, R.~P. and {Schlafly}, E.~F. and {Smartt}, S.~J. and {Sweeney}, W. and {Wainscoat}, R.~J. and {Burgett}, W.~S. and {Chastel}, S. and {Grav}, T. and {Heasley}, J.~N. and {Hodapp}, K.~W. and {Jedicke}, R. and {Kaiser}, N. and {Kudritzki}, R. -P. and {Luppino}, G.~A. and {Lupton}, R.~H. and {Monet}, D.~G. and {Morgan}, J.~S. and {Onaka}, P.~M. and {Shiao}, B. and {Stubbs}, C.~W. and {Tonry}, J.~L. and {White}, R. and {Ba{\~n}ados}, E. and {Bell}, E.~F. and {Bender}, R. and {Bernard}, E.~J. and {Boegner}, M. and {Boffi}, F. and {Botticella}, M.~T. and {Calamida}, A. and {Casertano}, S. and {Chen}, W. -P. and {Chen}, X. and {Cole}, S. and {Deacon}, N. and {Frenk}, C. and {Fitzsimmons}, A. and {Gezari}, S. and {Gibbs}, V. and {Goessl}, C. and {Goggia}, T. and {Gourgue}, R. and {Goldman}, B. and {Grant}, P. and {Grebel}, E.~K. and {Hambly}, N.~C. and {Hasinger}, G. and {Heavens}, A.~F. and {Heckman}, T.~M. and {Henderson}, R. and {Henning}, T. and {Holman}, M. and {Hopp}, U. and {Ip}, W. -H. and {Isani}, S. and {Jackson}, M. and {Keyes}, C.~D. and {Koekemoer}, A.~M. and {Kotak}, R. and {Le}, D. and {Liska}, D. and {Long}, K.~S. and {Lucey}, J.~R. and {Liu}, M. and {Martin}, N.~F. and {Masci}, G. and {McLean}, B. and {Mindel}, E. and {Misra}, P. and {Morganson}, E. and {Murphy}, D.~N.~A. and {Obaika}, A. and {Narayan}, G. and {Nieto-Santisteban}, M.~A. and {Norberg}, P. and {Peacock}, J.~A. and {Pier}, E.~A. and {Postman}, M. and {Primak}, N. and {Rae}, C. and {Rai}, A. and {Riess}, A. and {Riffeser}, A. and {Rix}, H.~W. and {R{\"o}ser}, S. and {Russel}, R. and {Rutz}, L. and {Schilbach}, E. and {Schultz}, A.~S.~B. and {Scolnic}, D. and {Strolger}, L. and {Szalay}, A. and {Seitz}, S. and {Small}, E. and {Smith}, K.~W. and {Soderblom}, D.~R. and {Taylor}, P. and {Thomson}, R. and {Taylor}, A.~N. and {Thakar}, A.~R. and {Thiel}, J. and {Thilker}, D. and {Unger}, D. and {Urata}, Y. and {Valenti}, J. and {Wagner}, J. and {Walder}, T. and {Walter}, F. and {Watters}, S.~P. and {Werner}, S. and {Wood-Vasey}, W.~M. and {Wyse}, R.},
        title = "{The Pan-STARRS1 Surveys}",
      journal = {arXiv e-prints},
         year = 2016,
        month = dec,
          eid = {arXiv:1612.05560},
        pages = {arXiv:1612.05560},
          doi = {10.48550/arXiv.1612.05560},
archivePrefix = {arXiv},
       eprint = {1612.05560},
 primaryClass = {astro-ph.IM},
       adsurl = {https://ui.adsabs.harvard.edu/abs/2016arXiv161205560C}
}

@ARTICLE{Bianchi17,
       author = {{Bianchi}, Luciana and {Shiao}, Bernie and {Thilker}, David},
        title = "{Revised Catalog of GALEX Ultraviolet Sources. I. The All-Sky Survey: GUVcat\_AIS}",
      journal = {\apjs},
         year = 2017,
        month = jun,
       volume = {230},
       number = {2},
          eid = {24},
        pages = {24},
          doi = {10.3847/1538-4365/aa7053},
archivePrefix = {arXiv},
       eprint = {1704.05903},
 primaryClass = {astro-ph.GA},
       adsurl = {https://ui.adsabs.harvard.edu/abs/2017ApJS..230...24B}
}

@ARTICLE{Gaia21,
       author = {{Gaia Collaboration} and {Brown}, A.~G.~A. and {Vallenari}, A. and {Prusti}, T. and {de Bruijne}, J.~H.~J. and {Babusiaux}, C. and {Biermann}, M. and {Creevey}, O.~L. and {Evans}, D.~W. and {Eyer}, L. and {Hutton}, A. and {Jansen}, F. and {Jordi}, C. and {Klioner}, S.~A. and {Lammers}, U. and {Lindegren}, L. and {Luri}, X. and {Mignard}, F. and {Panem}, C. and {Pourbaix}, D. and {Randich}, S. and {Sartoretti}, P. and {Soubiran}, C. and {Walton}, N.~A. and {Arenou}, F. and {Bailer-Jones}, C.~A.~L. and {Bastian}, U. and {Cropper}, M. and {Drimmel}, R. and {Katz}, D. and {Lattanzi}, M.~G. and {van Leeuwen}, F. and {Bakker}, J. and {Cacciari}, C. and {Casta{\~n}eda}, J. and {De Angeli}, F. and {Ducourant}, C. and {Fabricius}, C. and {Fouesneau}, M. and {Fr{\'e}mat}, Y. and {Guerra}, R. and {Guerrier}, A. and {Guiraud}, J. and {Jean-Antoine Piccolo}, A. and {Masana}, E. and {Messineo}, R. and {Mowlavi}, N. and {Nicolas}, C. and {Nienartowicz}, K. and {Pailler}, F. and {Panuzzo}, P. and {Riclet}, F. and {Roux}, W. and {Seabroke}, G.~M. and {Sordo}, R. and {Tanga}, P. and {Th{\'e}venin}, F. and {Gracia-Abril}, G. and {Portell}, J. and {Teyssier}, D. and {Altmann}, M. and {Andrae}, R. and {Bellas-Velidis}, I. and {Benson}, K. and {Berthier}, J. and {Blomme}, R. and {Brugaletta}, E. and {Burgess}, P.~W. and {Busso}, G. and {Carry}, B. and {Cellino}, A. and {Cheek}, N. and {Clementini}, G. and {Damerdji}, Y. and {Davidson}, M. and {Delchambre}, L. and {Dell'Oro}, A. and {Fern{\'a}ndez-Hern{\'a}ndez}, J. and {Galluccio}, L. and {Garc{\'\i}a-Lario}, P. and {Garcia-Reinaldos}, M. and {Gonz{\'a}lez-N{\'u}{\~n}ez}, J. and {Gosset}, E. and {Haigron}, R. and {Halbwachs}, J. -L. and {Hambly}, N.~C. and {Harrison}, D.~L. and {Hatzidimitriou}, D. and {Heiter}, U. and {Hern{\'a}ndez}, J. and {Hestroffer}, D. and {Hodgkin}, S.~T. and {Holl}, B. and {Jan{\ss}en}, K. and {Jevardat de Fombelle}, G. and {Jordan}, S. and {Krone-Martins}, A. and {Lanzafame}, A.~C. and {L{\"o}ffler}, W. and {Lorca}, A. and {Manteiga}, M. and {Marchal}, O. and {Marrese}, P.~M. and {Moitinho}, A. and {Mora}, A. and {Muinonen}, K. and {Osborne}, P. and {Pancino}, E. and {Pauwels}, T. and {Petit}, J. -M. and {Recio-Blanco}, A. and {Richards}, P.~J. and {Riello}, M. and {Rimoldini}, L. and {Robin}, A.~C. and {Roegiers}, T. and {Rybizki}, J. and {Sarro}, L.~M. and {Siopis}, C. and {Smith}, M. and {Sozzetti}, A. and {Ulla}, A. and {Utrilla}, E. and {van Leeuwen}, M. and {van Reeven}, W. and {Abbas}, U. and {Abreu Aramburu}, A. and {Accart}, S. and {Aerts}, C. and {Aguado}, J.~J. and {Ajaj}, M. and {Altavilla}, G. and {{\'A}lvarez}, M.~A. and {{\'A}lvarez Cid-Fuentes}, J. and {Alves}, J. and {Anderson}, R.~I. and {Anglada Varela}, E. and {Antoja}, T. and {Audard}, M. and {Baines}, D. and {Baker}, S.~G. and {Balaguer-N{\'u}{\~n}ez}, L. and {Balbinot}, E. and {Balog}, Z. and {Barache}, C. and {Barbato}, D. and {Barros}, M. and {Barstow}, M.~A. and {Bartolom{\'e}}, S. and {Bassilana}, J. -L. and {Bauchet}, N. and {Baudesson-Stella}, A. and {Becciani}, U. and {Bellazzini}, M. and {Bernet}, M. and {Bertone}, S. and {Bianchi}, L. and {Blanco-Cuaresma}, S. and {Boch}, T. and {Bombrun}, A. and {Bossini}, D. and {Bouquillon}, S. and {Bragaglia}, A. and {Bramante}, L. and {Breedt}, E. and {Bressan}, A. and {Brouillet}, N. and {Bucciarelli}, B. and {Burlacu}, A. and {Busonero}, D. and {Butkevich}, A.~G. and {Buzzi}, R. and {Caffau}, E. and {Cancelliere}, R. and {C{\'a}novas}, H. and {Cantat-Gaudin}, T. and {Carballo}, R. and {Carlucci}, T. and {Carnerero}, M.~I. and {Carrasco}, J.~M. and {Casamiquela}, L. and {Castellani}, M. and {Castro-Ginard}, A. and {Castro Sampol}, P. and {Chaoul}, L. and {Charlot}, P. and {Chemin}, L. and {Chiavassa}, A. and {Cioni}, M. -R.~L. and {Comoretto}, G. and {Cooper}, W.~J. and {Cornez}, T. and {Cowell}, S. and {Crifo}, F. and {Crosta}, M. and {Crowley}, C. and {Dafonte}, C. and {Dapergolas}, A. and {David}, M. and {David}, P.},
        title = "{Gaia Early Data Release 3. Summary of the contents and survey properties}",
      journal = {\aap},
         year = 2021,
        month = may,
       volume = {649},
          eid = {A1},
        pages = {A1},
          doi = {10.1051/0004-6361/202039657},
archivePrefix = {arXiv},
       eprint = {2012.01533},
 primaryClass = {astro-ph.GA},
       adsurl = {https://ui.adsabs.harvard.edu/abs/2021A&A...649A...1G}
}

@ARTICLE{Abbott21,
       author = {{Abbott}, T.~M.~C. and {Adam{\'o}w}, M. and {Aguena}, M. and {Allam}, S. and {Amon}, A. and {Annis}, J. and {Avila}, S. and {Bacon}, D. and {Banerji}, M. and {Bechtol}, K. and {Becker}, M.~R. and {Bernstein}, G.~M. and {Bertin}, E. and {Bhargava}, S. and {Bridle}, S.~L. and {Brooks}, D. and {Burke}, D.~L. and {Carnero Rosell}, A. and {Carrasco Kind}, M. and {Carretero}, J. and {Castander}, F.~J. and {Cawthon}, R. and {Chang}, C. and {Choi}, A. and {Conselice}, C. and {Costanzi}, M. and {Crocce}, M. and {da Costa}, L.~N. and {Davis}, T.~M. and {De Vicente}, J. and {DeRose}, J. and {Desai}, S. and {Diehl}, H.~T. and {Dietrich}, J.~P. and {Drlica-Wagner}, A. and {Eckert}, K. and {Elvin-Poole}, J. and {Everett}, S. and {Evrard}, A.~E. and {Ferrero}, I. and {Fert{\'e}}, A. and {Flaugher}, B. and {Fosalba}, P. and {Friedel}, D. and {Frieman}, J. and {Garc{\'\i}a-Bellido}, J. and {Gaztanaga}, E. and {Gelman}, L. and {Gerdes}, D.~W. and {Giannantonio}, T. and {Gill}, M.~S.~S. and {Gruen}, D. and {Gruendl}, R.~A. and {Gschwend}, J. and {Gutierrez}, G. and {Hartley}, W.~G. and {Hinton}, S.~R. and {Hollowood}, D.~L. and {Honscheid}, K. and {Huterer}, D. and {James}, D.~J. and {Jeltema}, T. and {Johnson}, M.~D. and {Kent}, S. and {Kron}, R. and {Kuehn}, K. and {Kuropatkin}, N. and {Lahav}, O. and {Li}, T.~S. and {Lidman}, C. and {Lin}, H. and {MacCrann}, N. and {Maia}, M.~A.~G. and {Manning}, T.~A. and {Maloney}, J.~D. and {March}, M. and {Marshall}, J.~L. and {Martini}, P. and {Melchior}, P. and {Menanteau}, F. and {Miquel}, R. and {Morgan}, R. and {Myles}, J. and {Neilsen}, E. and {Ogando}, R.~L.~C. and {Palmese}, A. and {Paz-Chinch{\'o}n}, F. and {Petravick}, D. and {Pieres}, A. and {Plazas}, A.~A. and {Pond}, C. and {Rodriguez-Monroy}, M. and {Romer}, A.~K. and {Roodman}, A. and {Rykoff}, E.~S. and {Sako}, M. and {Sanchez}, E. and {Santiago}, B. and {Scarpine}, V. and {Serrano}, S. and {Sevilla-Noarbe}, I. and {Smith}, J. Allyn and {Smith}, M. and {Soares-Santos}, M. and {Suchyta}, E. and {Swanson}, M.~E.~C. and {Tarle}, G. and {Thomas}, D. and {To}, C. and {Tremblay}, P.~E. and {Troxel}, M.~A. and {Tucker}, D.~L. and {Turner}, D.~J. and {Varga}, T.~N. and {Walker}, A.~R. and {Wechsler}, R.~H. and {Weller}, J. and {Wester}, W. and {Wilkinson}, R.~D. and {Yanny}, B. and {Zhang}, Y. and {Nikutta}, R. and {Fitzpatrick}, M. and {Jacques}, A. and {Scott}, A. and {Olsen}, K. and {Huang}, L. and {Herrera}, D. and {Juneau}, S. and {Nidever}, D. and {Weaver}, B.~A. and {Adean}, C. and {Correia}, V. and {de Freitas}, M. and {Freitas}, F.~N. and {Singulani}, C. and {Vila-Verde}, G. and {Linea Science Server}},
        title = "{The Dark Energy Survey Data Release 2}",
      journal = {\apjs},
         year = 2021,
        month = aug,
       volume = {255},
       number = {2},
          eid = {20},
        pages = {20},
          doi = {10.3847/1538-4365/ac00b3},
archivePrefix = {arXiv},
       eprint = {2101.05765},
 primaryClass = {astro-ph.IM},
       adsurl = {https://ui.adsabs.harvard.edu/abs/2021ApJS..255...20A}
}

@ARTICLE{Lasker08,
       author = {{Lasker}, Barry M. and {Lattanzi}, Mario G. and {McLean}, Brian J. and {Bucciarelli}, Beatrice and {Drimmel}, Ronald and {Garcia}, Jorge and {Greene}, Gretchen and {Guglielmetti}, Fabrizia and {Hanley}, Christopher and {Hawkins}, George and {Laidler}, Victoria G. and {Loomis}, Charles and {Meakes}, Michael and {Mignani}, Roberto and {Morbidelli}, Roberto and {Morrison}, Jane and {Pannunzio}, Renato and {Rosenberg}, Amy and {Sarasso}, Maria and {Smart}, Richard L. and {Spagna}, Alessandro and {Sturch}, Conrad R. and {Volpicelli}, Antonio and {White}, Richard L. and {Wolfe}, David and {Zacchei}, Andrea},
        title = "{The Second-Generation Guide Star Catalog: Description and Properties}",
      journal = {\aj},
         year = 2008,
        month = aug,
       volume = {136},
       number = {2},
        pages = {735-766},
          doi = {10.1088/0004-6256/136/2/735},
archivePrefix = {arXiv},
       eprint = {0807.2522},
 primaryClass = {astro-ph},
       adsurl = {https://ui.adsabs.harvard.edu/abs/2008AJ....136..735L}
}

@INPROCEEDINGS{Zacharias04,
       author = {{Zacharias}, N. and {Monet}, D.~G. and {Levine}, S.~E. and {Urban}, S.~E. and {Gaume}, R. and {Wycoff}, G.~L.},
        title = "{The Naval Observatory Merged Astrometric Dataset (NOMAD)}",
    booktitle = {American Astronomical Society Meeting Abstracts},
         year = 2004,
       series = {American Astronomical Society Meeting Abstracts},
       volume = {205},
        month = dec,
          eid = {48.15},
        pages = {48.15},
       adsurl = {https://ui.adsabs.harvard.edu/abs/2004AAS...205.4815Z}
}

@ARTICLE{Ochsenbein00,
       author = {{Ochsenbein}, F. and {Bauer}, P. and {Marcout}, J.},
        title = "{The VizieR database of astronomical catalogues}",
      journal = {\aaps},
         year = 2000,
        month = apr,
       volume = {143},
        pages = {23-32},
          doi = {10.1051/aas:2000169},
archivePrefix = {arXiv},
       eprint = {astro-ph/0002122},
 primaryClass = {astro-ph},
       adsurl = {https://ui.adsabs.harvard.edu/abs/2000A&AS..143...23O}
}

@ARTICLE{Mayer24,
       author = {{Mayer}, Martin G.~F. and {Becker}, Werner},
        title = "{Searching for X-ray counterparts of unassociated Fermi-LAT sources and rotation-powered pulsars with SRG/eROSITA}",
      journal = {\aap},
         year = 2024,
        month = apr,
       volume = {684},
          eid = {A208},
        pages = {A208},
          doi = {10.1051/0004-6361/202348620},
archivePrefix = {arXiv},
       eprint = {2401.17295},
 primaryClass = {astro-ph.HE},
       adsurl = {https://ui.adsabs.harvard.edu/abs/2024A&A...684A.208M}
}

@ARTICLE{Takeuchi13,
       author = {{Takeuchi}, Y. and {Kataoka}, J. and {Maeda}, K. and {Takahashi}, Y. and {Nakamori}, T. and {Tahara}, M.},
        title = "{Multiband Diagnostics of Unidentified 1FGL Sources with Suzaku and Swift X-Ray Observations}",
      journal = {\apjs},
         year = 2013,
        month = oct,
       volume = {208},
       number = {2},
          eid = {25},
        pages = {25},
          doi = {10.1088/0067-0049/208/2/25},
archivePrefix = {arXiv},
       eprint = {1307.5581},
 primaryClass = {astro-ph.HE},
       adsurl = {https://ui.adsabs.harvard.edu/abs/2013ApJS..208...25T}
}

@ARTICLE{Zavlin06,
       author = {{Zavlin}, Vyacheslav E.},
        title = "{XMM-Newton Observations of Four Millisecond Pulsars}",
      journal = {\apj},
         year = 2006,
        month = feb,
       volume = {638},
       number = {2},
        pages = {951-962},
          doi = {10.1086/449308},
archivePrefix = {arXiv},
       eprint = {astro-ph/0507235},
 primaryClass = {astro-ph},
       adsurl = {https://ui.adsabs.harvard.edu/abs/2006ApJ...638..951Z}
}

@ARTICLE{Evans14,
       author = {{Evans}, P.~A. and {Osborne}, J.~P. and {Beardmore}, A.~P. and {Page}, K.~L. and {Willingale}, R. and {Mountford}, C.~J. and {Pagani}, C. and {Burrows}, D.~N. and {Kennea}, J.~A. and {Perri}, M. and {Tagliaferri}, G. and {Gehrels}, N.},
        title = "{1SXPS: A Deep Swift X-Ray Telescope Point Source Catalog with Light Curves and Spectra}",
      journal = {\apjs},
         year = 2014,
        month = jan,
       volume = {210},
       number = {1},
          eid = {8},
        pages = {8},
          doi = {10.1088/0067-0049/210/1/8},
archivePrefix = {arXiv},
       eprint = {1311.5368},
 primaryClass = {astro-ph.HE},
       adsurl = {https://ui.adsabs.harvard.edu/abs/2014ApJS..210....8E}
}

@article{Kaur23,
doi = {10.3847/1538-4357/ac8b80},
url = {https://dx.doi.org/10.3847/1538-4357/ac8b80},
year = {2023},
month = {feb},
publisher = {The American Astronomical Society},
volume = {943},
number = {2},
pages = {167},
author = {Kaur, Amanpreet and Kerby, Stephen and Falcone, Abraham D.},
title = {Using Neural Networks to Differentiate Newly Discovered BL Lacertae Objects and FSRQs among the 4FGL Unassociated Sources Employing Gamma-Ray, X-Ray, UV/Optical, and IR Data},
journal = {The Astrophysical Journal}
}

@ARTICLE{Marelli12,
       author = {{Marelli}, Martino},
        title = "{The X-ray behaviour of Fermi/LAT pulsars}",
      journal = {arXiv e-prints},
         year = 2012,
        month = may,
          eid = {arXiv:1205.1748},
        pages = {arXiv:1205.1748},
          doi = {10.48550/arXiv.1205.1748},
archivePrefix = {arXiv},
       eprint = {1205.1748},
 primaryClass = {astro-ph.HE},
       adsurl = {https://ui.adsabs.harvard.edu/abs/2012arXiv1205.1748M},
    note = {Thesis on Fermi pulsars}
}

@ARTICLE{zhang23,
       author = {{Zhang}, Bing-Kai and {Tang}, Wei-Feng and {Wang}, Chun-Xiao and {Wu}, Qi and {Jin}, Min and {Dai}, Ben-Zhong and {Zhu}, Feng-Rong},
        title = "{The optical spectral features of 27 Fermi blazars}",
      journal = {\mnras},
         year = 2023,
        month = mar,
       volume = {519},
       number = {4},
        pages = {5263-5270},
          doi = {10.1093/mnras/stac3795},
archivePrefix = {arXiv},
       eprint = {2212.12331},
 primaryClass = {astro-ph.HE},
       adsurl = {https://ui.adsabs.harvard.edu/abs/2023MNRAS.519.5263Z}
}

@ARTICLE{kaur21,
       author = {{Kaur}, Amanpreet and {Falcone}, Abraham D. and {Stroh}, Michael C.},
        title = "{Classifying Blazar Candidates from the 3FGL Unassociated Catalog into BL Lacertae Objects and Flat Spectrum Radio Quasars Using Swift and WISE Data}",
      journal = {\apj},
         year = 2021,
        month = feb,
       volume = {908},
       number = {2},
          eid = {177},
        pages = {177},
          doi = {10.3847/1538-4357/abd324},
archivePrefix = {arXiv},
       eprint = {2012.06587},
 primaryClass = {astro-ph.HE},
       adsurl = {https://ui.adsabs.harvard.edu/abs/2021ApJ...908..177K}
}

@ARTICLE{ducci23,
       author = {{Ducci}, L. and {Malacaria}, C. and {Romano}, P. and {Bozzo}, E. and {Berton}, M. and {Santangelo}, A. and {Congiu}, E.},
        title = "{X-ray flashes from the low-mass X-ray binary IGR J17407{\ensuremath{-}}2808}",
      journal = {\aap},
         year = 2023,
        month = jun,
       volume = {674},
          eid = {A100},
        pages = {A100},
          doi = {10.1051/0004-6361/202346284},
archivePrefix = {arXiv},
       eprint = {2304.08816},
 primaryClass = {astro-ph.HE},
       adsurl = {https://ui.adsabs.harvard.edu/abs/2023A&A...674A.100D}
}

@ARTICLE{romano16,
       author = {{Romano}, P. and {Bozzo}, E. and {Esposito}, P. and {Sbarufatti}, B. and {Haberl}, F. and {Ponti}, G. and {D'Avanzo}, P. and {Ducci}, L. and {Segreto}, A. and {Jin}, C. and {Masetti}, N. and {Del Santo}, M. and {Campana}, S. and {Mangano}, V.},
        title = "{Searching for supergiant fast X-ray transients with Swift}",
      journal = {\aap},
         year = 2016,
        month = sep,
       volume = {593},
          eid = {A96},
        pages = {A96},
          doi = {10.1051/0004-6361/201628808},
archivePrefix = {arXiv},
       eprint = {1606.09261},
 primaryClass = {astro-ph.HE},
       adsurl = {https://ui.adsabs.harvard.edu/abs/2016A&A...593A..96R}
}

@ARTICLE{romano11,
       author = {{Romano}, P. and {Barthelmy}, S.~D. and {Krimm}, H.~A. and {Esposito}, P. and {Mangano}, V. and {Siegel}, M. and {Burrows}, D.~N. and {Evans}, P.~A. and {Farinelli}, R. and {Kennea}, J.~A. and {Palmer}, D.~M. and {Vercellone}, S. and {Gehrels}, N.},
        title = "{Swift detects an outburst of IGR J17407-2808/CXOU J174042.0-280724}",
      journal = {The Astronomer's Telegram},
         year = 2011,
        month = oct,
       volume = {3685},
        pages = {1},
       adsurl = {https://ui.adsabs.harvard.edu/abs/2011ATel.3685....1R}
}

@ARTICLE{heinke09,
       author = {{Heinke}, C.~O. and {Tomsick}, J.~A. and {Yusef-Zadeh}, F. and {Grindlay}, J.~E.},
        title = "{Two Rapidly Variable Galactic X-ray Transients Observed with Chandra, XMM-Newton, and Suzaku}",
      journal = {\apj},
         year = 2009,
        month = aug,
       volume = {701},
       number = {2},
        pages = {1627-1635},
          doi = {10.1088/0004-637X/701/2/1627},
archivePrefix = {arXiv},
       eprint = {0906.4743},
 primaryClass = {astro-ph.HE},
       adsurl = {https://ui.adsabs.harvard.edu/abs/2009ApJ...701.1627H}
}

@ARTICLE{filocomo23,
       author = {{Fil{\'o}como}, A. and {Albacete-Colombo}, J.~F. and {Mestre}, E. and {Pellizza}, L.~J. and {Combi}, J.~A.},
        title = "{{\ensuremath{\gamma}}-ray detection from occasional flares in T Tauri stars of NGC 2071 - I. Observational connection}",
      journal = {\mnras},
         year = 2023,
        month = oct,
       volume = {525},
       number = {2},
        pages = {1726-1730},
          doi = {10.1093/mnras/stad2029},
archivePrefix = {arXiv},
       eprint = {2308.12142},
 primaryClass = {astro-ph.HE},
       adsurl = {https://ui.adsabs.harvard.edu/abs/2023MNRAS.525.1726F}
}

@MANUAL{capalbi05,
       author = {{Capalbi}, M. and {Perri}, M. and {Saija}, B. and {Tamburelli}, F. and {Angelini}, Lorella},
        title = "{The SWIFT XRT Data Reduction Guide}",
         year = 2005,
        month = April,
        organization = {ASI Science Data Center and HEASARC},
        version = {1.2},
        note = {Available at \url{https://www.swift.ac.uk/analysis/xrt/files/xrt_swguide_v1_2.pdf}}
}

@ARTICLE{romani95,
       author = {{Romani}, Roger W. and {Yadigaroglu}, I. -A.},
        title = "{Gamma-Ray Pulsars: Emission Zones and Viewing Geometries}",
      journal = {\apj},
         year = 1995,
        month = jan,
       volume = {438},
        pages = {314},
          doi = {10.1086/175076},
archivePrefix = {arXiv},
       eprint = {astro-ph/9401045},
 primaryClass = {astro-ph},
       adsurl = {https://ui.adsabs.harvard.edu/abs/1995ApJ...438..314R}
}

@ARTICLE{sikora94,
       author = {{Sikora}, Marek and {Begelman}, Mitchell C. and {Rees}, Martin J.},
        title = "{Comptonization of Diffuse Ambient Radiation by a Relativistic Jet: The Source of Gamma Rays from Blazars?}",
      journal = {\apj},
         year = 1994,
        month = jan,
       volume = {421},
        pages = {153},
          doi = {10.1086/173633},
       adsurl = {https://ui.adsabs.harvard.edu/abs/1994ApJ...421..153S}
}

@ARTICLE{aydi22,
       author = {{Aydi}, E. and {Sokolovsky}, K.~V. and {Bright}, J.~S. and {Tremou}, E. and {Nyamai}, M.~M. and {Evans}, A. and {Strader}, J. and {Chomiuk}, L. and {Myers}, G. and {Hambsch}, F. -J. and {Page}, K.~L. and {Buckley}, D.~A.~H. and {Woodward}, C.~E. and {Walter}, F.~M. and {Mr{\'o}z}, P. and {Vallely}, P.~J. and {Geballe}, T.~R. and {Banerjee}, D.~P.~K. and {Gehrz}, R.~D. and {Fender}, R.~P. and {Gromadzki}, M. and {Kawash}, A. and {Knigge}, C. and {Mukai}, K. and {Munari}, U. and {Orio}, M. and {Ribeiro}, V.~A.~R.~M. and {Sokoloski}, J.~L. and {Starrfield}, S. and {Udalski}, A. and {Woudt}, P.~A.},
        title = "{The 2019 Outburst of the 2005 Classical Nova V1047 Cen: A Record Breaking Dwarf Nova Outburst or a New Phenomenon?}",
      journal = {\apj},
         year = 2022,
        month = nov,
       volume = {939},
       number = {1},
          eid = {6},
        pages = {6},
          doi = {10.3847/1538-4357/ac913b},
archivePrefix = {arXiv},
       eprint = {2108.07868},
 primaryClass = {astro-ph.SR},
       adsurl = {https://ui.adsabs.harvard.edu/abs/2022ApJ...939....6A}
}

@ARTICLE{liller05,
       author = {{Liller}, W. and {Jacques}, C. and {Pimentel}, E. and {Aguiar}, J.~G. De S. and {Shida}, R.~Y.},
        title = "{V1047 Centauri = Nova Centauri 2005}",
      journal = {\iaucirc},
         year = 2005,
        month = sep,
       volume = {8596},
        pages = {1},
       adsurl = {https://ui.adsabs.harvard.edu/abs/2005IAUC.8596....1L}
}

@ARTICLE{mroz19,
       author = {{Mroz}, P. and {Udalski}, A.},
        title = "{Rebrightening of V1047 Cen (Nova Cen 2005)}",
      journal = {The Astronomer's Telegram},
         year = 2019,
        month = jun,
       volume = {12876},
        pages = {1},
       adsurl = {https://ui.adsabs.harvard.edu/abs/2019ATel12876....1M}
}

@ARTICLE{torres14,
       author = {{Torres}, M.~A.~P. and {Jonker}, P.~G. and {Britt}, C.~T. and {Johnson}, C.~B. and {Hynes}, R.~I. and {Greiss}, S. and {Steeghs}, D. and {Maccarone}, T.~J. and {{\"O}zel}, F. and {Bassa}, C. and {Nelemans}, G.},
        title = "{Identification of 23 accreting binaries in the Galactic Bulge Survey}",
      journal = {\mnras},
         year = 2014,
        month = may,
       volume = {440},
       number = {1},
        pages = {365-386},
          doi = {10.1093/mnras/stu170},
archivePrefix = {arXiv},
       eprint = {1310.0224},
 primaryClass = {astro-ph.HE},
       adsurl = {https://ui.adsabs.harvard.edu/abs/2014MNRAS.440..365T}
}

@ARTICLE{dudus13,
       author = {{Dubus}, Guillaume},
        title = "{Gamma-ray binaries and related systems}",
      journal = {\aapr},
         year = 2013,
        month = aug,
       volume = {21},
          eid = {64},
        pages = {64},
          doi = {10.1007/s00159-013-0064-5},
archivePrefix = {arXiv},
       eprint = {1307.7083},
 primaryClass = {astro-ph.HE},
       adsurl = {https://ui.adsabs.harvard.edu/abs/2013A&ARv..21...64D}
}

@ARTICLE{shara09,
       author = {{Shara}, Michael M. and {Moffat}, Anthony F.~J. and {Gerke}, Jill and {Zurek}, David and {Stanonik}, Kathryn and {Doyon}, Ren{\'e} and {Artigau}, Etienne and {Drissen}, Laurent and {Villar-Sbaffi}, Alfredo},
        title = "{A Near-Infrared Survey of the Inner Galactic Plane for Wolf-Rayet Stars. I. Methods and First Results: 41 New WR Stars}",
      journal = {\aj},
         year = 2009,
        month = aug,
       volume = {138},
       number = {2},
        pages = {402-420},
          doi = {10.1088/0004-6256/138/2/402},
archivePrefix = {arXiv},
       eprint = {0905.1967},
 primaryClass = {astro-ph.GA},
       adsurl = {https://ui.adsabs.harvard.edu/abs/2009AJ....138..402S}
}

@ARTICLE{binder15,
       author = {{Binder}, B. and {Williams}, B.~F. and {Eracleous}, M. and {Plucinsky}, P.~P. and {Gaetz}, T.~J. and {Anderson}, S.~F. and {Skillman}, E.~D. and {Dalcanton}, J.~J. and {Kong}, A.~K.~H. and {Weisz}, D.~R.},
        title = "{The Chandra Local Volume Survey. I. The X-Ray Point Source Populations of NGC 55, NGC 2403, and NGC 4214}",
      journal = {\aj},
         year = 2015,
        month = sep,
       volume = {150},
       number = {3},
          eid = {94},
        pages = {94},
          doi = {10.1088/0004-6256/150/3/94},
archivePrefix = {arXiv},
       eprint = {1507.01999},
 primaryClass = {astro-ph.HE},
       adsurl = {https://ui.adsabs.harvard.edu/abs/2015AJ....150...94B}
}

@ARTICLE{mineo12,
       author = {{Mineo}, S. and {Gilfanov}, M. and {Sunyaev}, R.},
        title = "{X-ray emission from star-forming galaxies - I. High-mass X-ray binaries}",
      journal = {\mnras},
         year = 2012,
        month = jan,
       volume = {419},
       number = {3},
        pages = {2095-2115},
          doi = {10.1111/j.1365-2966.2011.19862.x},
archivePrefix = {arXiv},
       eprint = {1105.4610},
 primaryClass = {astro-ph.HE},
       adsurl = {https://ui.adsabs.harvard.edu/abs/2012MNRAS.419.2095M}
}

@ARTICLE{stroh13,
       author = {{Stroh}, Michael C. and {Falcone}, Abe D.},
        title = "{Swift X-Ray Telescope Monitoring of Fermi-LAT Gamma-Ray Sources of Interest}",
      journal = {\apjs},
         year = 2013,
        month = aug,
       volume = {207},
       number = {2},
          eid = {28},
        pages = {28},
          doi = {10.1088/0067-0049/207/2/28},
archivePrefix = {arXiv},
       eprint = {1305.4949},
 primaryClass = {astro-ph.HE},
       adsurl = {https://ui.adsabs.harvard.edu/abs/2013ApJS..207...28S}
}

@ARTICLE{hiemstra11,
       author = {{Hiemstra}, Beike and {M{\'e}ndez}, Mariano and {Done}, Chris and {D{\'\i}az Trigo}, Mar{\'\i}a and {Altamirano}, Diego and {Casella}, Piergiorgio},
        title = "{A strong and broad Fe line in the XMM-Newton spectrum of the new X-ray transient and black hole candidate XTE J1652-453}",
      journal = {\mnras},
         year = 2011,
        month = feb,
       volume = {411},
       number = {1},
        pages = {137-150},
          doi = {10.1111/j.1365-2966.2010.17661.x},
archivePrefix = {arXiv},
       eprint = {1004.4442},
 primaryClass = {astro-ph.HE},
       adsurl = {https://ui.adsabs.harvard.edu/abs/2011MNRAS.411..137H}
}

@ARTICLE{markwardt09b,
       author = {{Markwardt}, C.~B. and {Beardmore}, A.~P. and {Miller}, J. and {Swank}, J.~H.},
        title = "{Updated Position, Flux and Spectrum of XTE J1652-453}",
      journal = {The Astronomer's Telegram},
         year = 2009,
        month = jul,
       volume = {2120},
        pages = {1},
       adsurl = {https://ui.adsabs.harvard.edu/abs/2009ATel.2120....1M}
}

@ARTICLE{markwardt09a,
       author = {{Markwardt}, C.~B. and {Swank}, J.~H. and {Krimm}, H.~A. and {Pereira}, D. and {Strohmayer}, T.~E.},
        title = "{A New X-ray Transient: XTE J1652-453}",
      journal = {The Astronomer's Telegram},
         year = 2009,
        month = jul,
       volume = {2107},
        pages = {1},
       adsurl = {https://ui.adsabs.harvard.edu/abs/2009ATel.2107....1M}
}

@ARTICLE{hog00,
       author = {{H{\o}g}, E. and {Fabricius}, C. and {Makarov}, V.~V. and {Urban}, S. and {Corbin}, T. and {Wycoff}, G. and {Bastian}, U. and {Schwekendiek}, P. and {Wicenec}, A.},
        title = "{The Tycho-2 catalogue of the 2.5 million brightest stars}",
      journal = {\aap},
         year = 2000,
        month = mar,
       volume = {355},
        pages = {L27-L30},
       adsurl = {https://ui.adsabs.harvard.edu/abs/2000A&A...355L..27H}
}

@ARTICLE{romani96,
       author = {{Romani}, Roger W.},
        title = "{Gamma-Ray Pulsars: Radiation Processes in the Outer Magnetosphere}",
      journal = {\apj},
         year = 1996,
        month = oct,
       volume = {470},
        pages = {469},
          doi = {10.1086/177878},
       adsurl = {https://ui.adsabs.harvard.edu/abs/1996ApJ...470..469R}
}

@ARTICLE{urry95,
       author = {{Urry}, C. Megan and {Padovani}, Paolo},
        title = "{Unified Schemes for Radio-Loud Active Galactic Nuclei}",
      journal = {\pasp},
         year = 1995,
        month = sep,
       volume = {107},
        pages = {803},
          doi = {10.1086/133630},
archivePrefix = {arXiv},
       eprint = {astro-ph/9506063},
 primaryClass = {astro-ph},
       adsurl = {https://ui.adsabs.harvard.edu/abs/1995PASP..107..803U}
}

@ARTICLE{kerby21b,
       author = {{Kerby}, Stephen and {Kaur}, Amanpreet and {Falcone}, Abraham D. and {Eskenasy}, Ryan and {Hancock}, Fredric and {Stroh}, Michael C. and {Ferrara}, Elizabeth C. and {Ray}, Paul S. and {Kennea}, Jamie A. and {Grove}, Eric},
        title = "{Multiwavelength Spectral Analysis and Neural Network Classification of Counterparts to 4FGL Unassociated Sources}",
      journal = {\apj},
         year = 2021,
        month = dec,
       volume = {923},
       number = {1},
          eid = {75},
        pages = {75},
          doi = {10.3847/1538-4357/ac2e91},
archivePrefix = {arXiv},
       eprint = {2110.04100},
 primaryClass = {astro-ph.HE},
       adsurl = {https://ui.adsabs.harvard.edu/abs/2021ApJ...923...75K}
}

@ARTICLE{ballet23,
       author = {{Ballet}, J. and {Bruel}, P. and {Burnett}, T.~H. and {Lott}, B. and {The Fermi-LAT collaboration}},
        title = "{Fermi Large Area Telescope Fourth Source Catalog Data Release 4 (4FGL-DR4)}",
      journal = {arXiv e-prints},
         year = 2023,
        month = jul,
          eid = {arXiv:2307.12546},
        pages = {arXiv:2307.12546},
          doi = {10.48550/arXiv.2307.12546},
archivePrefix = {arXiv},
       eprint = {2307.12546},
 primaryClass = {astro-ph.HE},
       adsurl = {https://ui.adsabs.harvard.edu/abs/2023arXiv230712546B}
}

@ARTICLE{astropy22,
       author = {{Astropy Collaboration} and {Price-Whelan}, Adrian M. and {Lim}, Pey Lian and {Earl}, Nicholas and {Starkman}, Nathaniel and {Bradley}, Larry and {Shupe}, David L. and {Patil}, Aarya A. and {Corrales}, Lia and {Brasseur}, C.~E. and {N{\"o}the}, Maximilian and {Donath}, Axel and {Tollerud}, Erik and {Morris}, Brett M. and {Ginsburg}, Adam and {Vaher}, Eero and {Weaver}, Benjamin A. and {Tocknell}, James and {Jamieson}, William and {van Kerkwijk}, Marten H. and {Robitaille}, Thomas P. and {Merry}, Bruce and {Bachetti}, Matteo and {G{\"u}nther}, H. Moritz and {Aldcroft}, Thomas L. and {Alvarado-Montes}, Jaime A. and {Archibald}, Anne M. and {B{\'o}di}, Attila and {Bapat}, Shreyas and {Barentsen}, Geert and {Baz{\'a}n}, Juanjo and {Biswas}, Manish and {Boquien}, M{\'e}d{\'e}ric and {Burke}, D.~J. and {Cara}, Daria and {Cara}, Mihai and {Conroy}, Kyle E. and {Conseil}, Simon and {Craig}, Matthew W. and {Cross}, Robert M. and {Cruz}, Kelle L. and {D'Eugenio}, Francesco and {Dencheva}, Nadia and {Devillepoix}, Hadrien A.~R. and {Dietrich}, J{\"o}rg P. and {Eigenbrot}, Arthur Davis and {Erben}, Thomas and {Ferreira}, Leonardo and {Foreman-Mackey}, Daniel and {Fox}, Ryan and {Freij}, Nabil and {Garg}, Suyog and {Geda}, Robel and {Glattly}, Lauren and {Gondhalekar}, Yash and {Gordon}, Karl D. and {Grant}, David and {Greenfield}, Perry and {Groener}, Austen M. and {Guest}, Steve and {Gurovich}, Sebastian and {Handberg}, Rasmus and {Hart}, Akeem and {Hatfield-Dodds}, Zac and {Homeier}, Derek and {Hosseinzadeh}, Griffin and {Jenness}, Tim and {Jones}, Craig K. and {Joseph}, Prajwel and {Kalmbach}, J. Bryce and {Karamehmetoglu}, Emir and {Ka{\l}uszy{\'n}ski}, Miko{\l}aj and {Kelley}, Michael S.~P. and {Kern}, Nicholas and {Kerzendorf}, Wolfgang E. and {Koch}, Eric W. and {Kulumani}, Shankar and {Lee}, Antony and {Ly}, Chun and {Ma}, Zhiyuan and {MacBride}, Conor and {Maljaars}, Jakob M. and {Muna}, Demitri and {Murphy}, N.~A. and {Norman}, Henrik and {O'Steen}, Richard and {Oman}, Kyle A. and {Pacifici}, Camilla and {Pascual}, Sergio and {Pascual-Granado}, J. and {Patil}, Rohit R. and {Perren}, Gabriel I. and {Pickering}, Timothy E. and {Rastogi}, Tanuj and {Roulston}, Benjamin R. and {Ryan}, Daniel F. and {Rykoff}, Eli S. and {Sabater}, Jose and {Sakurikar}, Parikshit and {Salgado}, Jes{\'u}s and {Sanghi}, Aniket and {Saunders}, Nicholas and {Savchenko}, Volodymyr and {Schwardt}, Ludwig and {Seifert-Eckert}, Michael and {Shih}, Albert Y. and {Jain}, Anany Shrey and {Shukla}, Gyanendra and {Sick}, Jonathan and {Simpson}, Chris and {Singanamalla}, Sudheesh and {Singer}, Leo P. and {Singhal}, Jaladh and {Sinha}, Manodeep and {Sip{\H{o}}cz}, Brigitta M. and {Spitler}, Lee R. and {Stansby}, David and {Streicher}, Ole and {{\v{S}}umak}, Jani and {Swinbank}, John D. and {Taranu}, Dan S. and {Tewary}, Nikita and {Tremblay}, Grant R. and {Val-Borro}, Miguel de and {Van Kooten}, Samuel J. and {Vasovi{\'c}}, Zlatan and {Verma}, Shresth and {de Miranda Cardoso}, Jos{\'e} Vin{\'\i}cius and {Williams}, Peter K.~G. and {Wilson}, Tom J. and {Winkel}, Benjamin and {Wood-Vasey}, W.~M. and {Xue}, Rui and {Yoachim}, Peter and {Zhang}, Chen and {Zonca}, Andrea and {Astropy Project Contributors}},
        title = "{The Astropy Project: Sustaining and Growing a Community-oriented Open-source Project and the Latest Major Release (v5.0) of the Core Package}",
      journal = {\apj},
         year = 2022,
        month = aug,
       volume = {935},
       number = {2},
          eid = {167},
        pages = {167},
          doi = {10.3847/1538-4357/ac7c74},
archivePrefix = {arXiv},
       eprint = {2206.14220},
 primaryClass = {astro-ph.IM},
       adsurl = {https://ui.adsabs.harvard.edu/abs/2022ApJ...935..167A}
}

@article{Chawla2002,
   author = {Nitesh V Chawla and Kevin W Bowyer and Lawrence O Hall and W Philip Kegelmeyer},
   journal = {Journal of Artificial Intelligence Research},
   pages = {321-357},
   title = {SMOTE: Synthetic Minority Over-sampling Technique},
   volume = {16},
   year = {2002},
}

@article{Ghisellini2008,
   author = {G. Ghisellini and F. Tavecchio},
   doi = {10.1111/j.1365-2966.2008.13360.x},
   issn = {00358711},
   issue = {4},
   journal = {Monthly Notices of the Royal Astronomical Society},
   month = {7},
   pages = {1669-1680},
   title = {The blazar sequence: A new perspective},
   volume = {387},
   year = {2008},
}

@article{Ghisellini2017,
   author = {G. Ghisellini and C. Righi and L. Costamante and F. Tavecchio},
   doi = {10.1093/mnras/stx806},
   month = {2},
   title = {The Fermi blazar sequence},
   url = {http://dx.doi.org/10.1093/mnras/stx806},
   year = {2017},
}

@ARTICLE{Kaur2019,
       author = {{Kaur}, Amanpreet and {Falcone}, Abraham D. and {Stroh}, Michael D. and {Kennea}, Jamie A. and {Ferrara}, Elizabeth C.},
        title = "{Classification of New X-Ray Counterparts for Fermi Unassociated Gamma-Ray Sources Using the Swift X-Ray Telescope}",
      journal = {\apj},
         year = 2019,
        month = dec,
       volume = {887},
       number = {1},
          eid = {18},
        pages = {18},
          doi = {10.3847/1538-4357/ab4ceb},
archivePrefix = {arXiv},
       eprint = {1910.06317},
 primaryClass = {astro-ph.HE},
       adsurl = {https://ui.adsabs.harvard.edu/abs/2019ApJ...887...18K}
}

@ARTICLE{Kingma2014,
       author = {{Kingma}, Diederik P. and {Ba}, Jimmy},
        title = "{Adam: A Method for Stochastic Optimization}",
      journal = {arXiv e-prints},
         year = 2014,
        month = dec,
          eid = {arXiv:1412.6980},
        pages = {arXiv:1412.6980},
archivePrefix = {arXiv},
       eprint = {1412.6980},
 primaryClass = {cs.LG},
       adsurl = {https://ui.adsabs.harvard.edu/abs/2014arXiv1412.6980K}
}

@ARTICLE{collaboration2019,
       author = {{Abdollahi}, S. and {Acero}, F. and {Ackermann}, M. and {Ajello}, M. and {Atwood}, W.~B. and {Axelsson}, M. and {Baldini}, L. and {Ballet}, J. and {Barbiellini}, G. and {Bastieri}, D. and {Becerra Gonzalez}, J. and {Bellazzini}, R. and {Berretta}, A. and {Bissaldi}, E. and {Blandford}, R.~D. and {Bloom}, E.~D. and {Bonino}, R. and {Bottacini}, E. and {Brandt}, T.~J. and {Bregeon}, J. and {Bruel}, P. and {Buehler}, R. and {Burnett}, T.~H. and {Buson}, S. and {Cameron}, R.~A. and {Caputo}, R. and {Caraveo}, P.~A. and {Casandjian}, J.~M. and {Castro}, D. and {Cavazzuti}, E. and {Charles}, E. and {Chaty}, S. and {Chen}, S. and {Cheung}, C.~C. and {Chiaro}, G. and {Ciprini}, S. and {Cohen-Tanugi}, J. and {Cominsky}, L.~R. and {Coronado-Bl{\'a}zquez}, J. and {Costantin}, D. and {Cuoco}, A. and {Cutini}, S. and {D'Ammando}, F. and {DeKlotz}, M. and {de la Torre Luque}, P. and {de Palma}, F. and {Desai}, A. and {Digel}, S.~W. and {Di Lalla}, N. and {Di Mauro}, M. and {Di Venere}, L. and {Dom{\'\i}nguez}, A. and {Dumora}, D. and {Fana Dirirsa}, F. and {Fegan}, S.~J. and {Ferrara}, E.~C. and {Franckowiak}, A. and {Fukazawa}, Y. and {Funk}, S. and {Fusco}, P. and {Gargano}, F. and {Gasparrini}, D. and {Giglietto}, N. and {Giommi}, P. and {Giordano}, F. and {Giroletti}, M. and {Glanzman}, T. and {Green}, D. and {Grenier}, I.~A. and {Griffin}, S. and {Grondin}, M. -H. and {Grove}, J.~E. and {Guiriec}, S. and {Harding}, A.~K. and {Hayashi}, K. and {Hays}, E. and {Hewitt}, J.~W. and {Horan}, D. and {J{\'o}hannesson}, G. and {Johnson}, T.~J. and {Kamae}, T. and {Kerr}, M. and {Kocevski}, D. and {Kovac'evic'}, M. and {Kuss}, M. and {Landriu}, D. and {Larsson}, S. and {Latronico}, L. and {Lemoine-Goumard}, M. and {Li}, J. and {Liodakis}, I. and {Longo}, F. and {Loparco}, F. and {Lott}, B. and {Lovellette}, M.~N. and {Lubrano}, P. and {Madejski}, G.~M. and {Maldera}, S. and {Malyshev}, D. and {Manfreda}, A. and {Marchesini}, E.~J. and {Marcotulli}, L. and {Mart{\'\i}-Devesa}, G. and {Martin}, P. and {Massaro}, F. and {Mazziotta}, M.~N. and {McEnery}, J.~E. and {Mereu}, I. and {Meyer}, M. and {Michelson}, P.~F. and {Mirabal}, N. and {Mizuno}, T. and {Monzani}, M.~E. and {Morselli}, A. and {Moskalenko}, I.~V. and {Negro}, M. and {Nuss}, E. and {Ojha}, R. and {Omodei}, N. and {Orienti}, M. and {Orlando}, E. and {Ormes}, J.~F. and {Palatiello}, M. and {Paliya}, V.~S. and {Paneque}, D. and {Pei}, Z. and {Pe{\~n}a-Herazo}, H. and {Perkins}, J.~S. and {Persic}, M. and {Pesce-Rollins}, M. and {Petrosian}, V. and {Petrov}, L. and {Piron}, F. and {Poon}, H. and {Porter}, T.~A. and {Principe}, G. and {Rain{\`o}}, S. and {Rando}, R. and {Razzano}, M. and {Razzaque}, S. and {Reimer}, A. and {Reimer}, O. and {Remy}, Q. and {Reposeur}, T. and {Romani}, R.~W. and {Saz Parkinson}, P.~M. and {Schinzel}, F.~K. and {Serini}, D. and {Sgr{\`o}}, C. and {Siskind}, E.~J. and {Smith}, D.~A. and {Spandre}, G. and {Spinelli}, P. and {Strong}, A.~W. and {Suson}, D.~J. and {Tajima}, H. and {Takahashi}, M.~N. and {Tak}, D. and {Thayer}, J.~B. and {Thompson}, D.~J. and {Tibaldo}, L. and {Torres}, D.~F. and {Torresi}, E. and {Valverde}, J. and {Van Klaveren}, B. and {van Zyl}, P. and {Wood}, K. and {Yassine}, M. and {Zaharijas}, G.},
        title = "{Fermi Large Area Telescope Fourth Source Catalog}",
      journal = {\apjs},
         year = 2020,
        month = mar,
       volume = {247},
       number = {1},
          eid = {33},
        pages = {33},
          doi = {10.3847/1538-4365/ab6bcb},
archivePrefix = {arXiv},
       eprint = {1902.10045},
 primaryClass = {astro-ph.HE},
       adsurl = {https://ui.adsabs.harvard.edu/abs/2020ApJS..247...33A}
}

@article{Ackermann2015,
   author = {M. Ackermann and M. Ajello and W. B. Atwood and L. Baldini and J. Ballet and G. Barbiellini and D. Bastieri and J. Becerra Gonzalez and R. Bellazzini and E. Bissaldi and R. D. Blandford and E. D. Bloom and R. Bonino and E. Bottacini and T. J. Brandt and J. Bregeon and R. J. Britto and P. Bruel and R. Buehler and S. Buson and G. A. Caliandro and R. A. Cameron and M. Caragiulo and P. A. Caraveo and B. Carpenter and J. M. Casandjian and E. Cavazzuti and C. Cecchi and E. Charles and A. Chekhtman and C. C. Cheung and J. Chiang and G. Chiaro and S. Ciprini and R. Claus and J. Cohen-Tanugi and L. R. Cominsky and J. Conrad and S. Cutini and R. D'Abrusco and F. D'Ammando and A. De Angelis and R. Desiante and S. W. Digel and L. Di Venere and P. S. Drell and C. Favuzzi and S. J. Fegan and E. C. Ferrara and J. Finke and W. B. Focke and A. Franckowiak and L. Fuhrmann and Y. Fukazawa and A. K. Furniss and P. Fusco and F. Gargano and D. Gasparrini and N. Giglietto and P. Giommi and F. Giordano and M. Giroletti and T. Glanzman and G. Godfrey and I. A. Grenier and J. E. Grove and S. Guiriec and J. W. Hewitt and A. B. Hill and D. Horan and R. Itoh and G. Jóhannesson and A. S. Johnson and W. N. Johnson and J. Kataoka and T. Kawano and F. Krauss and M. Kuss and G. La Mura and S. Larsson and L. Latronico and C. Leto and J. Li and L. Li and F. Longo and F. Loparco and B. Lott and M. N. Lovellette and P. Lubrano and G. M. Madejski and M. Mayer and M. N. Mazziotta and J. E. McEnery and P. F. Michelson and T. Mizuno and A. A. Moiseev and M. E. Monzani and A. Morselli and I. V. Moskalenko and S. Murgia and E. Nuss and M. Ohno and T. Ohsugi and R. Ojha and N. Omodei and M. Orienti and E. Orlando and A. Paggi and D. Paneque and J. S. Perkins and M. Pesce-Rollins and F. Piron and G. Pivato and T. A. Porter and S. Rainò and R. Rando and M. Razzano and S. Razzaque and A. Reimer and O. Reimer and R. W. Romani and D. Salvetti and M. Schaal and F. K. Schinzel and A. Schulz and C. Sgrò and E. J. Siskind and K. V. Sokolovsky and F. Spada and G. Spandre and P. Spinelli and L. Stawarz and D. J. Suson and H. Takahashi and T. Takahashi and Y. Tanaka and J. G. Thayer and J. B. Thayer and L. Tibaldo and D. F. Torres and E. Torresi and G. Tosti and E. Troja and Y. Uchiyama and G. Vianello and B. L. Winer and K. S. Wood and S. Zimmer},
   doi = {10.1088/0004-637X/810/1/14},
   issn = {15384357},
   issue = {1},
   journal = {Astrophysical Journal},
   month = {9},
   publisher = {Institute of Physics Publishing},
   title = {THE THIRD CATALOG OF ACTIVE GALACTIC NUCLEI DETECTED BY THE FERMI LARGE AREA TELESCOPE},
   volume = {810},
   year = {2015},
}

@article{Cash1976,
   author = {W Cash},
   journal = {Astronomy \& Astrophysics},
   pages = {307-308},
   title = {Generation of Confidence Intervals for Model Parameters in X-ray Astronomy},
   volume = {52},
   year = {1976},
}

@article{Wilms2000,
   author = {J Wilms and A Allen and R Mccray},
   issue = {2},
   journal = {Astrophysical Journal},
   pages = {914-924},
   title = {ON THE ABSORPTION OF X-RAYS IN THE INTERSTELLAR MEDIUM},
   volume = {542},
   year = {2000},
}

@article{Arnaud1996,
   author = {K. A. Arnaud},
   journal = {ASP Conference Series},
   title = {XSPEC: The First Ten Years},
   volume = {101},
   year = {1996},
}

@article{SazParkinson2016,
       author = {{Saz Parkinson}, P.~M. and {Xu}, H. and {Yu}, P.~L.~H. and
         {Salvetti}, D. and {Marelli}, M. and {Falcone}, A.~D.},
        title = "{Classification and Ranking of Fermi LAT Gamma-ray Sources from the 3FGL Catalog using Machine Learning Techniques}",
      journal = {\apj},
         year = 2016,
        month = mar,
       volume = {820},
       number = {1},
          eid = {8},
        pages = {8},
          doi = {10.3847/0004-637X/820/1/8},
archivePrefix = {arXiv},
       eprint = {1602.00385},
 primaryClass = {astro-ph.HE},
       adsurl = {https://ui.adsabs.harvard.edu/abs/2016ApJ...820....8S}
}

@ARTICLE{Roming2005,
       author = {{Roming}, Peter W.~A. and {Kennedy}, Thomas E. and {Mason}, Keith O. and
         {Nousek}, John A. and {Ahr}, Lindy and {Bingham}, Richard E. and
         {Broos}, Patrick S. and {Carter}, Mary J. and {Hancock}, Barry K. and
         {Huckle}, Howard E. and {Hunsberger}, S.~D. and {Kawakami}, Hajime and
         {Killough}, Ronnie and {Koch}, T. Scott and {McLelland}, Michael K. and
         {Smith}, Kelly and {Smith}, Philip J. and {Soto}, Juan Carlos and
         {Boyd}, Patricia T. and {Breeveld}, Alice A. and {Holland}, Stephen T. and
         {Ivanushkina}, Mariya and {Pryzby}, Michael S. and {Still}, Martin D. and
         {Stock}, Joseph},
        title = "{The Swift Ultra-Violet/Optical Telescope}",
      journal = {\ssr},
         year = 2005,
        month = oct,
       volume = {120},
       number = {3-4},
        pages = {95-142},
          doi = {10.1007/s11214-005-5095-4},
archivePrefix = {arXiv},
       eprint = {astro-ph/0507413},
 primaryClass = {astro-ph},
       adsurl = {https://ui.adsabs.harvard.edu/abs/2005SSRv..120...95R}
}

@ARTICLE{Burrows2005,
       author = {{Burrows}, David N. and {Hill}, J.~E. and {Nousek}, J.~A. and
         {Kennea}, J.~A. and {Wells}, A. and {Osborne}, J.~P. and
         {Abbey}, A.~F. and {Beardmore}, A. and {Mukerjee}, K. and
         {Short}, A.~D.~T. and {Chincarini}, G. and {Campana}, S. and
         {Citterio}, O. and {Moretti}, A. and {Pagani}, C. and
         {Tagliaferri}, G. and {Giommi}, P. and {Capalbi}, M. and
         {Tamburelli}, F. and {Angelini}, L. and {Cusumano}, G. and
         {Br{\"a}uninger}, H.~W. and {Burkert}, W. and {Hartner}, G.~D.},
        title = "{The Swift X-Ray Telescope}",
      journal = {\ssr},
         year = 2005,
        month = oct,
       volume = {120},
       number = {3-4},
        pages = {165-195},
          doi = {10.1007/s11214-005-5097-2},
archivePrefix = {arXiv},
       eprint = {astro-ph/0508071},
 primaryClass = {astro-ph},
       adsurl = {https://ui.adsabs.harvard.edu/abs/2005SSRv..120..165B}
}

@ARTICLE{Gehrels2004,
       author = {{Gehrels}, N. and {Chincarini}, G. and {Giommi}, P. and {Mason}, K.~O. and
         {Nousek}, J.~A. and {Wells}, A.~A. and {White}, N.~E. and
         {Barthelmy}, S.~D. and {Burrows}, D.~N. and {Cominsky}, L.~R. and
         {Hurley}, K.~C. and {Marshall}, F.~E. and {M{\'e}sz{\'a}ros}, P. and
         {Roming}, P.~W.~A. and {Angelini}, L. and {Barbier}, L.~M. and
         {Belloni}, T. and {Campana}, S. and {Caraveo}, P.~A. and
         {Chester}, M.~M. and {Citterio}, O. and {Cline}, T.~L. and
         {Cropper}, M.~S. and {Cummings}, J.~R. and {Dean}, A.~J. and
         {Feigelson}, E.~D. and {Fenimore}, E.~E. and {Frail}, D.~A. and
         {Fruchter}, A.~S. and {Garmire}, G.~P. and {Gendreau}, K. and
         {Ghisellini}, G. and {Greiner}, J. and {Hill}, J.~E. and
         {Hunsberger}, S.~D. and {Krimm}, H.~A. and {Kulkarni}, S.~R. and
         {Kumar}, P. and {Lebrun}, F. and {Lloyd-Ronning}, N.~M. and
         {Markwardt}, C.~B. and {Mattson}, B.~J. and {Mushotzky}, R.~F. and
         {Norris}, J.~P. and {Osborne}, J. and {Paczynski}, B. and
         {Palmer}, D.~M. and {Park}, H. -S. and {Parsons}, A.~M. and {Paul}, J. and
         {Rees}, M.~J. and {Reynolds}, C.~S. and {Rhoads}, J.~E. and
         {Sasseen}, T.~P. and {Schaefer}, B.~E. and {Short}, A.~T. and
         {Smale}, A.~P. and {Smith}, I.~A. and {Stella}, L. and
         {Tagliaferri}, G. and {Takahashi}, T. and {Tashiro}, M. and
         {Townsley}, L.~K. and {Tueller}, J. and {Turner}, M.~J.~L. and
         {Vietri}, M. and {Voges}, W. and {Ward}, M.~J. and {Willingale}, R. and
         {Zerbi}, F.~M. and {Zhang}, W.~W.},
        title = "{The Swift Gamma-Ray Burst Mission}",
      journal = {\apj},
         year = 2004,
        month = aug,
       volume = {611},
       number = {2},
        pages = {1005-1020},
          doi = {10.1086/422091},
archivePrefix = {arXiv},
       eprint = {astro-ph/0405233},
 primaryClass = {astro-ph},
       adsurl = {https://ui.adsabs.harvard.edu/abs/2004ApJ...611.1005G}
}

@ARTICLE{Last2017,
       author = {{Last}, Felix and {Douzas}, Georgios and {Bacao}, Fernando},
        title = "{Oversampling for Imbalanced Learning Based on K-Means and SMOTE}",
      journal = {arXiv e-prints},
         year = 2017,
        month = nov,
          eid = {arXiv:1711.00837},
        pages = {arXiv:1711.00837},
archivePrefix = {arXiv},
       eprint = {1711.00837},
 primaryClass = {cs.LG},
       adsurl = {https://ui.adsabs.harvard.edu/abs/2017arXiv171100837L}
}

@article{Evans2020,
	doi = {10.3847/1538-4365/ab7db9},
	url = {https://doi.org/10.3847%2F1538-4365%2Fab7db9},
	year = 2020,
	month = {apr},
	publisher = {American Astronomical Society},
	volume = {247},
	number = {2},
	pages = {54},
	author = {P. A. Evans and K. L. Page and J. P. Osborne and A. P. Beardmore and R. Willingale and D. N. Burrows and J. A. Kennea and M. Perri and M. Capalbi and G. Tagliaferri and S. B. Cenko},
	title = {2SXPS: An Improved and Expanded Swift X-Ray Telescope Point-source Catalog},
	journal = {The Astrophysical Journal Supplement Series}
}

@ARTICLE{Zyuzin2018,
       author = {{Zyuzin}, Dmitry A. and {Karpova}, Anna V. and {Shibanov}, Yuriy A.},
        title = "{X-ray counterpart candidates for six new {\ensuremath{\gamma}}-ray pulsars}",
      journal = {\mnras},
         year = 2018,
        month = may,
       volume = {476},
       number = {2},
        pages = {2177-2185},
          doi = {10.1093/mnras/sty359},
archivePrefix = {arXiv},
       eprint = {1802.03985},
 primaryClass = {astro-ph.HE},
       adsurl = {https://ui.adsabs.harvard.edu/abs/2018MNRAS.476.2177Z}
}

@article{Abdollahi2020,
	doi = {10.3847/1538-4365/ab6bcb},
	url = {https://doi.org/10.3847/1538-4365/ab6bcb},
	year = 2020,
	month = {mar},
	publisher = {American Astronomical Society},
	volume = {247},
	number = {1},
	pages = {33},
	author = {S. Abdollahi and F. Acero and M. Ackermann and M. Ajello and W. B. Atwood and M. Axelsson and L. Baldini and J. Ballet and G. Barbiellini and D. Bastieri and J. Becerra Gonzalez and R. Bellazzini and A. Berretta and E. Bissaldi and R. D. Blandford and E. D. Bloom and R. Bonino and E. Bottacini and T. J. Brandt and J. Bregeon and P. Bruel and R. Buehler and T. H. Burnett and S. Buson and R. A. Cameron and R. Caputo and P. A. Caraveo and J. M. Casandjian and D. Castro and E. Cavazzuti and E. Charles and S. Chaty and S. Chen and C. C. Cheung and G. Chiaro and S. Ciprini and J. Cohen-Tanugi and L. R. Cominsky and J. Coronado-Bl{\'{a}}zquez and D. Costantin and A. Cuoco and S. Cutini and F. D'Ammando and M. DeKlotz and P. de la Torre Luque and F. de Palma and A. Desai and S. W. Digel and N. Di Lalla and M. Di Mauro and L. Di Venere and A. Dom{\'{\i}}nguez and D. Dumora and F. Fana Dirirsa and S. J. Fegan and E. C. Ferrara and A. Franckowiak and Y. Fukazawa and S. Funk and P. Fusco and F. Gargano and D. Gasparrini and N. Giglietto and P. Giommi and F. Giordano and M. Giroletti and T. Glanzman and D. Green and I. A. Grenier and S. Griffin and M.-H. Grondin and J. E. Grove and S. Guiriec and A. K. Harding and K. Hayashi and E. Hays and J. W. Hewitt and D. Horan and G. J{\'{o}}hannesson and T. J. Johnson and T. Kamae and M. Kerr and D. Kocevski and M. Kovac'evic' and M. Kuss and D. Landriu and S. Larsson and L. Latronico and M. Lemoine-Goumard and J. Li and I. Liodakis and F. Longo and F. Loparco and B. Lott and M. N. Lovellette and P. Lubrano and G. M. Madejski and S. Maldera and D. Malyshev and A. Manfreda and E. J. Marchesini and L. Marcotulli and G. Mart{\'{\i}}-Devesa and P. Martin and F. Massaro and M. N. Mazziotta and J. E. McEnery and I. Mereu and M. Meyer and P. F. Michelson and N. Mirabal and T. Mizuno and M. E. Monzani and A. Morselli and I. V. Moskalenko and M. Negro and E. Nuss and R. Ojha and N. Omodei and M. Orienti and E. Orlando and J. F. Ormes and M. Palatiello and V. S. Paliya and D. Paneque and Z. Pei and H. Pe{\~{n}}a-Herazo and J. S. Perkins and M. Persic and M. Pesce-Rollins and V. Petrosian and L. Petrov and F. Piron and H. Poon and T. A. Porter and G. Principe and S. Rain{\`{o}} and R. Rando and M. Razzano and S. Razzaque and A. Reimer and O. Reimer and Q. Remy and T. Reposeur and R. W. Romani and P. M. Saz Parkinson and F. K. Schinzel and D. Serini and C. Sgr{\`{o}} and E. J. Siskind and D. A. Smith and G. Spandre and P. Spinelli and A. W. Strong and D. J. Suson and H. Tajima and M. N. Takahashi and D. Tak and J. B. Thayer and D. J. Thompson and L. Tibaldo and D. F. Torres and E. Torresi and J. Valverde and B. Van Klaveren and P. van Zyl and K. Wood and M. Yassine and G. Zaharijas},
	title = {Fermi Large Area Telescope Fourth Source Catalog},
	journal = {The Astrophysical Journal Supplement Series}
}

@ARTICLE{Kerby2021a,
       author = {{Kerby}, Stephen and {Kaur}, Amanpreet and {Falcone}, Abraham D. and {Stroh}, Michael C. and {Ferrara}, Elizabeth C. and {Kennea}, Jamie A. and {Colosimo}, Joseph},
        title = "{X-Ray Spectra and Multiwavelength Machine Learning Classification for Likely Counterparts to Fermi 3FGL Unassociated Sources}",
      journal = {\aj},
         year = 2021,
        month = apr,
       volume = {161},
       number = {4},
          eid = {154},
        pages = {154},
          doi = {10.3847/1538-3881/abda53},
archivePrefix = {arXiv},
       eprint = {2101.04128},
 primaryClass = {astro-ph.HE},
       adsurl = {https://ui.adsabs.harvard.edu/abs/2021AJ....161..154K}
}

@INPROCEEDINGS{Falcone2011,
       author = {{Falcone}, Abraham and {Stroh}, M. and {Ferrara}, E. and {Grove}, E. and {Saz Parkinson}, P. and {Burrows}, D. and {Cheung}, C.~C. and {Donato}, D. and {Gehrels}, N. and {Kennea}, J.},
        title = "{Systematic Search for X-ray Counterparts of Fermi-LAT Unassociated Sources Using Swift Observations}",
    booktitle = {AAS/High Energy Astrophysics Division \#12},
         year = 2011,
       series = {AAS/High Energy Astrophysics Division},
       volume = {12},
        month = sep,
          eid = {4.03},
        pages = {4.03},
       adsurl = {https://ui.adsabs.harvard.edu/abs/2011HEAD...12.0403F}
}

@INPROCEEDINGS{breeveld2011updated,
       author = {{Breeveld}, A.~A. and {Landsman}, W. and {Holland}, S.~T. and {Roming}, P. and {Kuin}, N.~P.~M. and {Page}, M.~J.},
        title = "{An Updated Ultraviolet Calibration for the Swift/UVOT}",
    booktitle = {Gamma Ray Bursts 2010},
         year = 2011,
       editor = {{McEnery}, J.~E. and {Racusin}, J.~L. and {Gehrels}, N.},
       series = {American Institute of Physics Conference Series},
       volume = {1358},
        month = aug,
    publisher = {AIP},
        pages = {373-376},
          doi = {10.1063/1.3621807},
archivePrefix = {arXiv},
       eprint = {1102.4717},
 primaryClass = {astro-ph.IM},
       adsurl = {https://ui.adsabs.harvard.edu/abs/2011AIPC.1358..373B}
}

@ARTICLE{Bessell1998,
       author = {{Bessell}, M.~S. and {Castelli}, F. and {Plez}, B.},
        title = "{Model atmospheres broad-band colors, bolometric corrections and temperature calibrations for O - M stars}",
      journal = {\aap},
         year = 1998,
        month = may,
       volume = {333},
        pages = {231-250},
       adsurl = {https://ui.adsabs.harvard.edu/abs/1998A&A...333..231B}
}

@ARTICLE{Matplotlib,
       author = {{Hunter}, John D.},
        title = "{Matplotlib: A 2D Graphics Environment}",
      journal = {Computing in Science and Engineering},
         year = 2007,
        month = may,
       volume = {9},
       number = {3},
        pages = {90-95},
          doi = {10.1109/MCSE.2007.55},
       adsurl = {https://ui.adsabs.harvard.edu/abs/2007CSE.....9...90H}
}

@INPROCEEDINGS{FTOOLS,
       author = {{Blackburn}, J.~K.},
        title = "{FTOOLS: A FITS Data Processing and Analysis Software Package}",
    booktitle = {Astronomical Data Analysis Software and Systems IV},
         year = 1995,
       editor = {{Shaw}, R.~A. and {Payne}, H.~E. and {Hayes}, J.~J.~E.},
       series = {Astronomical Society of the Pacific Conference Series},
       volume = {77},
        month = jan,
        pages = {367},
       adsurl = {https://ui.adsabs.harvard.edu/abs/1995ASPC...77..367B}
}
\bibliographystyle{aasjournal}


\begin{longrotatetable}
\begin{deluxetable*}{ccccccccccccc}
     \tablecaption{Parameters for singlet X-ray sources detected with \Swift-XRT within 4FGL-DR4 unassociated source uncertainty ellipses. Data is acquired using publicly available \Fermi-LAT, \Swift-XRT, and \Swift-UVOT data. Gamma-ray fluxes are from 100~MeV to 100~GeV \citep[	exttt{Energy\_Flux100;}][]{Abdollahi2020}. Derivation of X-ray (unabsorbed X-ray flux, $F_X$, and X-ray photon index, $\alpha_X$) and UV/optical ($V$-band magnitude, $m_V$) parameters are described in Sec.~\ref{sec:Analysis}. UVOT filter describes the filter used to calculate $m_V$. If a source has UVOT observations but no detections, a limiting magnitude is calculated for $m_V$ using the labeled filter's observation. If no data is given for filter or $m_V$, the source lacks UVOT observations. SwXF4 describes the alternative alias of the X-ray sources described in \citet{kerby21b} where a dash represents a new X-ray source. Full machine-readable tables are included. \label{tab:xrt_single}}
     \tablewidth{\columnwidth}
     \tabletypesize{\scriptsize}
\tablehead{
     \colhead{Target} & \colhead{$\log(F_\gamma)$} & \colhead{$\Gamma_\gamma$} & \colhead{Vari. Ind.} & \colhead{$\mathrm{Sig_{curv}}$} & \colhead{XRT Source} & \colhead{UVOT Source} & \colhead{S/N} & \colhead{$\log(F_X)$} & \colhead{$\alpha_X$} & \colhead{Filter} & \colhead{$m_V$} & \colhead{SwXF4} \\ 
     \colhead{4FGL} & \colhead{$[\mathrm{erg~s^{-1}cm^{-2}}]$} &  & \colhead{log} & \colhead{log} & \colhead{SwXF4-DR4} & \colhead{SwUVF4-DR4} & \colhead{XRT} & \colhead{$[\mathrm{erg~s^{-1}cm^{-2}}]$} &  & \colhead{UVOT} &  & \colhead{\citealt{kerby21b}}
     }
     \startdata \hline
J0004.4-4001 & -11.60 & 2.24 & 1.59 & --- & J000434.3-400032 & J000434.2-400035 & 8.62 & -12.40 & 2.48 & uu & 17.20 & J000434.1-400036 \\ \hline
J0022.0-5921 & -11.45 & 2.42 & 1.30 & --- & J002127.6-591947 & J002127.5-591948 & 4.65 & -12.58 & 2.79 & uu & 19.17 & --- \\ \hline
J0023.6-4209 & -12.02 & 2.15 & 1.45 & --- & J002303.6-420508 & J002303.8-420509 & 5.23 & -11.94 & 1.46 & uu & 17.06 & --- \\ \hline
J0025.4-4838 & -11.72 & 2.37 & 1.40 & --- & J002536.6-483807 & J002536.9-483810 & 6.79 & -12.27 & 2.34 & uu & 19.92 & J002536.8-483808 \\ \hline
J0026.1-0732 & -11.68 & 1.89 & 1.10 & 0.14 & J002611.7-073116 & J002611.6-073115 & 21.48 & -11.39 & 2.17 & uu & 19.50 & J002611.9-073115 \\ \hline
J0027.0-1134 & -12.11 & 1.82 & 1.17 & --- & J002710.0-113639 & J002710.1-113638 & 5.79 & -11.91 & 2.08 & uu & 17.62 & --- \\ \hline
J0031.5-5648 & -11.83 & 1.90 & 1.25 & 0.20 & J003135.7-564639 & J003135.4-564641 & 6.02 & -12.37 & 2.35 & vv & 19.36 & J003135.1-564640 \\ \hline
J0037.2-2653 & -11.87 & 2.67 & 0.98 & --- & J003729.6-265043 & J003729.5-265045 & 4.83 & -12.30 & 1.65 & uu & 19.55 & J003729.6-265043 \\
\enddata
  \end{deluxetable*}
\end{longrotatetable}
\begin{longrotatetable}
\begin{deluxetable*}{ccccccccccccc}
     \tablecaption{Parameters for X-ray sources detected with \Swift-XRT within 4FGL-DR4 unassociated source uncertainty ellipses. These X-ray sources are not the only X-ray sources within the 4FGL uncertainty ellipse. Data is acquired using publicly available \Fermi-LAT, \Swift-XRT, and \Swift-UVOT data. Gamma-ray fluxes are from 100~MeV to 100~GeV \citep[\texttt{Energy\_Flux100;}][]{Abdollahi2020}. Derivation of X-ray and UV/optical parameters are described in Sec.~\ref{sec:Analysis}. UVOT filter describes the filter used to calculate $m_V$. If a source has UVOT observations but no detections, a limiting magnitude is calculated for $m_V$ using the labeled filter's observation. If no data is given for filter or $m_V$, the source lacks UVOT observations. SwXF4 describes the alternative alias of the X-ray sources described in \citet{kerby21b} where a dash represents a new X-ray source. Full machine-readable tables are included. \label{tab:xrt_multi}}
     \tablewidth{\columnwidth}
     \tabletypesize{\scriptsize}
\tablehead{
     \colhead{Target} & \colhead{$\log(F_\gamma)$} & \colhead{$\Gamma_\gamma$} & \colhead{Vari. Ind.} & \colhead{$\mathrm{Sig_{curv}}$} & \colhead{XRT Source} & \colhead{UVOT Source} & \colhead{S/N} & \colhead{$\log(F_X)$} & \colhead{$\alpha_X$} & \colhead{Filter} & \colhead{$m_V$} & \colhead{SwXF4} \\ 
     \colhead{4FGL} & \colhead{$[\mathrm{erg~s^{-1}cm^{-2}}]$} &  & \colhead{log} & \colhead{log} & \colhead{SwXF4-DR4} & \colhead{SwUVF4-DR4} & \colhead{XRT} & \colhead{$[\mathrm{erg~s^{-1}cm^{-2}}]$} &  & \colhead{UVOT} &  & \colhead{\citealt{kerby21b}}
     }
     \startdata \hline
J0017.1-4605 & -11.67 & 2.87 & 1.13 & --- & J001708.6-460607 & J001708.6-460606 & 7.20 & -12.43 & 1.74 & uu & 18.08 & J001708.5-460612 \\
 &  &  &  &  & J001750.9-460440 & J001750.6-460438 & 5.14 & -12.32 & 0.73 & uu & 18.16 & --- \\ \hline
J0040.2-2725 & -11.95 & 2.00 & 0.78 & --- & J003954.2-272516 & --- & 4.40 & -12.98 & 0.83 & vv & $>20.05$ & --- \\
 &  &  &  &  & J004023.7-272254 & --- & 6.10 & -12.66 & 0.46 & m2 & 21.29 & --- \\ \hline
J0045.8-1324 & -12.01 & 1.92 & 1.12 & --- & J004539.4-132509 & --- & 4.45 & -13.11 & 1.70 & uu* & $>22.15$ & --- \\
 &  &  &  &  & J004602.8-132420 & J004602.8-132422 & 7.39 & -12.72 & 2.08 & uu & 18.91 & --- \\ \hline
J0126.3-6746 & -12.09 & 1.82 & 1.11 & 0.27 & J012610.9-674744 & J012610.8-674743 & 6.73 & -12.33 & 2.53 & uu & 18.78 & J012612.0-674746 \\
 &  &  &  &  & J012621.9-674624 & --- & 4.19 & -12.64 & 1.96 & uu & 20.61 & J012621.6-674638 \\
\enddata
  \end{deluxetable*}
\end{longrotatetable}

\startlongtable
\begin{deluxetable*}{ccccccccc}
\tablecaption{Neural network classification of X-ray singlets utilizing the blazar probability parameter, $P_\mathrm{bzr}$. For X-ray sources with a $V$-magnitude lower limit, $P_\mathrm{lim}$ represents the classification results if $m_V = 30$. X-ray sources shared with \citet{kerby21b} have blazar probabilities from said study, $P_\mathrm{K21}$. Gamma-ray sources flagged as SGUs in 4FGL-DR4 \citep{ballet23} are given the SGU flag. Potential SIMBAD counterparts are listed with their offset in arcseconds and literature classification. X-ray sources with multiple potential SIMBAD counterparts within 5\arcsec\xspace are marked with an asterisk, *, on their SIMBAD name. The table is separated into likely pulsars, likely blazars, and ambiguous sources from our classification. \label{tab:class_single}}
\tablewidth{0pt}
\tablehead{
\colhead{4FGL} & \colhead{SGU} & \colhead{SwXF4-DR4} & \colhead{$P_\mathrm{bzr}$} & \colhead{$P_\mathrm{lim}$} & \colhead{$P_\mathrm{K21}$} & \colhead{SIMBAD Name} & \colhead{Offset (\arcsec)} & \colhead{Classification}
}
     \startdata
\hline
    \multicolumn{9}{c}{Likely Pulsars $(P_\mathrm{bzr} < 0.01)$} \\ \hline
\hline 
J0752.0-2931 &  & J075209.9-293013 & 0.0001 & --- & --- & --- & --- & --- \\ \hline
J0844.9-4117 & * & J084408.8-411912 & 0.0002 & --- & --- & UCAC4 244-028737 & 2.84 & YSO Candidate \\ \hline
J1357.3-6123 & * & J135657.5-612314 & 0.0003 & --- & 0.9985 & 2MASS J13565717-6123149 & 0.26 & LongPeriodV* Candidate \\ \hline
J1429.8-0739 &  & J142949.4-073303 & 0.0029 & --- & 0.9901 & LEDA 1017101 & 0.70 & Galaxy \\ \hline
J1612.1+1407 &  & J161205.0+140908 & 0.0000 & --- & --- & --- & --- & --- \\ \hline
J1700.2-4237c & * & J165956.2-423019 & 0.0010 & 0.9976 & 0.0205 & --- & --- & --- \\ \hline
J1733.4+2235 &  & J173330.3+223614 & 0.0044 & --- & 0.5335 & --- & --- & --- \\
\enddata

\end{deluxetable*}

\startlongtable
\begin{deluxetable*}{ccccccccc}
\tablecaption{Neural network classification of X-ray multiplets utilizing the blazar probability parameter, $P_\mathrm{bzr}$. For X-ray sources with a $V$-magnitude lower limit, $P_\mathrm{lim}$ represents the classification results if $m_V = 30$. X-ray sources shared with \citet{kerby21b} have blazar probabilities from said study, $P_\mathrm{K21}$. Potential SIMBAD counterparts are listed with their offset in arcseconds and literature classification. X-ray sources with multiple potential SIMBAD counterparts within 5\arcsec\xspace are marked with an asterisk, *, on their SIMBAD name. The table is organized by 4FGL source names. \label{tab:class_multi}}
\tablewidth{0pt}
\tablehead{
\colhead{4FGL} & \colhead{SGU} & \colhead{SwXF4-DR4} & \colhead{$P_\mathrm{bzr}$} & \colhead{$P_\mathrm{lim}$} & \colhead{$P_\mathrm{K21}$} & \colhead{SIMBAD Name} & \colhead{Offset (\arcsec)} & \colhead{Classification}
}
     \startdata
\hline
J0017.1-4605 &  & J001708.6-460607 & 1.0000 & --- & 0.9993 & --- & --- & --- \\
 &  & J001750.9-460440 & 1.0000 & --- & --- & --- & --- & --- \\ \hline
J0040.2-2725 &  & J003954.2-272516 & 1.0000 & 1.0000 & --- & [VV2006] J003954.1-272514 & 2.43 & QSO \\
 &  & J004023.7-272254 & 1.0000 & --- & --- & --- & --- & --- \\ \hline
J0045.8-1324 &  & J004539.4-132509 & 1.0000 & 1.0000 & --- & --- & --- & --- \\
 &  & J004602.8-132420 & 1.0000 & --- & --- & --- & --- & --- \\ \hline
J0126.3-6746 &  & J012610.9-674744 & 1.0000 & --- & 0.9991 & --- & --- & --- \\
 &  & J012621.9-674624 & 1.0000 & --- & 0.9968 & --- & --- & --- \\
\enddata

\end{deluxetable*}

\startlongtable
\begin{deluxetable*}{ccccccc}
\tablecaption{Table of potential radio counterparts to X-ray singlets. X-ray sources are listed with their estimated $V$-band magnitudes. Radio counterparts are listed with their survey name, offset in arcseconds, and radio flux density. Radio-loudness parameter, $R$, calculated with optical flux density in units of $\mathrm{erg~s^{-1}~cm^{-2}~Hz^{-1}}$. \label{tab:radio_single}}
\tablewidth{0pt}
\tablehead{
\colhead{4FGL} & \colhead{SwXF4-DR4} & \colhead{$m_V$} & \colhead{Radio Source} & \colhead{Offset (\arcsec)} & \colhead{$F_\mathrm{Rad}$ (mJys)} & \colhead{$R$}
}
     \startdata
\hline
J0004.4-4001 & J000434.3-400032 & 17.20 & VLASS1QLCIR J000434.19-400033.6 & 1.57 & 32.45 & 67 \\ \hline
J0022.0-5921 & J002127.6-591947 & 19.17 & RACSJ002127.4-591949 & 1.57 & 93.43 & 1200 \\ \hline
J0023.6-4209 & J002303.6-420508 & 17.06 & --- & --- & --- & --- \\ \hline
J0025.4-4838 & J002536.6-483807 & 19.92 & RACSJ002536.8-483810 & 0.48 & 4.98 & 128 \\ \hline
J0026.1-0732 & J002611.7-073116 & 19.50 & VLASS1QLCIR J002611.63-073115.5 & 0.26 & 6.41 & 111 \\ \hline
J0027.0-1134 & J002710.0-113639 & 17.62 & VLASS1QLCIR J002710.12-113637.5 & 0.81 & 19.88 & 61 \\ \hline
J0031.5-5648 & J003135.7-564639 & 19.36 & RACSJ003135.4-564641 & 0.37 & 7.27 & 111 \\ \hline
J0037.2-2653 & J003729.6-265043 & 19.55 & RACSJ003729.6-265045 & 2.19 & 2.67 & 49 \\
\enddata

\end{deluxetable*}

\startlongtable
\begin{deluxetable*}{ccccccc}
\tablecaption{Table of potential radio counterparts to X-ray multiplets. X-ray sources are listed with their estimated $V$-band magnitudes. Radio counterparts are listed with their survey name, offset in arcseconds, and radio flux density. Radio-loudness parameter, $R$, calculated with optical flux density in units of $\mathrm{erg~s^{-1}~cm^{-2}~Hz^{-1}}$. \label{tab:radio_multi}}
\tablewidth{0pt}
\tablehead{
\colhead{4FGL} & \colhead{SwXF4-DR4} & \colhead{$m_V$} & \colhead{Radio Source} & \colhead{Offset (\arcsec)} & \colhead{$F_\mathrm{Rad}$ (mJys)} & \colhead{$R$}
}
     \startdata
\hline
J0017.1-4605 & J001708.6-460607 & 18.08 & --- & --- & --- & --- \\
 & J001750.9-460440 & 18.16 & --- & --- & --- & --- \\ \hline
J0040.2-2725 & J003954.2-272516 & $>20.05$ & --- & --- & --- & --- \\
 & J004023.7-272254 & 21.29 & --- & --- & --- & --- \\ \hline
J0045.8-1324 & J004539.4-132509 & $>22.15$ & VLASS1QLCIR J004539.47-132508.0 & 2.12 & 1.72 & $>344$ \\
 & J004602.8-132420 & 18.91 & VLASS1QLCIR J004602.82-132422.1 & 0.10 & 8.15 & 82 \\ \hline
J0126.3-6746 & J012610.9-674744 & 18.78 & RACSJ012610.7-674743 & 0.48 & 8.39 & 75 \\
 & J012621.9-674624 & 20.61 & --- & --- & --- & --- \\
\enddata

\end{deluxetable*}

\begin{deluxetable*}{lccl}
\tablecaption{Descriptive table for supplementary machine-readable tables. \label{tab:mrt}}
\tablewidth{0pt}
\tablehead{
\lcolhead{Label} & \colhead{Format} & \colhead{Units} & \lcolhead{Description}
}
     \startdata
RA & F7.3 & deg & Right ascension$^{a}$ \\ 
Dec & F7.3 & deg & Declination$^{a}$ \\ 
4FGL & A13 &  & Target 4FGL identifier \\ 
logFG & F6.2 & [$\mathrm{erg/s/cm^2}$] & $\log F_\gamma$, Gamma-ray flux from 0.1-100 GeV flux$^{b}$ \\ 
dlogFG & F4.2 & [$\mathrm{erg/s/cm^2}$] & $\log F_\gamma$ uncertainty$^{b}$ \\ 
PL\_Index & F5.3 &  & $\Gamma_\gamma$, Gamma-ray photon index when fitting with PowerLaw$^{b}$ \\ 
Variability\_Index & F7.3 &  & Year-to-year gamma-ray variability$^{b}$ \\ 
PLEC\_SigCurv & F5.3 &  & Curvature significance of PLEC model$^{b}$ \\
\enddata
\tablecomments{\textbf{Notes} \\
\noindent $^a$ XRT Source RA and Declination are given in J2000 epoch. \\
\noindent $^b$ Parameters are from the data columns in 4FGL-DR4 (Ballet+ 23). \\
\noindent $^c$ Background-subtracted X-ray counts of XRT source. \\
\noindent $^d$ Magnitude uncertainties of -1.00 are detection-limiting magnitudes. \\
\noindent $^e$ Filters with an asterik used an optical spectral slope of 1.5. \\
\noindent $^f$ nan values imply no viable overlapping observation of target. \\
\noindent $^g$ nan values imply UVOT source was detected. \\
\noindent $^h$ Sources marked with an asterisk have multiple nearby potential sources. \\
\noindent $^i$ Based on VLASS and RACS catalogs with preference for VLASS if source detected in both. \\
\noindent $^j$ VLASS band is 2 - 4 GHz, and RACS band is 0.75 - 1.5 GHz. \\
}
\end{deluxetable*}



\end{document}